\documentclass[11pt]{article}
\usepackage{jheppubmod}
\usepackage[showonlyrefs]{mathtools}
\usepackage{psfrag}
\usepackage{array}
\usepackage{amssymb}
\usepackage{graphicx}
\usepackage{caption}
\usepackage[labelsep=quad]{subcaption}
\usepackage{epstopdf}
	
\usepackage{epsfig}
\usepackage[punctsep]{collref}
\collectsep[]{;}	
	
\def\cn{{\mathcal N}}
\def\tr{{\rm Tr}}

\def\wa{\widetilde{a}}

\def\ep{\epsilon}
\def\lvac{\langle 0 |}
\def\rvac{|0 \rangle}

\def\Or[#1]{{\text{O}}\left({#1}\right)}
\def\dotl[#1,#2]{\left\langle #1,\, #2 \right\rangle}
\def\dotlb[#1,#2]{\left\langle #1,\, #2 \right\rangle}
\def\dotlm[#1,#2]{\left[ #1,\, #2 \right]}
\def\dotp[#1,#2]{(\vect{#1} \cdot\vect{#2})}
\def\aff[#1,#2]{\hat{#1}(#2)}
\def\n4sym{{\cal N}=4 SYM}
\def\>{\rangle}
\def\<{\langle}
\def\weight[#1,#2,#3]{\{(#1),#2,#3\}}
\def\ads[#1]{$\text{AdS}_{#1}$}
\def\cft[#1]{$\text{CFT}_{#1}$}
\def\front{\rm I}
\def\black{\rm II}
\def\other{\rm III}
\def\white{\rm IV}

\newcommand{\be}{\begin{equation}}
\newcommand{\ee}{\end{equation}}
\newcommand{\ba}{\begin{align}}
\newcommand{\ea}{\end{align}}
\newcommand{\bs}{\begin{split}}
\def\sess\end{split}
\def\zhor{z_0}
\newcommand{\hatbb}[1]{\hat{#1}}
\newcommand{\vect}[1]{{\boldsymbol{#1}}}
\newcommand{\norm}[1]{|{\boldsymbol{#1}}|}

\newcommand{\state}[1]{{|#1\rangle}}

\newcommand{\normfact}{\Upsilon}
\title{An Infalling Observer in AdS/CFT}
\author[a,b]{Kyriakos Papadodimas}
\author[c,d,e]{and Suvrat Raju}
\affiliation[a]{Centre for Theoretical Physics, University of Groningen, Nijenborgh 4, The Netherlands.}
\affiliation[b]{Theory Group, Physics Department, CERN, CH-1211 Geneva 23,
Switzerland.}
\affiliation[c]{International Centre for Theoretical Sciences, IISc Campus, Bengaluru 560012, India.}
\affiliation[d]{Harish-Chandra Research Institute, Chatnag Marg, Jhunsi,
Allahabad 211019, India.}
\affiliation[e]{School of Natural Sciences, Institute for Advanced Study,
Princeton, NJ 08540, USA.}
\emailAdd{k.papadodimas@rug.nl}
\emailAdd{suvrat@icts.res.in}

\date{}

\abstract{We describe the experience of an observer falling into a
  black hole using the AdS/CFT correspondence.  In order to do this,  we
  reconstruct the local bulk operators measured by the observer along
  his trajectory outside the black hole. We then extend our
  construction beyond the black hole horizon. We show that this is
 possible because of an effective doubling of the observables in the
  boundary theory, when it is in a pure state that is close to the
  thermal state. Our construction allows us to rephrase questions
  about information-loss
and the structure of the metric at the horizon in terms of more
familiar CFT correlators. It suggests that to precisely
identify black-hole microstates, the observer would need to conduct
measurements to an accuracy of $e^{-S_{\text{BH}}}$. This appears to be
inconsistent with the ``fuzzball'' proposal, and other recent proposals
in which pure states in the ensemble of the black hole are represented
by macroscopically distinct geometries. Furthermore, our description of the black hole interior
in terms of CFT operators provides a natural realization of
black hole complementarity and a method of preserving
unitarity without ``firewalls.''
}
\preprint{\parbox{3cm}{HRI/ST/1107 \\  ICTS/2012/10}}
\begin{document}
\maketitle

\section{Introduction}
Even though quantum gravity has attracted theoretical interest for decades, several aspects of the theory continue to be actively debated. This includes, among other questions, the issue of how to define ``local operators'' in a theory of quantum gravity. Where black holes are involved, the situation seems to be even more puzzling. What is the nature of spacetime behind the horizon of the black hole? What about the horizon itself? Even though the principle of equivalence suggests that there is nothing special about the horizon of a large black hole, there have been several speculations that quantum gravity effects cause the interior to be modified into a fuzzball \cite{Mathur:2012np,Mathur:2009hf,Mathur:2008nj,Lunin:2002bj,Lunin:2001fv}, and more recently, that the horizon of an ``old black hole'' 
is replaced by a firewall \cite{Almheiri:2012rt}. (See also \cite{Bousso:2012as, Nomura:2012sw, Mathur:2012jk, Susskind:2012rm, Bena:2012zi, Giveon:2012kp, Banks:2012nn, Ori:2012jx, Brustein:2012jn, Susskind:2012uw, Marolf:2012xe,  Hossenfelder:2012mr, Nomura:2012cx, Hwang:2012nn, Larjo:2012jt,braunstein2009v1, Avery:2012tf, Chowdhury:2012vd}.)  These latter proposals originate, not from direct calculations in quantum gravity, but rather in arguments that the information 
paradox (in various incarnations) cannot be solved without modifying the geometry at, or behind the horizon.

While quantum gravity is a mysterious subject, the AdS/CFT correspondence \cite{Maldacena:1997re} provides us with a setting where we can examine these ideas within a perfectly well defined theory. To this end, in this paper we would like to examine the following conceptual questions:
\begin{itemize}
\item
Is it possible to describe the results of local experiments in AdS, at least within perturbation theory in ${1 \over N}$, using the boundary field theory?
\item
In the presence of a black hole, can we also describe local experiments {\em behind} the black hole horizon in the boundary theory?
\item
Does our construction of the degrees of freedom behind the horizon shed any light on the information puzzle?
\end{itemize}
More precisely, we will imagine that within anti-de Sitter space, we start with some matter that then collapses to form a large black hole. We will then try and reconstruct local operators outside this black hole, and behind the horizon. 

In fact, if we imagine an observer who lives outside this black hole for a while and then dives in, then we would require answers to all the questions above to describe his experience; this is the reason for the title of our paper.

We will consider some strongly coupled CFT in $d$ dimensions, which has the properties that would allow it to have a bulk dual, without restricting ourselves to any specific example of the AdS/CFT correspondence. We will place the CFT in a pure initial state that thermalizes after a while i.e. it evolves to a state that is almost indistinguishable from a thermal state.  In this state, we will then show how to reorganize all the operators that are accessible in the CFT to a low-energy observer, into fields
 that are labeled by points in the semi-classical geometry of a big black hole in \ads[d+1]. We emphasize that these are still CFT operators, although rather than being labeled by a boundary point, they are labeled by a bulk point. We will further show, to lowest order in the ${1 \over N}$ expansion and argue to higher orders, that the correlators of the operators that we have constructed are the same as the correlators of perturbative fields on this geometry. Second, we will 
push this construction {\em past the horizon} and show how to construct perturbative fields behind the horizon in terms of CFT operators. This will give us important clues about how to resolve the firewall paradox, as we explore in section \ref{sec:applications}. 

Our construction follows important work by several other authors \cite{Banks:1998dd, Balasubramanian:1999ri, Bena:1999jv, Hamilton:2006fh,Hamilton:2006az,Hamilton:2005ju,Hamilton:2007wj,VanRaamsdonk:2009ar,VanRaamsdonk:2010pw,VanRaamsdonk:2011zz,Czech:2012bh,Heemskerk:2012mn}. However, as we describe in more detail below, we have been able to make some technical improvements on the construction of holographic operators outside the horizon. Our construction of the black hole interior in terms of CFT data is new, and to our knowledge has not been explored before.  

We now quickly summarize our results. Before considering the black hole, we start by considering empty AdS. To construct local operators in this space, we consider a generalized free-field \cite{ElShowk:2011ag} ${\cal O}(t, \vect{x})$ when the CFT is in the vacuum. Working with the modes ${\cal O}_{\omega,\vect{k}}$ of ${\cal O}(t, \vect{x})$ in  momentum space, we are able to write a CFT operator that is labeled by a point in the bulk of the AdS Poincare patch
\be
\label{cftpure}
\phi_{\text{CFT}} (t,\vect{x}, z) = \int_{\omega>0} {d \omega d^{d-1} \vect{k} \over (2 \pi)^d} \,\left[ {\cal O}_{\omega, \vect{k}} \xi_{\omega, \vect{k}}(t, \vect{x}, z) + \text{h.c}.\right]
\ee
When the mode functions $\xi$ are appropriately chosen, the operator on the left has the same correlators as a free-field propagating in \ads[d+1]: for example, its commutator at two points that are spacelike separated in \ads[d+1] vanishes. We show how these operators can be continued beyond the Poincare patch onto all of global AdS. This serves as a warm up for our next task of looking beyond the black hole horizon. 

We then consider generalized free-fields ${\cal O}(t,\vect{x})$, but in a CFT state that, although pure, is ``close'' to the thermal state. We will refer to this state as $\state{\Psi}$ below. Now, we find that we have to write
\be
\label{cftthermal}
\phi_{\text{CFT}} (t,\vect{x}, z) = \int_{\omega>0} {d \omega  d^{d-1} \vect{k} \over (2 \pi)^d}\,\left[ {\cal O}_{\omega, \vect{k}} f_{\omega, \vect{k}}(t, \vect{x}, z) + \text{h.c}.\right]
\ee
Although \eqref{cftthermal} looks deceptively similar to \eqref{cftpure}, there are several differences. First, the mode functions $f$ are different from the ones that we encountered above. Another important difference is that, while in \eqref{cftpure}, we can set the mode functions for all cases where $\omega^2 < \vect{k}^2$ to zero, we cannot do so in \eqref{cftthermal}. Nevertheless, with $f$ chosen appropriately, \eqref{cftthermal} gives a good  and local description of fields in front of the black hole horizon.

Next, we point out that in this pure state $\state{\Psi}$, after it has settled down, for each such operator ${\cal O}$, there must necessarily exist operators $\widetilde{\cal O}$ that have the properties that they (a) commute with ${\cal O}$ and (b) that, in the state $\state{\Psi}$, measurements of $\widetilde{\cal O}$ are completely (anti)-correlated with measurements of ${\cal O}$. (We make this more precise in section \ref{sec:behind}.) For us, these operators $\widetilde{\cal O}$ play the role that operators in the ``second copy'' of the CFT would have played, had we been dealing with an eternal black hole.  Using these operators $\widetilde{\cal O}$ we now construct operators behind the horizon: 
\[ 
\phi_{\text{CFT}}(t,\vect{x},z) =
\int_{\omega>0} {d\omega d^{d-1}\vect{k}  \over (2 \pi)^d}\, \left[ {\cal O}_{\omega,\vect{k}} g_{\omega,\vect{k}}^{(1)}(t,\vect{x},z) + \widetilde{\cal O}_{\omega,\vect{k}} g_{\omega,\vect{k}}^{(2)}(t,\vect{x},z)+ \text{h.c.}
\right]
 \]
where $g^{(1)}$ and $g^{(2)}$ are again functions that are chosen to make 
this operator local.  

Our construction is perfectly regular as we cross the horizon. This appears to be in contradiction with both the fuzzball and the firewall proposals.  In section \ref{sec:applications} we first show that it is not possible to pinpoint the microstate of the CFT by measuring correlators of light operators to any given fixed order in the ${1 \over N}$ expansion. Our construction then implies that by doing experiments that are limited to some finite order in ${1 \over N}$, either at low or high energies, the bulk observer cannot distinguish the microstates of the black hole. Since the fuzzball proposal solves the information paradox by postulating that the observer can detect the microstate by doing ``low energy experiments'', our proposal appears to be inconsistent with this resolution.

Turning to the firewall paradox, in this paper we provide only indirect evidence for the absence of firewalls at the horizon. This is because our construction works in detail for a big black hole that does not evaporate (except over the Poincare recurrence time), and so this leaves us with the theoretical possibility that small black holes in AdS could have firewalls near the horizon. However, our description of the degrees of freedom in the interior of the black hole also provides us with several lessons that we can use to understand the information paradox that appear to make firewalls superfluous. 

In particular, our construction can, roughly, be interpreted as showing how, if we expand our space of observables to include operators that give us finer and finer information about the CFT microstate, then eventually we reach a stage where the operators behind the horizon are no longer independent of 
the operators in front of the horizon. Such a description provides a natural realization of black-hole ``complementarity'' \cite{'tHooft:1990fr,Susskind:1993if} and, as we explore below, removes the necessity of firewalls.

We should also point out that while this might naively imply a violation of causality, this causality violation is visible only when we measure very high point correlators of operators (where the number of insertions scales with $N$) or equivalently measure a low point correlator to exponential accuracy in $N$. At this level of accuracy, we do not believe that a semi-classical spacetime description makes sense at all and so this putative causality-violation is not a cause for concern.

There have been many earlier attempts to study the interior of the black hole using AdS/CFT and the literature on the information paradox is vast. Besides the papers  mentioned above, a very incomplete list of references which were relevant to our work include  \cite{Kiem:1995iy,Balasubramanian:1999zv,Giddings:2001pt,Maldacena:2001kr,Hubeny:2002dg,Kraus:2002iv,Levi:2003cx,Fidkowski:2003nf,Barbon:2003aq,Kaplan:2004qe,Balasubramanian:2004zu,Festuccia:2005pi,Balasubramanian:2005mg,
Balasubramanian:2005qu,Festuccia:2006sa,Balasubramanian:2007qv,Marolf:2008mf,Balasubramanian:2008da,deBoer:2009un,Horowitz:2009wm,Balasubramanian:2011dm,Avery:2011nb,Simon:2011zza}. 

A brief overview of this paper is as follows.  In section \ref{sec:emptyads}, we describe the construction of local operators in empty AdS using boundary operators in a flat space CFT in a pure state.  We work in the Poincare patch and then show how our construction can be continued past the Poincare horizon into global AdS.  In section \ref{sec:outside}, we repeat this construction in front of the horizon of a big black hole in AdS. In section \ref{sec:behind}, we push this construction past the black-hole horizon and write down local bulk operators behind the horizon in terms of CFT operators.  In section \ref{sec:applications}, we discuss the implications of these results, with a particular view to the information paradox and various unconventional proposals for resolving it --- including the ``firewall'' paradox and the fuzzball proposal. We also discuss a qubit toy model that is surprisingly effective at realizing many of these ideas. In section \ref{sec:subtleties}, we address some common worries 
and 
show that our construction is not destabilized by tiny effects like 
Poincare recurrence on the boundary. We also describe schematically how it may be extended to higher orders in the ${1 \over N}$ expansion, and discuss some other subtleties. We conclude in section \ref{sec:conclusion}. The appendices provide some technical details and also examine the specific case of two-dimensional thermal correlators and the BTZ-black-hole background.

%%% Local Variables: 
%%% mode: latex
%%% TeX-master: "infalling_paper"
%%% End: 

\section{Reconstructing local bulk observables in empty AdS \label{sec:emptyads}}
In this section, we develop the methodology of constructing local 
operators from boundary operators in empty AdS. 

Consider an \ads[d+1]/\cft[d] duality, where we have some {\em generalized free fields} ${\cal O}$ that live on the boundary. By {\em generalized free fields}, we mean that the correlators of ${\cal O}$ factorize to leading order in some parameter, which we denote by ${1 \over N}$.\footnote{In this paper, by $N$ we are not referring to the rank of the gauge-group on the boundary. The reader may prefer to think of $N$ as the coefficient of the two-point function of the stress tensor 
 (which actually scales like $K^2$ in the ${\cn = 4}$ SYM theory with gauge group SU(K)) but more generally, $N$ can be taken
to any parameter that controls the factorization of correlators.}
\be
\label{factorization}
\lvac {\cal O}(\vect{x_1}) \ldots {\cal O}(\vect{x_{2n}}) \rvac= {1 \over 2^n} \sum_{\pi} \lvac{\cal O}(\vect{x_{\pi_1}}) {\cal O}(\vect{x_{\pi_2}}) \rvac \ldots \lvac {\cal O}(\vect{x_{\pi_{2n-1}}}) {\cal O}(\vect{x_{\pi_{2n}}})\rvac + \ldots,
\ee
where $\pi$ runs over all permutations and the dots denote terms subleading in $1/N$.

Even though the operators ${\cal O}$ look deceptively simple, they are, in fact, rather complicated as Heisenberg operators. For example, as we will discuss below, the same Heisenberg operators have rather different properties about a thermal state: for example, their  
2-point function will look very different from the 2-point function about the vacuum. This is because there are ${1 \over N}$ effects that we have neglected in \eqref{factorization} that become important in a thermal state where the energy density scales with some power of $N$. 

However, for now, we turn to the properties of these operators in the vacuum of the CFT, since that is what is pertinent to empty AdS.
\subsection{Properties of generalized free field modes about the vacuum}
\label{subsec:freevac}
It will be convenient for us to work in Fourier space, as opposed to the position space prescriptions of \cite{Banks:1998dd, Bena:1999jv, Hamilton:2005ju, Hamilton:2006fh,Hamilton:2006az,Hamilton:2007wj,Heemskerk:2012mn,Bousso:2012mh}.
It will also be convenient to work with correlators where we choose a prior ordering, as opposed to the more commonly considered time-ordered correlators for reasons that will become apparent shortly. 

We define the Fourier modes of ${\cal O}(x)$ as usual
\be
\label{okdef}
{\cal O}_{\omega,\vect{k}} =  \int dt d^{d-1}
\vect{x} \,\,{\cal O}(t,\vect{x}) \,\,e^{i\omega t - i\vect{k} \cdot \vect{x}} .
\ee
Here by the boldface $\vect{k},\vect{x}$ we denote the $(d-1)$-dimensional spatial components of the corresponding $d$-vectors. We are working in signature mostly plus. Let us determine a few properties of these modes. Since the only correlators at leading order in ${1 \over N}$ are 2-point functions, we can analyze the properties of these Fourier modes by studying the Wightman functions. 

Let us first point out the advantage of Wightman functions --- we use
this term synonymously with correlators where we pick an ordering ahead of time ---  over time-ordered correlators. The time-ordered correlator is defined by
\[
\lvac  T\{{\cal O}(t, \vect{x}) {\cal O}(t', \vect{x'}\}) \rvac = \theta(t - t') \lvac {\cal O}(t, \vect{x}) {\cal O}(t', \vect{x'}) \rvac + \theta(t' - t) \lvac {\cal O}(t', \vect{x'}) {\cal O}(t, \vect{x}) \rvac.
 \]
When we Fourier transform this expression, apart from the ${\cal O}_{\omega,\vect{k}}$ defined above, we also get a contribution from the Fourier transform of the $\theta$ function. So, to study the properties of ${\cal O}_{\omega,\vect{k}}$, it is simpler to consider the Wightman function.

\paragraph{An Analysis of the Wightman 2-Point Function}
We now proceed to analyze the 2-point Wightman function to study the properties of ${\cal O}_{\omega,\vect{k}}$. To compute this object, we start in {\em Euclidean space}, where the 2-point correlator is given by
\[
\langle {\cal O}(\tau, \vect{x}) {\cal O}(0,\vect{0}) \rangle = \left({1 \over \tau^2 + \vect{x}^2}\right)^{\Delta}.
 \]
Here $\Delta$ is the conformal dimension of ${\cal O}$. This form is fixed by scale invariance. 
Now, let us continue to Lorentzian space. When we continue to Lorentzian space, to get the {\em time-ordered correlator} we should take $\tau = i (1 - i \epsilon) t$.  The logic of this prescription is that we want to rotate the time contour but not rotate it all the way. In particular, as Euclidean time runs from $-i \infty$ to $i \infty$, we want
Lorentzian time to run from $-\infty (1 + i \epsilon)$ to $\infty(1 + i \epsilon)$. With this prescription, we find that the time-ordered correlator is given by
\[
\lvac T\left\{{\cal O}(t,\vect{x}), {\cal O}(0,\vect{0}) \right\} \rvac = \left({-1 \over  (1 - i \epsilon)^2 t^2-\vect{x}^2} \right)^{\Delta}  = \left({-1 \over t^2 - \vect{x}^2 - i \epsilon}\right)^{\Delta},
 \]
The time-ordered function coincides with the Wightman function for $t > 0$.  If we assume that ${\cal O}$ is a Hermitian operator then the Wightman function for $t < 0$ is simply the complex conjugate of the time-ordered correlator. 
\[
\lvac {\cal O}(t,\vect{x}) {\cal O}(0,\vect{0}) \rvac^* = \lvac {\cal O}(0,\vect{0}) {\cal O}(t,\vect{x}) \rvac = \lvac T\{{\cal O}(0,\vect{0}) {\cal O}(t,\vect{x}) \} \rvac, \quad \text{for} \quad t < 0,
 \]
This leads to
\be
\label{wightmanfuncvacuum}
\lvac {\cal O}(t,\vect{x}), {\cal O}(0,\vect{0}) \rvac  = \left({-1 \over t^2 - \vect{x}^2 - i \epsilon t}\right)^{\Delta} = \left({-1 \over (t - i \epsilon)^2 - \vect{x}^2} \right)^{\Delta}.
\ee

Let us understand this expression. We choose the branch cut of the function $f(x) = x^{\Delta}$ to lie along the negative $x$ axis and $f$ to be real for positive real $x$. Then, the $i \epsilon$ prescription tells us
how to pick up the phase of the answer. For any point on the $(t,\vect{x})$ plane let us define the following real number
\[
\xi \equiv \left|t^2 - \vect{x}^2\right|^{\Delta} .
 \]
Notice that this is an unambiguously defined positive real number (or zero, on the lightcone). The Wightman function is
\be
\label{wightman}
\lvac {\cal O}(t,\vect{x}) {\cal O}(0,\vect{0})\rvac = {e^{i {\cal Q}}\over \xi}
\ee
where the phase--${\cal Q}$ is as follows
\[
{\cal Q} = \left\{\begin{array}{ll}
0&\text{for}~\vect{x}^2 - t^2 > 0,\\
-\pi \Delta&\text{for}~t^2 - \vect{x}^2 > 0,~{\rm and}~t>0\\
 \pi \Delta&\text{for}~t^2 - \vect{x}^2 > 0,~{\rm and}~t<0\\
\end{array}\right.
 \]
We should emphasize that this is a perfectly Lorentz invariant prescription. For $\vect{x}^2 > t^2$, there is no ambiguity in the choice of phase. For $t^2 > \vect{x}^2$, a Lorentz transformation cannot change the sign of $t$.

\paragraph{Fourier Transform}
Now, we will Fourier transform the expression above to return to the properties of ${\cal O}_{\omega,\vect{k}}$.  The expression \eqref{wightmanfuncvacuum} is 
valid in any number of dimensions. We define the quantity $G(\omega,\vect{k})$ by
\[
\lvac {\cal O}_{\omega,\vect{k}} {\cal O}_{\omega',\vect{k}'}\rvac = G(\omega,\vect{k}) \delta(\omega+\omega')\delta^{d-1}(\vect{k}+\vect{k}')
\]
where we used the time and space translational invariance of the 2-point function \eqref{wightman}.
We have  
\be
\label{wvacuumdef}
G(\omega, \vect{k})=  \int dt d^{d-1}\vect{x}\,\,  \left({-1 \over (t - i \epsilon)^2 - \vect{x}^2} \right)^{\Delta} e^{i \omega t - i \vect{k} \cdot \vect{x}}
\ee
\begin{figure}
\label{wightmanfuncfig}
\begin{center}
\begin{subfigure}[b]{6cm}
\includegraphics[width=6cm]{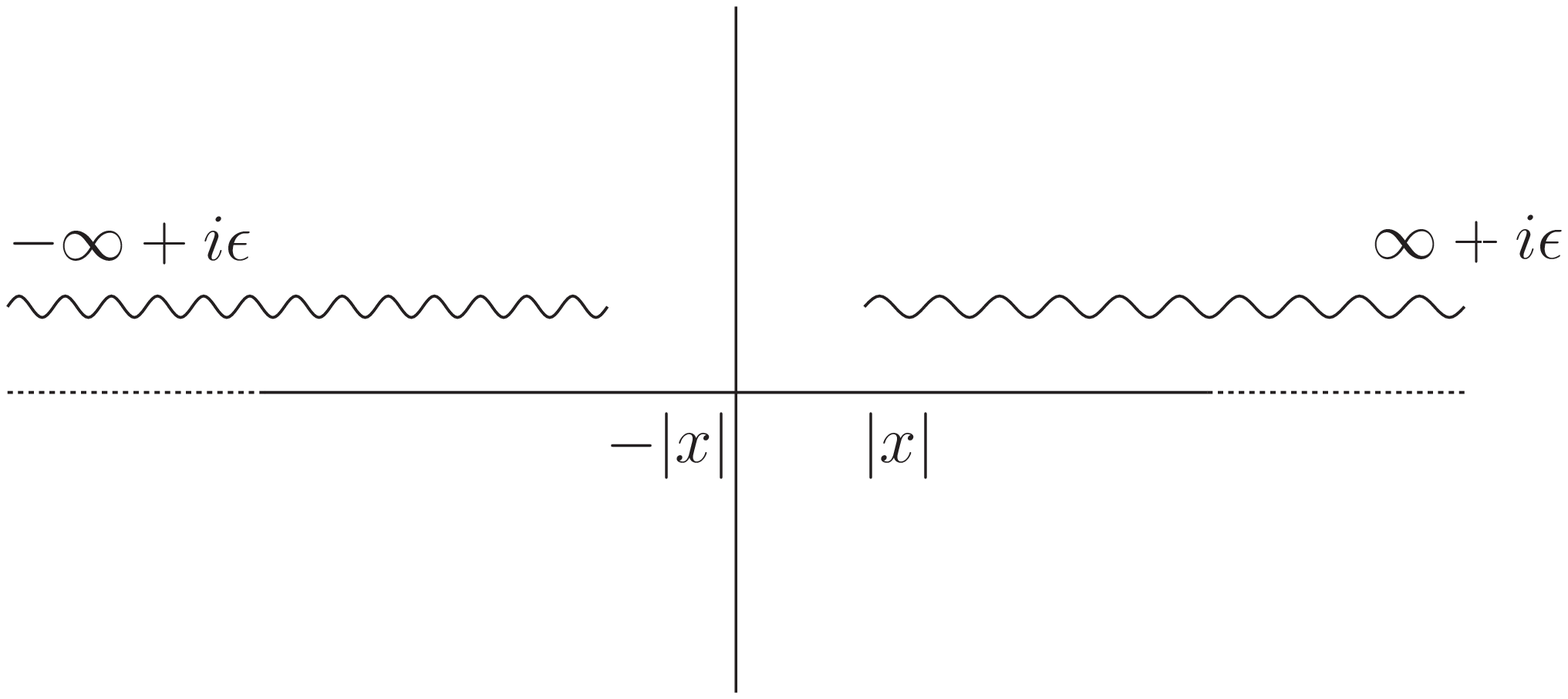}
\caption{Analytic Structure}
\label{analyticwightman}
\end{subfigure}
\qquad
\qquad
\begin{subfigure}[b]{6cm}
\includegraphics[width=6cm]{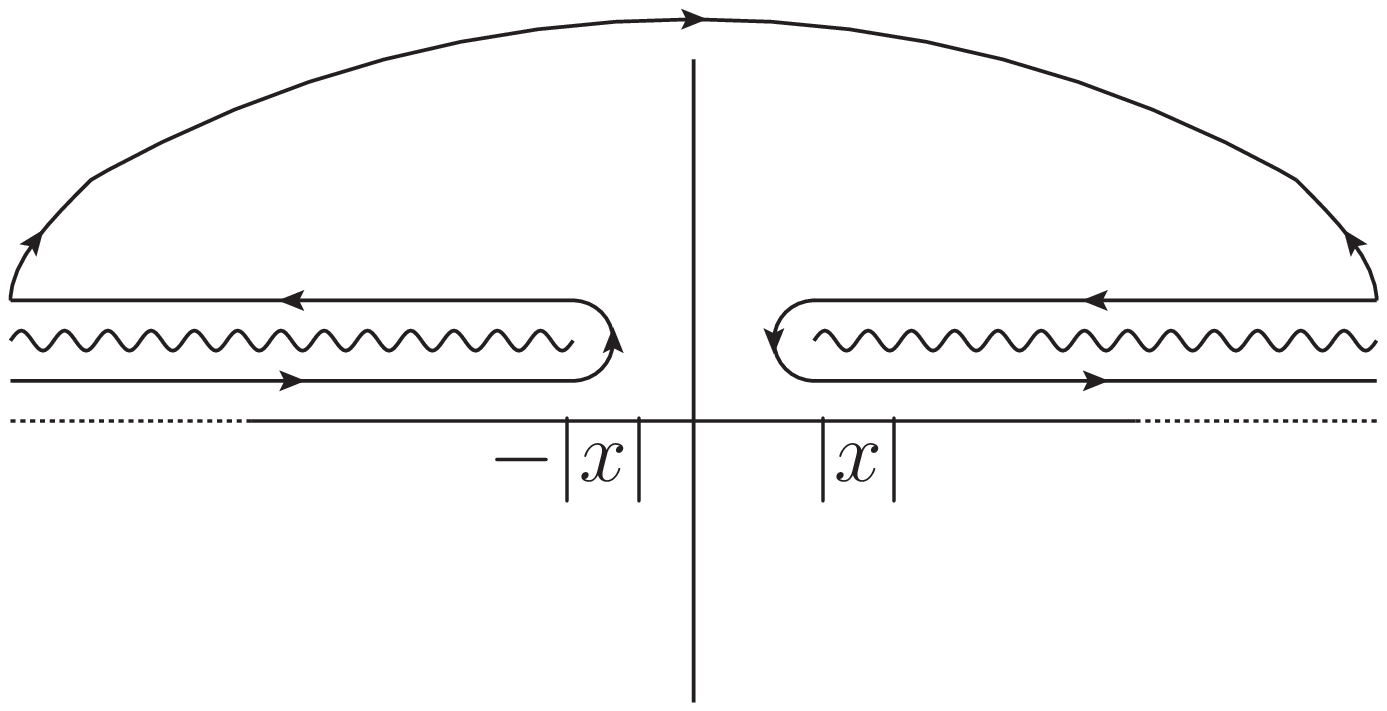}
\caption{Contour of Integration}
\label{wvacontour} 
\end{subfigure}
\caption{Wightman Function}
\end{center}
\end{figure}
The analytic structure of the integrand in the $t$-plane is shown in Figure \ref{analyticwightman}. Note that there are two branch cuts, both of which lie in the upper half plane.

Let us do the $t$ integral first. Note that this integral can be taken to have a branch cut running from $(i \epsilon + \norm{x}, i \ep + \infty)$ and another one from $(i \ep - \infty, i \ep - \norm{x})$. In the case where $\omega < 0$, we can close the $t$-contour in the {\em lower half plane}, and since there are no singularities in that region, we just get $0$. Now, by the remark above, \eqref{wvacuumdef} is invariant under Lorentz transformations that are continuously connected to the identity. If $\vect{k}^2 - \omega^2 > 0$, then by using such a transformation, we can make the time component of the momentum $d$-vector negative. This immediately tells us that we need to consider \eqref{wvacuumdef} only for vectors that have
\[
\omega > 0, \quad \text{and} \quad \omega^2 - \vect{k}^2 > 0
\]

We can now transform the $t$-integral in \eqref{wvacuumdef} by deforming the original contour, which runs along the real axis to the contour shown in \ref{wvacontour}. Considering the various phases along the legs of this contour carefully, we see that this is just
\be
\label{wightmanfuncvacuumb}
G(\omega, \vect{k} ) = 2 \left( e^{i \pi \Delta} - e^{-i \pi \Delta} \right) \int d^{d-1} \vect{x} \int_{t=|\vect{x}|}^{\infty} {e^{i \omega t - i \vect{k} \cdot \vect{x}} \over |t^2 - \vect{x}^2|^{\Delta}} dt,
\ee
We now change coordinates to: $\rho^2 = t^2 - \vect{x}^2$, and write $t = \rho \cosh \zeta, |\vect{x}| = \rho \sinh \zeta$. We also choose a frame where
the vector $(\omega, \vect{k}) \rightarrow (\sqrt{\omega^2 - \vect{k}^2}, 0, 0, \ldots)$. We can then rewrite \eqref{wightmanfuncvacuumb} as
\[
\begin{split}
G(\omega, \vect{k}) &= \theta(\omega) \theta(\omega^2 - \vect{k}^2) 2 \left( e^{i \pi \Delta} - e^{-i \pi \Delta} \right) {V_{d-2}} \int {e^{i \sqrt{\omega^2 - \vect{k}^2} \rho \cosh{\zeta}} \over \rho^{2 \Delta}} \, \rho^{d-1} d\rho \, \sinh(\zeta)^{d-1} d \zeta \\ 
&\equiv N_{\Delta,d} \,\theta(\omega) \theta(\omega^2 - \vect{k}^2) (\omega^2 - \vect{k}^2) ^{\Delta - d/2}
\end{split}
 \]
where $V_{d-2}$ is the volume of the $d-2$-sphere and $N_{\Delta,d}$
is an irrelevant non-zero numerical constant that comes from the
integral over $\zeta$ and $\rho$. 

Notice that this implies that the modes ${\cal O}_{\omega,\vect{k}}$ with positive $\omega$ annihilate the vacuum, while the modes ${\cal O}_{-\omega,-\vect{k}}$ and $\omega>0$, when acting on the vacuum, create excitations of energy $\omega$ and momentum $\vect{k}$. Moreover, {\em in the vacuum}, we have for the commutator
\[
\lvac[{\cal O}_{\omega, \vect{k}}, {\cal O}_{\omega', \vect{k}'}]
\rvac = 
G(|\omega|, \vect{k})\, \text{sgn}(\omega) \,\delta(\omega+\omega') \,\delta^{d-1}(\vect{k} + \vect{k}'),
 \]
In fact, large $N$ factorization \eqref{factorization} implies that the right hand side of the equation above is unchanged if replace the vacuum  by any normalized state $|{\cal S} \rangle$  that is created by the action of only a finite number of insertions of ${\cal O}$.

At subleading orders in $N$, the spacelike modes of ${\cal O}$ are relevant, and they do not satisfy the algebra above.  This implies that if we consider the Wightman two-point function in a state with an energy that scales with $N$ (like a big black hole), then these relations stop holding as we will find below.

However, the calculation above tells us that 
{\it
\begin{quote}
At leading order in ${1 \over N}$, while computing finite-point correlators of ${\cal O}_{\omega,\vect{k}}$ about the vacuum, we can neglect the spacelike modes of ${\cal O}_{\omega,\vect{k}}.$
\end{quote}}
\noindent Moreover, if we define the operators (for $\omega > 0$ and
$\norm{k}^2 > 0$)
\[
\begin{split}
\hat{\cal O}_{\omega, \vect{k}} &= {{\cal O}_{\omega, \vect{k}} \over 
\sqrt{G(\omega,\vect{k})}},
\\
\hat{\cal O}^{\dagger}_{\omega, \vect{k}} &= {{\cal O}_{-\omega, -\vect{k}} \over 
\sqrt{G(\omega,\vect{k})}},
\end{split}
 \]
then these operators (inserted between states $|{\cal S}\rangle$ made out of finite number of insertions of ${\cal O}$) just satisfy the algebra of free oscillators
\[
[\hat{O}_{\omega, \vect{k}}, \hat{O}_{\omega', \vect{k}'}^{\dagger}] = \delta(\omega-\omega') \delta^{d-1}(\vect{k} - \vect{k}').
 \]

Physically this means that the excitations created by the action of the generalized free field ${\cal O}$ have the structure of a freely generated Fock space. However, these excitations are qualitatively different from those created by an ordinary free field. In the case of an ordinary free field $\phi$ on the boundary, the excitations are simply labeled by the $d-1$ components of their spatial momentum $\vect{k}$, while their energy is determined by $\omega = \sqrt{\vect{k}^2+m^2}$. This dispersion relation follows from the {\it equation of motion} that the ordinary free field satisfies (for example $\Box\phi = m^2\phi$). In contrast, the excitations created by a generalized free field ${\cal O}$ are labeled by $d$ independent numbers, namely $\vect{k}$ {\it and} $\omega$. Except for the condition that $\omega^2 > \vect{k}^2$, there is no constraint among them, i.e. no dispersion relation, because the generalized free field ${\cal O}$ {\it does not satisfy any wave equation on the boundary}. 
In a sense that we will make precise in the next subsection, the excitations created by a generalized free field can be reorganized as the excitations of an ordinary free field living in a higher dimensional (AdS) spacetime.

From now on, we will take $\omega$ to be positive and will always work with positive frequency modes, that is, instead of writing ${\cal O}_{-\omega,-\vect{k}}$ we will write ${\cal O}_{\omega,\vect{k}}^\dagger$.

\subsection{Local operators on the Poincare patch}
Now, consider AdS$_{d+1}$ in the Poincare patch with geometry
\be
\label{poincarepatch}
ds^2 = {-dt^2 + d\vect{x}^2 + dz^2 \over z^2}.
\ee
Notice that we are working in units where the AdS radius is set to
one. We consider a free massive scalar field propagating on this background.
The equation of motion of this field is
\be
\label{poincareeom}
(\Box -m^2)\phi=0
\ee
It is quite easy to solve \eqref{poincareeom}. The standard solution
is in terms of Bessel functions and the {\em normalizable mode} is
\be
\label{normalizablemomentum}
\xi_{\omega, \vect{k}}(t,\vect{x},z) = e^{-i\omega t + i \vect{k} \cdot
  \vect{x}} {\Gamma(1+\nu)\over \normfact}\left({2\over \sqrt{\omega^2 - \vect{k}^2} }\right)^\nu z^{d/2}J_{\nu}(\sqrt{\omega^2 - \vect{k}^2} z) ,
 \ee
where $\nu = \sqrt{m^2 + d^2/4}$.
We have chosen the overall normalization of the mode such that it behaves 
like $\normfact^{-1}\times z^\Delta \times e^{-i\omega t + i \vect{k} \cdot \vect{x}}$ near the boundary $z=0$, and we have $\Delta= \nu + d/2$. We will also take 
\be
\label{normfactdef}
\normfact = (2 \pi)^{d \over 2} \sqrt{2 {\Gamma(\Delta - {d \over 2} + 1) \pi^{d \over 2}  \over \Gamma(\Delta)}},
\ee
for later convenience.

Notice that it is possible to find normalizable modes which do not blow up at the Poincare horizon only if $\omega^2\geq \vect{k}^2$ i.e. there are no normalizable modes with spacelike momentum along the boundary directions. This is consistent with the fact that at large $N$ the spacelike Fourier modes ${\cal O}_{\omega,\vect{k}}$ can not create any excitations when acting on the vacuum, as we found above.

We can now easily write down a (nonlocal) CFT operator, which behaves like a {\em local field} in the Poincare patch. This is simply given by
\be
\label{finalpoincare}
\boxed{
\phi_{\text{CFT}}(t, \vect{x}, z) = \int_{\omega>0} {d\omega  d^{d-1} \vect{k}  \over (2 \pi)^d}\left[{\cal O}_{\omega,\vect{k}}\,\, \xi_{\omega,\vect{k}}(t,\vect{x},z) + {\cal O}^{\dagger}_{\omega,\vect{k}}\,\, \xi^*_{\omega,\vect{k}}(t,\vect{x},z) \right]}
\ee 
We use the subscript CFT to emphasize that while $\phi_{\text{CFT}}$ seems to depend on all AdS coordinates $(t,\vect{x},z)$ it is still an operator in the conformal field theory, though clearly non-local. From the point of view of the CFT, the coordinate $z$ is in a sense ``auxiliary''. It simply parameterizes how exactly we have smeared the boundary operator ${\cal O}(t,\vect{x})$ to reproduce the nonlocal operator $\phi_{\text{CFT}}(t,\vect{x},z)$. 

The main point here is that since $\phi_{\text{CFT}}$ has exactly the same expansion as that of a free massive field in AdS, it necessarily behaves --- at large $N$ --- like a local field in the ``emergent'' AdS space, which is constructed by the boundary coordinates $t,\vect{x}$ together with the parameter $z$ and equipped with the metric \eqref{poincarepatch}. For example, with the choice in \eqref{normfactdef}, we have\footnote{Our unusual choice for the normalization of this field has to do with the fact that we wanted to remove factors of $2 \pi$ from our momentum mode commutators.}
\be
[\phi_{\text{CFT}}(t, \vect{x}, z), \dot{\phi}_{\text{CFT}}(t, \vect{x'}, z')] = {i \over (2 \pi)^d} \delta^{d-1}(\vect{x} - \vect{x'}) \delta(z - z') z^{d-1}.
\ee
Moreover, the action of the conformal transformations on ${\cal O}(t,\vect{x})$ generates just an action of the isometries of AdS on $\phi_{\text{CFT}}(t,\vect{x},z)$.

\subsection{Local operators behind the Poincare horizon}
So far what we have done is not surprising, and was already done in the works cited at the beginning of the subsection \ref{subsec:freevac}. We now show how it is possible to construct local fields on {\em global AdS} using these modes. 

\begin{figure}[!h]
%\label{poincaretoglobal}
\begin{center}
\begin{subfigure}[b]{6cm}
\includegraphics[width=6cm, height=5cm]{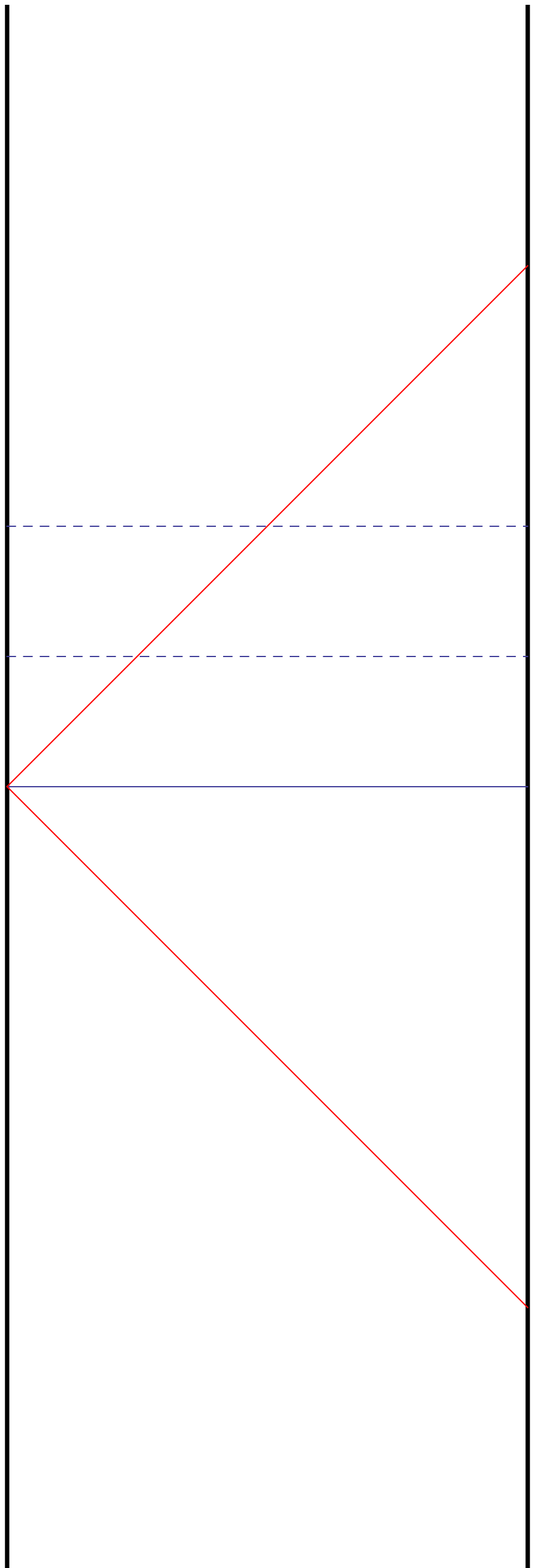}
\caption{Global time slices \label{globaltimecoord}}
\end{subfigure}
\qquad
\qquad
\begin{subfigure}[b]{6cm}
\includegraphics[width=6cm, height=5cm]{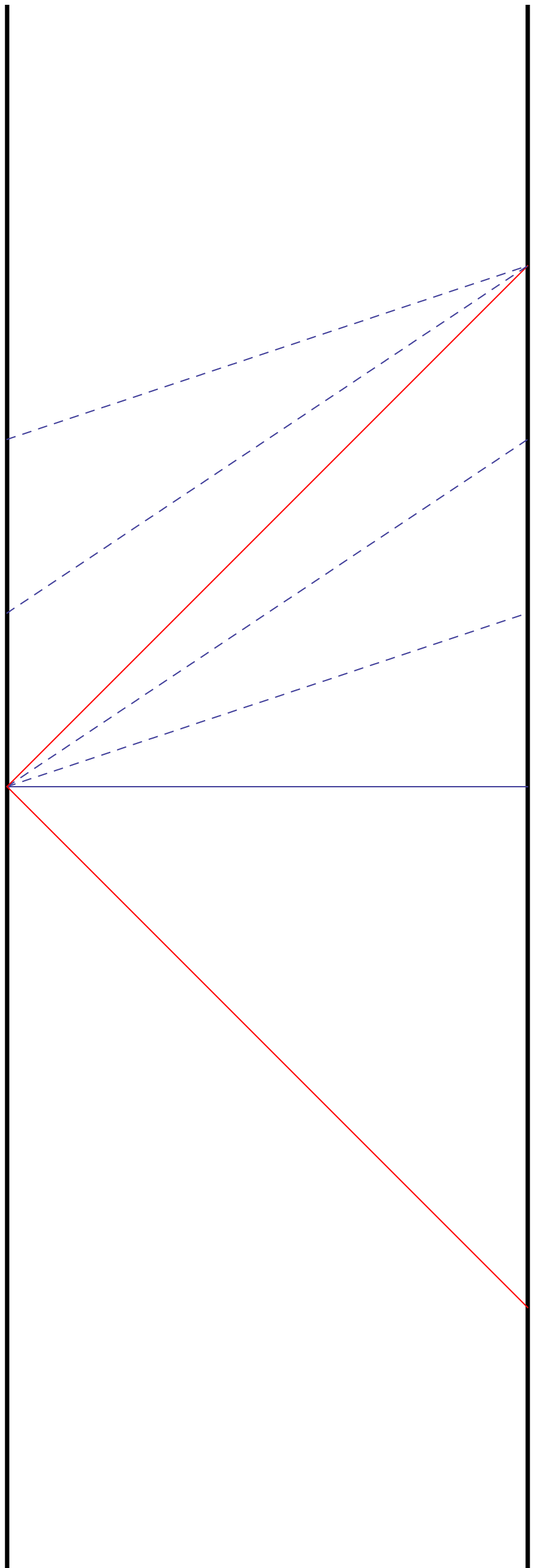}
\caption{Poincare time slices \label{pointimecoord}}
\end{subfigure}
\caption{Local operators in global AdS using the boundary of the Poincare patch \label{poincaretoglobal}}
\end{center}
\end{figure}

Figure \ref{poincaretoglobal} shows two ways in which this may be done. 
In the method indicated in \ref{globaltimecoord}, we first construct local operators on a Cauchy slice that lies entirely within the Poincare patch. We then use the equations of motion to evolve these operators forward in global AdS time beyond the Poincare patch. In practice, to write down an explicit local field like \eqref{finalpoincare}, we would need to do a Bogoliubov transform from the solutions to the wave equation given in \eqref{normalizablemomentum} that have a well defined energy with respect to Poincare time to solutions that have a well defined energy with respect to global AdS time. 

It is simpler to write down explicit operators using the method indicated
in figure \ref{pointimecoord}. We can quantize a field in global AdS
using the sequence of spacelike slices shown there. There is a natural
continuation of these slices beyond the future horizon. We now demonstrate this construction more precisely.

Consider global AdS written in coordinates
\[
ds^2 = -\cosh^2 \rho \,d t^2 + d \rho^2 + \sinh^2 \rho \,d \Omega_{d-1}^2
\]
The $d-1$ sphere that enters here can be parameterized in terms of $d-1$ angles and embedded in $R^d$ through $(\Omega_1, \Omega_2, \ldots \Omega_d) = (\cos \theta_1, \sin \theta_1 \cos \theta_2, \ldots \sin \theta_1 \sin \theta_2 ... \sin \theta_{d-1})$. (If $d$ is even, then the last entry has a $\cos \theta_d$ instead but this detail is not relevant below.) 

To {\em motivate our construction} of operators in this space, we write
\be
\label{globaltopoincare}
\begin{split}
t &=  \frac{\cosh \rho \sin \tau }{\cos \tau \cosh \rho + \cos \theta_1  \sinh \rho} \\
x_i &= \frac{\Omega_{i+1} \sinh\rho }{\cos \tau \cosh \rho + \cos \theta_1 \sinh\rho}, \quad i = 1 \ldots d-1 \\
z &= {1 \over \cos \tau\cosh \rho  + \cos \theta_1 \sinh \rho}
\end{split}
\ee
 The boundary, which is at $z = 0$ is clearly at $\rho = \infty$. The horizon, which is $z = \infty$ is defined by the equation
\[
\cos \tau\cosh \rho  + \sin \theta_1 \sinh \rho=0.
 \]
At the boundary, the coordinate change between $t,\vect{x}$ and $\tau, \Omega_i$ is
\[
t = {\sin \tau \over \cos \tau + \sin \theta_1}, \quad x_i = {\Omega_{i+1} \over \cos \tau  + \sin \theta_1 }.
 \]

Now we analytically continue our solutions from the Poincare patch to all of global AdS.  We take the solutions given in \eqref{normalizablemomentum} and substitute the coordinate transformation \eqref{globaltopoincare}. So, in global coordinates we write the solution
\[
\xi_{\omega,\vect{k}}(\tau, \rho, \Omega_i) = e^{-i \omega t(\tau, \rho, \Omega_i)} e^{i \vect{k} \cdot  \vect{x}(\tau, \rho, \Omega_i)}{\Gamma(1+\nu) \over \normfact}
\left({2\over \sqrt{\omega^2 - \vect{k}^2}}\right)^\nu z(\tau, \rho, \Omega_i))^{d \over 2} J_{\nu}(\sqrt{\omega^2 - \vect{k}^2} z(\tau, \rho, \Omega_i)),
\]
where the functions $t(\rho, \tau, \Omega_i), \vect{x}(\rho, \tau, \Omega_i), z(\rho, \tau, \Omega_i)$ are given by \eqref{globaltopoincare}. There is a small subtlety regarding a phase that we need to be careful about. 

When we cross $z = 0$, and go to negative $z$, we pick up a phase. This is because
\[
J_{\nu}(-x) = e^{i \pi \nu} J_{\nu}(x).
 \]
As we cross the next horizon, we should then pick up a net phase of $e^{2 \pi i \nu}$. 

With this convention, we can write down local operators in global AdS 
simply as
\be
\label{finalglobal}
\phi_{\text{CFT}}(\tau, \rho, \Omega_i) = \int_{\omega>0} {d\omega  d^{d-1} \vect{k}  \over (2 \pi)^d} \left[{\cal O}_{\omega,\vect{k}}\,\, \xi_{\omega,\vect{k}}(\tau, \rho, \Omega_i) + {\cal O}^{\dagger}_{\omega,\vect{k}}\,\, \xi^*_{\omega,\vect{k}}(\tau, \rho, \Omega_i) \right]
\ee 
The fields that we get by this construction are clearly periodic in the Poincare patch up to a phase. However, this is consistent with the fact that the equations of motion require $\phi$ to be periodic in global time, up to a phase. 

It is also easy to see that if we take the limit of this operator 
\[
\lim_{\rho \rightarrow \infty}  \left[\normfact\, e^{\rho \Delta}\,\phi_{\text{CFT}}(\tau, \rho, \Omega_i)\right] = {\cal O}^g(\tau, \Omega_i)
\]
where ${\cal O}^g$ is nothing but the continuation of the operator ${\cal O}$ from Minkowski space to its infinite sheeted covering that is conformal to  ${\mathbb S}^{d-1} \times {\mathbb R}$ and we remind the reader that 
the normalization factor $\normfact$ is given in \eqref{normfactdef}. In fact, many years ago Luscher and Mack \cite{Luscher:1974ez} showed that the correlation functions of the CFT on ${\mathbb R}^{d-1,1}$ could be continued to ${\mathbb S}^{d-1} \times {\mathbb R}$. This continuation can also be done, by first continuing to Euclidean space, and then continuing back to get the space ${\mathbb S}^{d-1} \times {\mathbb R}$. 

\paragraph{Continuing the CFT from ${\mathbb R}^{d-1,1}$ to ${\mathbb S}^{d-1} \times {\mathbb R}$}
For the benefit of the reader, we briefly review how CFTs can be
continued from Minkowski space to ${\mathbb S}^{d-1} \times {\mathbb R}$. The reader who is
already familiar with this topic can skip to the next subsection,
which has a ``discussion'' of the implications of our construction.

This extension is described very clearly in \cite{Aharony:1999ti}.  (see pp. 37--39.). We focus on the time coordinate $t$ and the spatial radial coordinate $r$ of Minkowski space. Let us compactify it down to the triangle by writing $r\pm t = \tan(\theta \pm \tau)$. This maps a triangle in the $\tau,\theta$ patch to the full $t,r$ plane. Writing $u_{\pm} = r \pm t$, we find that
\[
{\partial \over \partial \tau} = {1 \over 2} \left[(1 + u_{+}^2) {\partial \over \partial u_{+}} + (1 + u_{-}^2) {\partial \over \partial u_{-}} \right]]= {1 \over 2} \left(P_0 + K_0\right),
 \]
If we take our Hamiltonian to be ${1 \over 2}\left(P_0 + K_0 \right)$, then this Hamiltonian generates translations in $\tau$. In the $\tau, \theta$ plane we can continue correlators past  the edges of the triangle
onto the whole cylinder. 

There is another way to understand why it is this quantity that should
be used as the Hamiltonian when we go from ${\mathbb R}^{d-1,1}$ to ${\mathbb S}^{d-1} \times {\mathbb R}$. The basic point is that we can understand the action of the conformal algebra on the Hilbert space of the {\em Lorentzian theory}, by cutting up the Euclidean path integral that defines the theory in different ways. The boundary Euclidean path integral can be performed on ${\mathbb R}^d$ (or after adding a point at infinity on ${\mathbb S}^d$). We can cut this space either as ${\mathbb R}^{d-1} \times {\mathbb R}$ or as ${\mathbb S}^{d-1} \times {\mathbb R}$.\footnote{Note that both Minkowski space and the space ${\mathbb S}^{d-1} \times {\mathbb R}$ give the same space upon Euclidean continuation.} 

The Hilbert space that we get in these two ways is different, although
the theories are isomorphic. In the ${\mathbb R}^{d-1} \times {\mathbb
  R}$ slicing, states are defined by inserting an operator at
Euclidean time $-\infty$ and doing the path integral up to $0$. In the
${\mathbb S}^{d-1} \times {\mathbb R}$ slicing, states are defined by
inserting an operator at the origin and doing the path integral out to
unit radius. 

The dual of a state is defined, in the flat-space slicing, by inserting an operator at Euclidean time $+\infty$ and doing the path integral back to $0$; in the ${\mathbb S}^{d-1} \times {\mathbb R}$ slicing by inserting an operator at $\infty$ and do the path integral back to the unit sphere.

Consequently, the definition of the adjoint is different as well. In the flat-space slicing, the adjoint operation clearly leads to the hermiticity conditions
\[
P_i^\dagger = P_i; \quad P_0^{\dagger} = -P_0; \quad K_i^\dagger =
K_i; \quad K_0^{\dagger} = -K_0. 
 \]
The minus sign for the time component comes, in this language, because the mapping from bras to kets involves taking $t_{\text{euclidean}} \rightarrow -t_{\text{euclidean}}$. Here, the $i$ component runs only over spatial indices. The algebra that we get in this manner from the Euclidean path integral is isomorphic to the group $SO(d,2)$, with the usual hermiticity conditions.

On the other hand, in the ${\mathbb S}^{d-1} \times {\mathbb R}$ slicing, the mapping from bras to kets involves taking $|x| \rightarrow {1 \over |x|}$. So, the adjoint of a translation involves an inversion followed by a translation and another inversion. This leads to the Hermiticity conditions
\[
P_{\mu}^\dagger = K_{\mu}.
 \]
Hence, we get two different relations for the adjoint by changing the inner-product.

Now, instead of changing the inner product we can instead do a 
similarity transform on the operators, and keep the inner product fixed. This similarity transform is given in detail in \cite{Minwalla:1997ka}. (See section 2.1.) This gives rise to the relation that the dilatation operator in the ${\mathbb S}^{d-1} \times {\mathbb R}$ slicing is related to the operators in the ${\mathbb R}^{d-1} \times {\mathbb R}$ slicing through
\[
D^{{\mathbb S}^{d-1}} = {i \over 2} (P_0^{{\mathbb R}^{d-1}} + K_0^{{\mathbb R}^{d-1}}),
 \]
This operator is anti-hermitian, just like $P_0$ above but when we Wick rotate, we get the hermitian Hamiltonian above.

\paragraph{Discussion}
Note that, from the point of a view of a CFT that lives on the boundary of Poincare AdS, this construction seems a little surprising. A common, but incorrect, belief
is that this CFT has only partial information about global AdS. To the contrary, as we have shown above, the CFT gives us {\em all bulk correlators} in global AdS. 

We have also encountered the objection that ``if someone were to turn on a source of change the Hamiltonian somewhere in global AdS beyond the Poincare patch, how would 
the flat-space CFT know about this''?  Indeed, the CFT would not be
able to account for such a ``divine intervention'' but we emphasize
that this objection could also have been raised in the original
Poincare patch. If we were to change the Hamiltonian somewhere in the bulk, this would simply change the theory away from the original CFT and without further information, we would not be able to compute these new correlators. 
  
So, our flat-space CFT describes a {\em specific} theory on the Poincare patch, and a {\em specific} theory on global AdS. It can also describe small deformations away from this theory, such as those obtained by turning on sources in the bulk since the effect of these sources can be captured by a power series expansion in the original correlators. It cannot account for arbitrary changes to the Hamiltonian but this is not unusual and holds for all examples of the AdS/CFT correspondence.
 
%%% Local Variables: 
%%% mode: latex
%%% TeX-master: "infalling_paper"
%%% End: 

\section{Black holes in AdS: setup and review}
\label{semiblack}

Just as we did for the 
CFT in its vacuum state, we would now like to re-organize the degrees of
freedom of the CFT, in a heavy typical pure state, into degrees of freedom
that resemble perturbative fields propagating on an AdS black hole
background. 

As we have explained in the introduction, our setup is that we start with
the CFT in a pure state and then allow it to ``settle down'' so that
it resembles a thermal state more and more. When this happens, we 
find that the CFT describes fields propagating in, what is called, the
``eternal AdS black hole.'' 

 Maldacena \cite{Maldacena:2001kr} explained that the eternal AdS black hole has a holographic description in terms of two copies of a CFT in a specific entangled quantum state. In this paper, the eternal black hole geometry will emerge as an auxiliary device for computations done {\it in a single} CFT!

In fact, this is not so surprising from the naive semi-classical perspective since it is indeed true that quantum fields
on a collapsing star start behaving like those in an eternal black hole background, if we probe the
 geometry ``late enough.''  We review these ideas from semi-classical General Relativity in some detail below and we also review the basic formalism of quantizing fields in a black hole background.  The reader who is familiar with
these topics, or is willing to accept our claims, can jump directly to section \ref{sec:outside}.

We wish to emphasize an important logical point. In our construction in sections \ref{sec:outside} and \ref{sec:behind}, we will {\em not assume} any of the claims that we are making in this section. Rather one of the points of our paper is that we independently find a picture in the conformal field theory which is consistent with the
expectations of conventional semi-classical physics. Our review below is meant to (a) remind the reader what these expectations are and (b) serve as a guide --- although not as a logical crutch --- for our later construction.

\subsection{Collapsing stars and eternal black holes}

\begin{figure}
\begin{center}
\begin{subfigure}[t]{6cm}
\psfrag{Aads}{\rm A}
\psfrag{Bads}{\rm B}
\psfrag{tnot}{$t_0$}
\includegraphics[width=6cm]{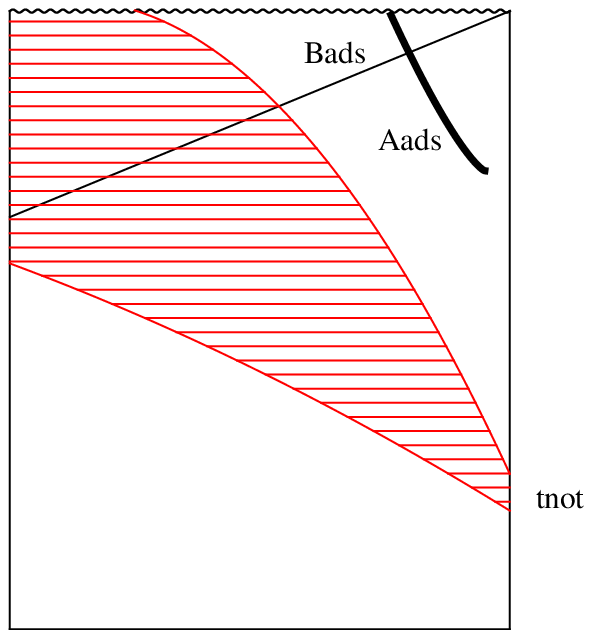}
\caption{\parbox{5cm}{Collapse of a star (red) to form a black hole in AdS. The matter is injected from the boundary at some time $t_0$. A local observer (black line) dives in much later.} \label{fig:collapsinggeometry}}
\end{subfigure}
\qquad
\qquad
\begin{subfigure}[t]{6cm}
\psfrag{frontads}{$\front$}
\psfrag{backads}{$\other$}
\psfrag{blackads}{$\black$}
\psfrag{whiteads}{\white}
\includegraphics[width=6cm]{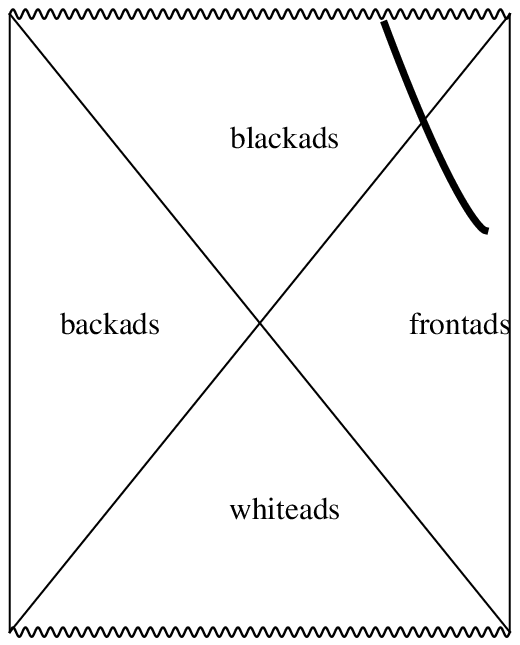}
\caption{\parbox{5cm}{AdS eternal black hole , reproduces the measurements of the observer at late times. The quantum fields are placed in the AdS-Hartle-Hawking vacuum.}\label{fig:eternalads}}
\end{subfigure}
\caption{Collapse vs eternal black hole in AdS}
\label{collapseads}
\end{center}
\end{figure}

In the first part of this section, we review the semi-classical expectation that the details of a collapsing star cease to matter both in front and behind the horizon for ``late enough'' times.
More specifically, and referring to figure \ref{collapseads} 
we have 
\vskip10pt
\noindent {\bf Semi-Classical Expectation:} {\it  Late time bulk correlators in region A and region B of the collapsing star geometry, can be well approximated by correlators in region $\front$ and region $\black$ respectively of the eternal black hole.}
\vskip10pt
Here when we say late, we mean late enough so that all the matter has fallen into the black hole and that the fluctuations of the horizon (quasi-normal modes) have mostly decayed away, but not so late that quantum mechanical effects become important. The two timescales have different parametric dependence on $N$ (or $\hbar$) so they can be clearly separated.

For asymptotically flat black holes this claim holds when the eternal black hole is taken in the Unruh vacuum. For black holes formed in AdS --- for instance, by throwing in matter from infinity as in figure \ref{fig:collapsinggeometry} ---, the claim holds when the AdS eternal black hole is taken in the Hartle-Hawking vacuum.\footnote{
The boundary of AdS acts like a reflecting wall, so the radiation coming out of the black hole eventually turns around and falls back in. The Hartle-Hawking state describes an equilibrium configuration where there is no net flux of energy. There is no AdS analogue of the Unruh vacuum.} These results are of course very well known in the context of flat space \cite{Unruh:1976db,birrell1984quantum,Hayden:2007cs} and we believe that they extend naturally
to the case of AdS.

\begin{figure}
\begin{center}
\psfrag{early}{E}
\psfrag{late}{L}
\includegraphics[width=5cm]{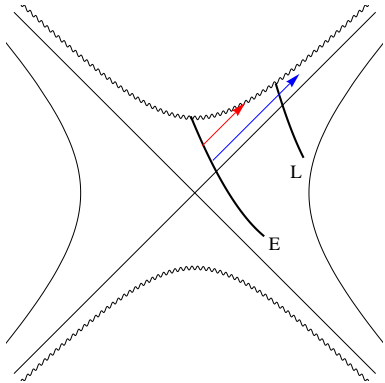}
\caption{Kruskal diagram of AdS eternal black hole. As the observer (L) dives in later, it gets increasingly difficult for any signal from an earlier observer (E) to reach him. }
\label{fig:kruskal}
\end{center}
\end{figure}

The claim that the geometry outside the black-hole ``settles down'' is probably familiar to most readers; the claim that, at late times, we can replace the entire history of the collapse in region B in figure \ref{fig:collapsinggeometry} by an effective Kruskal geometry is probably less familiar. 
However, the intuitive justification for these statements is the same and can be seen in the Kruskal diagram in figure \ref{fig:kruskal}. 

Within geometric optics, an early-time observer (Observer E) can influence the late-time observer (Observer L) only by emitting a photon or another particle that travels along a trajectory that intersects the world-line of L. However, as L dives in later and later, the window and the solid angle within which E must send his signal becomes smaller and smaller.\footnote{Among other places, this fact was discussed in the works \cite{Susskind:1993mu,Hayden:2007cs,Sekino:2008he} in relation to the consistency between black hole complementarity and the absence of quantum cloning.}

This geometric-optics observation can easily be extrapolated to classical wave mechanics. If we consider a source that emits energy within some solid angle then if we keep the power of the source fixed, its influence at late times diminishes. So, perturbatively, it is clear that a disturbance in the Kruskal geometry at early times cannot influence the late time physics either in front or behind the horizon. The argument that the details of the collapse itself do not matter at late times is an extrapolation from this perturbative argument. 

Notice that if we fix the time at which the observer L falls in, it is always possible to turn on some matter-excitations which influences the observer. However, these configurations --- while they may be solutions to the equations of motion --- are not relevant as approximations to the geometry of the collapsing black hole.  In our case, as we show in the figure, we create a black hole by turning on a source on the boundary, wait a ``sufficient'' time, and throw the observer L in. By the arguments above, the semi-classical expectation is that this observer should just perceive the geometry of an eternal black hole all along his world line.

One subtle point is that the existence of region $\other$ cannot be neglected in region $\black$. This is because, while classically no influence can propagate from region $\other$ to region $\black$ and influence the late time observer in figure \ref{fig:eternalads}, when we {\em quantize} the field in region $\black$ then, even at late times, it has both ``left-moving modes'' that are analytic continuations of modes from region $\front$ and ``right moving modes'' that are analytic continuations of modes from region $\other$. We return to this in section \ref{modesbrane}.

We emphasize again that in our construction, we will not assume either this feature of the quantum mechanics on the eternal AdS spacetime, or the semi-classical general relativity expectation above. Rather we will find both independently in the dual conformal field theory.

\subsection{Classical Properties of the AdS eternal black hole \label{secadseternal}}

The AdS eternal black hole that we have drawn in figure \ref{fig:eternalads} is a maximal continuation of the AdS-Schwarzschild black hole, just like the Kruskal geometry is a continuation of the Schwarzschild geometry. In this section, we review the metric and geometry of this space in some more detail.
We will  work with the planar version of AdS black holes (i.e. branes). It is straightforward to rewrite everything in terms of AdS black holes with spherical event horizons --- this is even necessary if we wish to address questions related to Poincare recurrence, and other finite volume effects. In this paper, we did not do so since it was more convenient to work with momenta ${\bf k}$ rather than spherical harmonics. 

The metric of the eternal AdS black brane is given by
\begin{equation}
\label{threebrane}
ds^2 = {\ell^2 \over z^2} \left[-h(z) dt^2 + {1 \over h(z)} dz^2 + d\vect{x}^2 \right],
\end{equation}
where \[
h(z) = 1-{z^d \over \zhor^d}.
\]
The horizon is at $z=\zhor$, the boundary at $z=0$ and $\vect{x}$ is a $(d-1)$-dimensional vector. There is no flux turned on here. The metric \eqref{threebrane} is a solution to the equations of motion for the action
\[
S = {-1 \over 16 \pi G_N} \int \sqrt{-g} \left[R + {d(d-1) \over \ell^2} \right].
\]
We have displayed the AdS radius $\ell$ explicitly here because it will make a brief appearance in our discussion of the temperature below.  However, in what follows, and everywhere else,  we will set $\ell = 1$.

We introduce the tortoise coordinate defined by
$
{dz_* \over dz} =- h^{-1}(z)
$. Now the horizon is at $z_*\rightarrow-\infty$. We fix the overall additive ambiguity in the definition of $z_*$ by requiring $z_*\rightarrow 0$ as $z\rightarrow 0$. The metric takes the form
\[
ds^2 = {h(z) \over z^2}(-dt^2 +dz_*^2) +{d\vect{x}^2 \over z^2}
\]
We go to lightcone coordinates
\be
\label{lightconec}
u= t-z_*\qquad,\quad v= t+z_*
\ee
\[
ds^2 = -{h(z)\over z^2} du dv + {d\vect{x}^2 \over z^2}
\]
Here $z$ is defined implicitly via $u,v$ by the previous changes of coordinates. Finally we define
\be
\label{kruskalc}
U = -e^{-{d u \over 2 \zhor}}\qquad,\qquad V = e^{{d v \over 2 \zhor}}
\ee
to get
\begin{equation}
\label{kruskal}
ds^2 = {4 h(z)\over d^2 U V} \left({\zhor \over z}\right)^2
 dU dV +{d\vect{x}^2 \over z^2}
\end{equation}
This metric is originally defined in the region $U<0,V>0$. The future horizon is at $U\rightarrow 0, V={\rm constant}$ and the past horizon at $V\rightarrow 0, U={\rm constant}$. In the form \eqref{kruskal}, it is clear that the metric is smooth at both these horizons and can be smoothly extended past them.\footnote{Near the horizons we have $h \approx {d\over \zhor}(\zhor-z)$. The tortoise coordinate is $z_* \approx {\zhor \over d}\log\left({\zhor-z \over \zhor}\right)$. From the change of coordinates we find $h \approx d e^{d z_* \over \zhor} = d U V.$}
On the other hand there is a specific positive value of $UV$ for which $h$ blows up (in our conventions for $z_*$ this happens at $UV = e^{-\pi}$). These points represent the future and past singularities.

\paragraph{How we take the Large $N$ Limit \\}
Let us pause briefly to specify precisely what we mean by taking the large $N$ limit in the context of this bulk geometry. Although this is a point that seems to cause confusion at times, what we are doing is perfectly conventional. In taking the large $N$ limit, we {\em keep the solution \eqref{threebrane} fixed}. This means that the temperature of the gauge theory also remains fixed and does not scale with $N$. Indeed, the temperature of the black-brane solution \eqref{threebrane} can easily be calculated to be (with all factors restored) 
\[
T = {\hbar c d \over 4 \pi \zhor k_B},
\]
where $c$ is the speed of light and $k_B$ is the Boltzmann constant. Note that $G_N$ does not appear here.  On the other hand, with this fixed temperature, the  ADM mass-density of the black-brane will contain a factor of ${1 \over G_N}$. So the energy density of the boundary CFT does scale with $N$, as we take the large $N$ limit. 

We should point out that $\ell$ also does not appear in the formula for the temperature. This is because we are considering the black-brane solution in AdS. If we were to instead consider the AdS-Schwarzschild solution, then the temperature would depend on $\ell$. Instead of thinking of a black-brane, the reader may instead prefer to think of a {\em big black hole} in AdS i.e. one where the horizon size is larger than the AdS radius.  Such a black-hole is thermodynamically favoured, since the corresponding temperature is higher than the Hawking-Page transition temperature. So, for the conformal field theory on a sphere on radius $R$, our analysis is valid when we take the temperature to be any number larger than the phase transition temperature in units of  ${1 \over R}$, as long as the temperature {\em does not scale with $N$}.

\subsection{Quantization in an eternal AdS black hole \label{modesbrane}}

\begin{figure}
\begin{center}
\psfrag{frontads}{$\front$}
\psfrag{backads}{$\other$}
\psfrag{blackads}{$\black$}
\psfrag{whiteads}{$\white$}
\psfrag{uzero}{U=0}
\psfrag{vzero}{V=0}
\psfrag{s1}{$\Sigma_{\front}$}
\psfrag{s2}{$\Sigma_{\other}$}
\includegraphics[width=10cm]{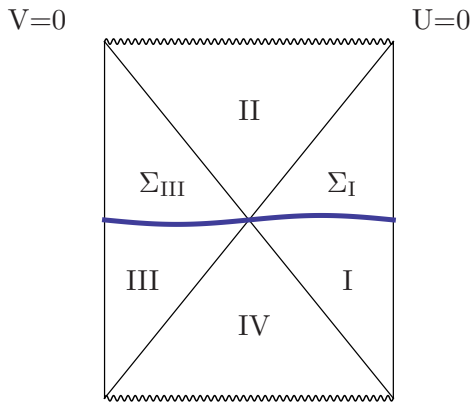}
\caption{Cauchy slice for the eternal AdS black brane geometry.}
\label{penroseadsb}
\end{center}
\end{figure}

We also need to remind the reader how to quantize a field on the background of an eternal AdS black hole. 
In figure \ref{penroseadsb} we see a Cauchy slice for the entire spacetime. 
It can be thought of as the union of two smaller slices, denoted as $\Sigma_{\front}$ and $\Sigma_{\other}$. The slice $\Sigma_{\front}$ is a complete Cauchy slice if we restrict ourselves to events taking place in region $\front$ and the slice $\Sigma_{\other}$ for events in region $\other$. However in order to describe regions $\black$ and $\white$ we need the entire slice $\Sigma_{\front} \oplus \Sigma_{\other}$. When we quantize the field in AdS we impose normalizable boundary conditions at infinity. This means that only the subleading mode (``the vev'') of the field can be turned on. For simplicity we avoid discussing the window of masses where two alternative quantizations are acceptable.

One way to find a complete set of solutions to be used for the quantization, is to first work with the wedge $\front$, whose Cauchy slice is $\Sigma_{\front}$ , and then with the wedge $\other$ and then put them together (the same thing as we do for the quantization of Rindler space or the flat space Schwarzschild solution). So we start with region $\front$.
We consider the black hole solution \eqref{threebrane} and a scalar field obeying
\[
(\Box - m^2) \phi =0 
\]
we consider a solution of the form
\[
f_{\omega,\vect{k}}(t,\vect{x},z) = e^{-i \omega t + i \vect{k} \cdot \vect{x}} \psi_{\omega,\vect{k}}(z)
\]
Plugging into the Klein-Gordon equation we get a second order ordinary differential equation for $\psi_{\omega,\vect{k}}(z)$. 
It has two linearly independent solutions. We impose normalizability of the solution near the boundary i.e.
\[
\psi_{\omega, \vect{k}} (z) \underset{z \rightarrow 0}{\longrightarrow} \normfact^{-1}\,z^{\Delta},
\]
which eliminates one linear combination of the solutions. So, for each choice of $(\omega,\vect{k})$ we have a unique normalizable solution.

We do not impose any boundary conditions at the horizon. Had we imposed ingoing boundary conditions we would have found solutions only for complex $\omega$ i.e. 
the quasinormal frequencies. For the quantization of the field we need to find a complete set of solutions of the wave equation without any restriction on the horizon. 
The solutions that we found are linear combinations of ingoing and outgoing modes. By an appropriate choice of the overall phase, the modes can be taken to be real. 
If we introduce the tortoise radial coordinate $z_*$, in which the horizon is at $z_*\rightarrow - \infty$ we find that the modes behave like
\be
\label{nearhpsi}
\psi_{\omega,\vect{k}}\underset{z \rightarrow z_0}{\longrightarrow} c(\omega,\vect{k})\left(e^{-i\delta_{\omega,\vect{k}}}e^{-i\omega z_*} + e^{i\delta_{\omega,\vect{k}}}e^{i\omega z_*}\right)
\ee
where $c(\omega,\vect{k})$ is a positive real constant.
The relative phase difference $e^{i\delta_{\omega,\vect{k}}}$ is physically meaningful and cannot be removed by changing the conventions.

It is also possible to normalize the modes, so that they are canonical
with respect to the Klein-Gordon norm. In this second normalization, we write
\[
\hat{f}_{\omega,\vect{k}}(t,\vect{x},z) = e^{-i \omega t + i \vect{k} \cdot \vect{x}} \hat{\psi}_{\omega,\vect{k}}(z),
\]
where
\[
\begin{split}
&\hat{\psi}_{\omega, \vect{k}} \underset{z \rightarrow z_0}{\longrightarrow} z_0^{d-1 \over 2} \times (e^{-i\delta_{\omega,\vect{k}}}e^{-i\omega z_*} + e^{i\delta_{\omega,\vect{k}}}e^{ i\omega z_* }),\\
&\hat{\psi}_{\omega, \vect{k}} \underset{z \rightarrow 0}{\longrightarrow} c(\omega,\vect{k})^{-1} z_0^{{d-1 \over 2}} z^{\Delta} {1 \over \normfact}.
\end{split}
\]
This notation, where modes that go like ``1'' near the boundary are denoted by  $f_{\omega,\vect{k}}(t,\vect{x},z)$ and those that go like ``1'' near the horizon are denoted by $\hat{f}_{\omega,\vect{k}}(t,\vect{x},z)$ will be used below.

In any case, we have found a complete set of solutions of the Klein-Gordon equation for region $\front$ which can be used to expand the quantum field $\phi$ in region $\front$ in creation and annihilation modes
\[
\phi(t,\vect{x},z) = \int_{\omega>0} {d\omega d^{d-1}\vect{k} \over  (2 \pi)^d}\,\,{1\over \sqrt{2\omega}}\left[a_{\omega,\vect{k}} \hat{f}_{\omega,\vect{k}}(t,\vect{x},z) + \,\,{\rm h.c.}\right]
\]
The modes satisfy the standard commutation relations
\[
[a_{\omega,\vect{k}},a_{\omega',\vect{k}'}^\dagger] = \delta(\omega-\omega')\delta^{d-1}(\vect{k}-\vect{k}')\]
with all other commutators vanishing. 

Notice that the spectrum in $\omega$ is continuous. This does not have to do with the non-compactness of the spatial directions, even if we consider a black hole in global AdS we still 
find a continuous spectrum in $\omega$ --- even though the boundary theory lives on a compact space ${\mathbb S}^{d-1}$. The continuum in $\omega$ is a large $N$ artifact and related
to the approximately continuous spectrum of the dual large $N$ gauge theory in the deconfined phase, see \cite{Festuccia:2005pi, Festuccia:2006sa} for more details.

Following the same analysis in region $\other$ we get another set of creation and annihilation modes that we denote by $\widetilde{a}_{\omega,\vect{k}}$. These modes satisfy an identical-oscillator type algebra among themselves, and commute with all the modes $a_{\omega, \vect{k}}$. If we have the expansion of the field in a complete basis both in regions I and III, it is straightforward to extend it to regions II and IV. 

While it should be obvious from figure \ref{penroseadsb}, we would like to emphasize again that in order to describe a local field in region II (inside the black hole) it is necessary to use {\it both} the operators $a_{\omega,\vect{k}}$ which are visible outside the horizon {\it and} the operators $\widetilde{a}_{\omega,\vect{k}}$ which seem to come from region III. 

Finally let us mention that the natural vacuum for a quantum field in AdS in the presence of a big AdS black hole is the analogue of the Hartle-Hawking vacuum. The Hawking radiation from the black holes
is reflected by the AdS potential and an equilibrium state like that of the flat-space Hartle Hawking state is reached (there is no analogue of the Unruh vacuum). In terms of our oscillators
the Hartle Hawking state is characterized by thermal occupation levels
\be
\label{hhocup}
\langle a_{\omega,\vect{k}} \,a^\dagger_{\omega',\vect{k}'}\rangle_{\rm HH} = {e^{\beta \omega}\over e^{\beta \omega}-1} \delta(\omega-\omega')\delta^{d-1}(\vect{k}-\vect{k}')
\ee
\be
\label{hhocupb}
\langle a_{\omega,\vect{k}}^\dagger \,a_{\omega',\vect{k}'}\rangle_{\rm HH} = {1\over e^{\beta \omega}-1} \delta(\omega-\omega')\delta^{d-1}(\vect{k}-\vect{k}')
\ee
and similar for the modes $\widetilde{a}_{\omega,\vect{k}}$. Here $\beta$ is the Hawking temperature of the black hole \eqref{threebrane}.

%%% Local Variables: 
%%% mode: latex
%%% TeX-master: "infalling_paper"
%%% End: 

\section{Reconstructing the black hole from the boundary, outside the horizon \label{sec:outside}}

We will now try to reconstruct the black hole geometry from the boundary. 
In AdS/CFT we have the following identifications between the boundary CFT and the bulk
\[
\begin{split}
{\rm Thermalization\,\,of\,\,pure\,\,state} &\qquad\Leftrightarrow\qquad 
{\rm Black\,\, hole\,\, formation\,\, by\,\, gravitational\,\, collapse}\cr
{\rm Thermal\,\, density\,\, matrix}  & \qquad\Leftrightarrow\qquad {\rm Eternal\,\, black\,\, hole}
\end{split}
 \]
In the same way that the collapsing star can be approximated  {\it for certain questions} by the eternal black hole, we expect that late time\footnote{As in the bulk, by late we mean after the thermalization has occurred, but not too late so that Poincare recurrences and other finite $N$ effects become important.} CFT correlation functions on a heavy pure state can be approximated by correlation functions on a thermal density matrix. 

Hence, we will first focus on thermal correlators and explain how to represent local bulk operators in the case where the boundary theory is in a thermal density matrix. We will then carry over this definition of the operator (at large $N$) to the case where we have a typical pure state. We will discuss the validity of this approach and possible non-trivial sensitivity on the specific pure state in section \ref{sec:subtleties}.

\subsection{Boundary thermal correlators}

\subsubsection{Large $N$ factorization at finite $T$}
\label{factorcav}

Let us consider a scalar conformal primary operator ${\cal O}$. For simplicity we assume that its thermal 1-point function ${\rm Tr}(\rho \,{\cal O})$ vanishes,  
where $\rho = e^{-\beta H}$. This can be ensured by considering a theory with a ${\mathbb Z}_2$ symmetry, under which ${\cal O}$ is odd.

A central assumption in what follows is that large $N$ factorization holds for thermal correlation functions i.e. that we have the analogue of \eqref{factorization}
\[
{\rm Tr}(\rho \,{\cal O}(x_1)...{\cal O}(x_{2n})) = {1\over 2^n} \sum_\pi
{\rm Tr}(\rho \,{\cal O}(x_{\pi_1}) {\cal O}(x_{\pi_2}))...{\rm Tr}(\rho \,{\cal O}(x_{\pi_{2n-1}}) {\cal O}(x_{\pi_{2n}})) + \ldots,
\]
where the dots are terms that are subleading in ${1 \over N}$. 
Of course the 2-point function ${\rm Tr}(\rho\, {\cal O}(x_1)\, {\cal O}(x_2))$ in which the thermal correlators factorize into is completely different 
from the zero temperature 2-point function $\lvac {\cal O}(x_1) {\cal O}(x_2) \rvac$. Also, we need to stress that ---as in the zero temperature case---
there are several caveats about the validity of factorization. We should be careful to {\it first} fix all other parameters/numbers that enter the correlator and {\it then} take $N$ to infinity. For example, factorization can fail if

\begin{itemize}
\item We scale the number of operators (i.e. $n$ in the formula above) with some power of $N$. Scaling the number of external legs with powers of $N$ invalidates the naive 't Hooft counting and a priori we have no reason to expect the correlator to still factorize.

\item We scale the conformal dimension of ${\cal O}$ in an $N$-dependent way --- the same comments as above apply.

\item We scale some of the distances $|x_i-x_k|$ to be too small, in an $N$-dependent way (this can be thought of as a high energy scattering, where higher and higher loops in $1/N$ become important).

\item We scale the temperature in a $N$-dependent way. (See the paragraph in section \ref{secadseternal} for more details.)

\item We scale (some of) the distances $|x_i-x_j|$ to {\it increase} in an $N$-dependent fashion. This will be particularly important when we consider the operators evaluated on
typical pure states vs the thermal ensemble.

\end{itemize}
Of course this list is not supposed to be exhaustive. In general we have to be careful about how various factors in the problem scale, when taking the large $N$ limit.

\subsubsection{Analytic structure of thermal 2-point functions}
\label{therman}

We now discuss general properties on finite temperature Wightman functions. Since correlators at large $N$ factorize to products of 2-point functions, we will focus on the latter. 

Consider two local operators ${\cal O}_1,{\cal O}_2$. From time and space translational invariance of the thermal ensemble we only need
\[
F_{12}(t) \equiv Z_{\beta}^{-1}{\rm Tr}\left(e^{-\beta H} {\cal O}_1(t,\vect{x}) \, {\cal O}_2(0,\vect{0})\right),
\]
where we did not indicate the $\vect{x}$ dependence on the LHS. Here 
\[
Z_{\beta} = {\rm Tr}\left[e^{-\beta H}\right],
\]
 is the partition function which will appear frequently below. Since the operators do not generally commute, we also have a related function
\[
F_{21}(t) \equiv Z_{\beta}^{-1}{\rm Tr}\left(e^{-\beta H} {\cal O}_2(0,\vect{0}) \, {\cal O}_1(t,\vect{x})\right).
\]
Let us keep $\vect{x}$ fixed and analytically continue $t$ to complex values. By inserting a complete set of states and using the positivity of the energy spectrum we find that the analytically continued function $F_{12}(t)$ is meromorphic in the strip
\[
-\beta<{\rm Im}\,t <0.
\]
However, in general, it cannot be analytically continued to positive values of ${\rm Im}\,t$.  The function $F_{21}(t)$ can be analytically continued to positive imaginary values of $t$ and is meromorphic in the strip 
\[
0<{\rm Im} \,t<\beta.
\]  

The value of the function $F_{12}(t)$ for real $t$ is equal to the limit of the meromorphic function in the strip $-\beta<{\rm Im}t <0$, as we take ${\rm Im} t \rightarrow 0^-$ from below. As we approach the other end of the strip from above (i.e. ${\rm Im}t \rightarrow -\beta^-$) and by inserting a complete set of states and using the cyclicity of the trace we derive the so called KMS condition
\be
\label{kms}
F_{12}(t-i\beta^-) = F_{21}(t).
\ee
So the correlator is periodic in imaginary time, up to an exchange in the order of the insertion of operators. 

The discussion so far has been very general and would apply to any quantum system at finite temperature, even in non-relativistic quantum mechanics. In relativistic QFT we have an important additional property: local operators at spacelike separation commute. This means that the functions $F_{12}(t)$ and $F_{21}(t)$ must have the same value along the segment of the real axis $-|\vect{x}|<t<|\vect{x}|$ and hence one must be the analytic continuation of the other! For $t>|\vect{x}|$ we generally have $\lim_{{\rm Im}\, t\rightarrow 0^-} F_{12}(t) \neq \lim_{{\rm Im}\, t\rightarrow 0^+} F_{21}(t)$ and this discontinuity is proportional to the (thermal expectation value of the) commutator of the two operators. Because of the KMS condition this discontinuity appears periodically along all semi-infinite lines $t = \pm (|\vect{x}| + {\cal R}^+) + i m \beta$ where ${\cal R}^+$ denotes positive real numbers and $m \in \mathbb{Z}$.

\begin{figure}
\begin{center}
\psfrag{beta}{$\beta$}
\psfrag{beta2}{$2\beta$}
\psfrag{beta3}{$-\beta$}
\psfrag{beta4}{$-2\beta$}
\psfrag{Ret}{${\rm Re}(t)$}
\psfrag{zero}{$0$}
\psfrag{xplus}{$|\vect{x}|$}
\psfrag{xminus}{$-|\vect{x}|$}
\psfrag{Imt}{${\rm Im}(t)$}
\includegraphics[width=9cm]{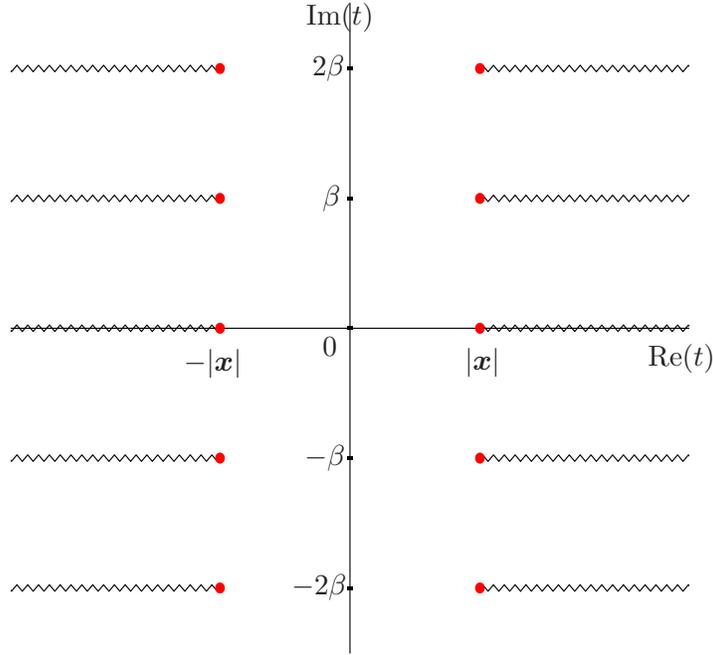}
\caption{Domain of analyticity of a thermal 2-point function of the two local operators ${\cal O}_1(t,\vect{x})$ and ${\cal O}_2(0,\vect{0})$. The domain ${\cal D}$ is defined as the complex plane with the indicated branch cuts removed. On this domain we have a holomorphic function ${\cal F}(z)$ which satisfies ${\cal F}(z+i\beta)= {\cal F}(z)$. The function ${\cal F}$ contains all information about the thermal 2-point function. For example, we have $\lim_{\epsilon\rightarrow 0^+}{\cal F}(t-i\epsilon) = Z_{\beta}^{-1}{\rm Tr}\left(e^{-\beta H} {\cal O}_1(t,\vect{x}) {\cal O}_2(0,\vect{0})\right)$ and also 
 $\lim_{\epsilon\rightarrow 0^+}{\cal F}(t+i\epsilon) = Z_{\beta}^{-1}{\rm Tr}\left(e^{-\beta H} {\cal O}_2(0,\vect{0}){\cal O}_1(t,\vect{x}) \right)$
. The discontinuity along the branch cut is proportional to the thermal expectation value of the commutator of the two operators.}
\label{analyticity}
\end{center}
\end{figure}

Hence, by combining the KMS condition with the spacelike commutativity of local fields we have arrived at the following important conclusion: consider the domain ${\cal D}$ of the complexified $t$-plane defined as $\mathbb{C}$ minus ``cuts'' starting from $\pm|\vect{x}| + i m \beta$ and extending all the way to infinity parallel to the real axis, as depicted in figure \ref{analyticity}. This domain ${\cal D}$ is a simply-connected domain. Then, there is a {\it holomorphic} function ${\cal F}(z)$ defined in the domain ${\cal D}$ with the property that
\[
{\cal F}(z+i\beta) = {\cal F}(z) ,
\]
for all $z\in {\cal D}$ and also
\begin{align}
\nonumber
\lim_{\epsilon \rightarrow 0^+} {\cal F}(t-i \epsilon) = F_{12}(t), \\
\nonumber
\lim_{\epsilon \rightarrow 0^+} {\cal F}(t+i \epsilon) = F_{21}(t).
\end{align}
The holomorphic function ${\cal F}$ contains all the information about the 2-point function of operators ${\cal O}_1,{\cal O}_2$ for both possible orderings.

We mentioned that generally $F_{12}(t)$ cannot  be analytically continued to complex $t$ with positive imaginary part by going through the part of the positive real axis with $t>|\vect{x}|$ (i.e. the analytic continuation is possible only by going around, and through the ``spacelike segment'' $-|\vect{x}|<t<|\vect{x}|$). While this is indeed  generic, it is not always true: there are quantum systems (for example a free QFT) whose thermal 2-point function {\it can} be analytically continued up through the part of the positive real axis with $t>|\vect{x}|$. However even in that case, the analytic continuation of the function through what used to be the ``cut'' of the domain ${\cal D}$ {\it does not lead to the same value} as the one that we would get by going through the ``spacelike segment'' which lies purely within the domain ${\cal D}$. In other words the correlation functions analytically continued to complex times are generically multivalued functions --- and we have to specify the ``sheet'' on 
which we are evaluating them.

We will define the ``principal sheet'' of the analytically continued thermal correlators, as the one defined by doing the continuation purely within the domain ${\cal D}$ described above. This will be very important when we discuss the analytic continuation behind the horizon. Let us emphasize this once more: whenever we write down a thermal correlator where some of the time arguments have been given an imaginary value, the {\it definition} of this analytically continued correlator is that one has to start with the correlator defined in ``real time'' and then continue it to the complex values by working only in the domain ${\cal D}$ and going ``around'' the cuts, through the Euclidean (spacelike) region. 

After these generalities, let us return to the case where we are looking at the thermal 2-point function of a conformal primary ${\cal O}$. From translational invariance in space and time we only need the function
\[
G_\beta(t,\vect{x}) = Z_{\beta}^{-1}{\rm Tr}\left(e^{-\beta H}{\cal O}(t,\vect{x})\,{\cal O}(0,\vect{0})\right),
\]
we will also consider its Fourier transform
\[
G_\beta(\omega,\vect{k}) = \int dt d^{d-1}\vect{x} \,\, e^{i \omega t - i \vect{k}
\vect{x}}\,G_{\beta}(t,\vect{x}).
\]
We have the following general properties. First, rotational invariance implies
\[
G_\beta(t,\vect{x}) = G_\beta(t,-\vect{x}).
\]
Second the KMS condition for the thermal trace implies
\[
G_{\beta}(t-i\beta,\vect{x}) = G_\beta(-t,-\vect{x}).
\]
In Fourier space the corresponding statements are
\[
G_\beta(\omega,\vect{k}) = G_\beta(\omega,-\vect{k}),
\label{rotfourier}
 \]
\[
G_\beta(-\omega,\vect{k}) = e^{-\beta \omega} G_\beta(\omega,\vect{k}).
\label{kmsfourier}
 \]
These are exact properties which hold in all dimensions and arbitrary coupling.\footnote{Clearly the discussion in this section was also independent of large $N$.} We can explicitly verify them in the case of 2d CFT thermal real-time correlation functions, which can be explicitly computed, as presented in appendix \ref{appendix2dthermal}. 

\subsubsection{Mode expansion of thermal 2-point function}

Now we proceed by expanding the boundary field in its Fourier modes. For simplicity we take the CFT to live in ${\mathbb R}^{d-1,1}$ at finite temperature.\footnote{We could also do a similar analysis for the finite temperature CFT on ${\mathbb S}^{d-1}\times {\rm time}$, where we would have to replace the ${\bf k}$ momenta with discrete spherical harmonic modes.} We have  
\be
\label{modesa}
{\cal O}(t,\vect{x}) = \int_{\omega>0} {d\omega d^{d-1}\vect{k} \over (2 \pi)^d} \,\,\left( {\cal O}_{\omega,\vect{k}} e^{-i\omega t +i \vect{k} \vect{x}} +  {\cal O}_{\omega,\vect{k}}^\dagger e^{i\omega t -i \vect{k} \vect{x}}\right) 
\ee
As before, this is the {\it definition} of the (non-local) operators ${\cal O}_{\omega,\vect{k}}$.

We can now consider the thermal 2-point function of ${\cal O}$ expanded in terms of its modes. From translational invariance we immediately conclude that
\[
Z_{\beta}^{-1}{\rm Tr}\left(e^{-\beta H} {\cal O}_{\omega,\vect{k}} {\cal O}_{\omega',\vect{k}'} \right) =  
Z_{\beta}^{-1}{\rm Tr}\left(e^{-\beta H} {\cal O}_{\omega,\vect{k}}^\dagger {\cal O}_{\omega',\vect{k}'}^\dagger \right) =0,
\]
and for the other combinations we have
\begin{align}
\nonumber 
&Z_{\beta}^{-1}{\rm Tr}\left(e^{-\beta H} {\cal O}_{\omega,\vect{k}} {\cal O}_{\omega',\vect{k}'}^\dagger \right)
= G_\beta(\omega,\vect{k}) \delta(\omega-\omega') \delta^{d-1}(\vect{k}-\vect{k}'), \\
\nonumber
&Z_{\beta}^{-1}{\rm Tr}\left(e^{-\beta H} {\cal O}_{\omega,\vect{k}}^\dagger {\cal O}_{\omega',\vect{k}'} \right)
= G_\beta(-\omega,-\vect{k}) \delta(\omega-\omega') \delta^{d-1}(\vect{k}-\vect{k}').
\end{align}
So for the thermal expectation value of the commutator we have
\[
Z_{\beta}^{-1}{\rm Tr}\left(e^{-\beta H} [{\cal O}_{\omega,\vect{k}}\,,\, {\cal O}_{\omega',\vect{k}'}^\dagger] \right)
= \left(G_\beta(\omega,\vect{k})- G_\beta(-\omega,-\vect{k}) \right)\delta(\omega-\omega') \,\delta^{d-1}(\vect{k}-\vect{k}').
\]

Actually, given large $N$ factorization of the thermal correlators we can derive a more general statement: the relations above hold even in the presence of other operators ${\cal T}_m$ even if they are ``composite'', as long as they are ``light'' and their scaling dimension does not scale with $N$:
\begin{align}
\nonumber
\tr\left(e^{-\beta H} {\cal T}_1 [{\cal O}_{\omega,\vect{k}}\,,\, {\cal O}_{\omega',\vect{k}'}] {\cal T}_2\right) =
&\tr\left(e^{-\beta H} {\cal T}_1[{\cal O}_{\omega,\vect{k}}^\dagger\,,\, {\cal O}_{\omega',\vect{k}'}^\dagger] {\cal T}_2\right)=0, \\
\label{thermalcoma}
\tr\left(e^{-\beta H} {\cal T}_1[{\cal O}_{\omega,\vect{k}}\,,\, {\cal O}_{\omega',\vect{k}'}^\dagger]  {\cal T}_2\right) 
= &\left(G_\beta(\omega,\vect{k})- G_\beta(-\omega,-\vect{k}) \right)\delta(\omega-\omega') \,\delta^{d-1}(\vect{k}-\vect{k}') \\ \nonumber &\times \tr\left(e^{-\beta H} {\cal T}_1 {\cal T}_2 \right).
 \end{align}

The commutator \eqref{thermalcoma} implies that the operators ${\cal O}_{\omega,\vect{k}}$ behave like the creation and annihilation operators of harmonic oscillators, though in an unconventional normalization. We can rescale them and define operators
\be
\label{rescaled}
\hatbb{\cal O}_{\omega,\vect{k}} = {{\cal O}_{\omega,\vect{k}} \over \left(
G_\beta(\omega,\vect{k})- G_\beta(-\omega,-\vect{k})\right)^{1\over 2}},
\ee
which have canonical commutation relations. If we compute the thermal expectation value of the ``occupation level'' $n_{\omega,\vect{k}}=\hatbb{\cal O}_{\omega,\vect{k}}^\dagger 
\hatbb{\cal O}_{\omega,\vect{k}} 
$ for each of these oscillators we have
\[
Z_{\beta}^{-1}{\rm Tr}\left( e^{-\beta H} \hatbb{\cal O}_{\omega,\vect{k}}^\dagger 
\hatbb{\cal O}_{\omega',\vect{k}'}\right) = { G_\beta(-\omega,
-\vect{k})\over G_\beta(\omega,\vect{k})- G_\beta(-\omega,-\vect{k})} \delta(\omega-\omega')\delta^{d-1}(\vect{k}-\vect{k}')
\]
and we find using \eqref{rotfourier} and \eqref{kmsfourier} that
\be
\label{thermaloca}
Z_{\beta}^{-1}{\rm Tr}\left( e^{-\beta H} \hatbb{\cal O}_{\omega,\vect{k}}^\dagger 
\hatbb{\cal O}_{\omega',\vect{k}'}\right)= {1\over e^{\beta \omega}-1} \delta(\omega-\omega')\delta^{d-1}(\vect{k}-\vect{k}'),
\ee
and similarly
\be
\label{thermalocab}
Z_{\beta}^{-1}{\rm Tr}\left( e^{-\beta H} \hatbb{\cal O}_{\omega,\vect{k}}
\hatbb{\cal O}_{\omega',\vect{k}'}^\dagger \right)= {e^{\beta \omega}\over e^{\beta \omega}-1} \delta(\omega-\omega')\delta^{d-1}(\vect{k}-\vect{k}'),
\ee
which is the standard occupation level for a harmonic oscillator of frequency $\omega$ when placed at temperature $\beta$. 

The physical interpretation of these results is that the modes \eqref{thermalcoma} are the CFT analogue of the AdS-Schwarzschild modes of a scalar field around a black hole in AdS that we called $a_{\omega,\vect{k}}$ in section \ref{modesbrane} hence we have the natural identification
\vskip5pt
\[
 \hatbb{\cal O}_{\omega,\vect{k}}\qquad\Leftrightarrow \qquad a_{\omega,\vect{k}}
\]
\vskip5pt
The CFT modes $\hatbb{\cal O}_{\omega,\vect{k}}$ seem to be thermally populated at the Hawking temperature of the black hole $\beta$, as we see in \eqref{thermaloca},\eqref{thermalocab}. This is the CFT analogue of the ``thermal atmosphere'' of the black hole that we discussed in the previous sections, for example in equations \eqref{hhocup}, \eqref{hhocupb}. In a sense, they can be thought of as a thermally excited gas of glueballs, hovering around the quark-gluon plasma --- though the interpretation
may be not fully accurate. More technically it means that, as expected, we get the occupation levels determined by the AdS-Hartle-Hawking vacuum for the dual scalar field. It is interesting that this result follows simply from general properties of thermal field theories {\it together with large $N$ factorization} and does not need any other assumptions about the coupling. In particular the same conclusion would hold for the regime of small $\lambda$ where 
the spacetime and the dual black hole would be highly stringy. On the other hand, at finite $N$ there is no sense in which the excitations of ${\cal O}_{\omega,\vect{k}}$ behave like a freely generated Fock space.

Also notice that, since these modes are ---in a sense--- fluctuations around a non Lorentz invariant background we have no reason to expect that they will exist only for timelike $(\omega,\vect{k})$. Indeed, as we see in appendix \ref{appendix2dthermal} for the special case of 2d CFTs where we can compute $G_\beta(\omega,\vect{k})$ analytically, we see that the modes exist for all values of $(\omega,\vect{k})$, even spacelike ones.

\subsection{Properties of thermal Fourier modes}
\label{subsec:thermalfourier}

As we mentioned above, in a thermal state, the spacelike Fourier modes of our generalized free fields do not vanish. The reader might be surprised by this, given that we argued in section \ref{sec:emptyads}, using nothing but large $N$ factorization and conformal invariance, that correlators involving the spacelike modes ${\cal O}_{\omega,\vect{k}}$ should vanish at leading order in ${1 \over N}$.  The resolution is the energy densities in the thermal state are $\Or[N^2]$, and so the previous argument breaks down. Nevertheless, as we will show below, even in the thermal state, it is true that 
{\em correlators involving operators with  large spacelike momenta die off exponentially.}

To see this, we note that
\[
\begin{split}
&G_{\beta}(\omega, \vect{k}) = \int  dt\,d^{d-1} \vect{x} \, e^{i \omega t - i \vect{k} \cdot x} G_{\beta}(t,\vect{x})\\
&= V_{d-3} \int dt d\theta d\norm{x} \, e^{i \omega t} G_{\beta}(t,\vect{x}) \norm{x}^{d-2} e^{i \norm{k} \norm{x} \cos \theta} (\sin(\theta))^{d - 3} , 
\end{split}
 \]
where $V_{d-3}$ is the volume of the $(d-3)$-sphere. Doing the $\theta$ integral, and using the rotational invariance of $G_{\beta}(t,\vect{x})$, we find that
\be
\label{thermalgfourier1}
G_{\beta}(\omega, \vect{k}) = \frac{\pi  2^{\frac{5}{2}-\frac{d}{2}} \Gamma (d-3) (\norm{k})^{\frac{3}{2}-\frac{d}{2}}
  }{\Gamma \left(\frac{d-3}{2}\right)}V_{d-3}\int dt d\norm{x}\,e^{i \omega t} G_{\beta}(t,\norm{x}) \norm{x}^{d-1 \over 2} J_{d - 3 \over 2}(\norm{k} \norm{x}) .
\ee
Using the identity
\[
\int_{0}^{\infty} J_{\nu} (p x) J_{\nu}(p y) p d p = {1 \over x} \delta(x - y),
 \]
we can rewrite \eqref{thermalgfourier1} as
\be
\label{thermalgfourier2}
\frac{\pi  2^{\frac{5}{2}-\frac{d}{2}} \Gamma (d-3) 
  }{\Gamma \left(\frac{d-3}{2}\right)}{V_{d-3}}  \norm{x}^{d-3 \over 2}  G_{\beta}(\omega, \norm{x}) =  \int_0^{\infty}  G_{\beta}(\omega, \norm{k}) J_{d - 3 \over 2}(\norm{k} \norm{x}) \norm{k}^{d - 1 \over 2} d \norm{k}.
\ee
We will now show below that if we consider the function $G_{\beta}(\omega, \norm{x})$ then we can continue $\norm{x}$ into the upper half plane till $\text{Im}(\norm{x}) = {\beta \over 2}$.  Since, for large $\norm{k}$ and $\text{Im}(\norm{x}) = {\beta \over 2}$ the Bessel function grows like $e^{\beta \norm{k} \over 2}$, we see that the integral in  \eqref{thermalgfourier2}  can only converge if $G_{\beta} (\omega, \norm{k})$ dies off as $e^{-{\beta \norm{k} \over 2}}$ for large $\norm{k}$ and fixed $\omega$. 

To see this analytic property, we write the thermal Green function as a sum over a complete set of states (indexed both by $|m\rangle$ and $|n\rangle$ below)
\be
\label{thermalgspectrumdecomp}
\begin{split}
 G_{\beta}(t,\norm{x}) &= \sum_{m, n} \langle m| e^{-\beta H} {\cal O}(t,\vect{x}) |n\rangle \langle n| {\cal O} (0,\vect{0})|m\rangle  \\
&= \sum_{m,n} e^{-\beta E_m} e^{-i (E_n-E_m)t + i (\vect{P}_n-\vect{P}_m) \cdot \vect{x} }|\langle m|{\cal O}|n\rangle|^2,
\end{split}
\ee
where we know, ahead of time, that the left hand side depends only on $\norm{x}$ although individual terms on the right hand side depend on $\vect{x}$. 
By Fourier transforming this expression in time, we find that  
\[
G_\beta(\omega,\norm{x}) = \int dt e^{i\omega t} G_{\beta}(t,\vect{x})  = \sum_{m,n} e^{-\beta E_m}  \delta(\omega-E_n+E_m)e^{i (\vect{P}_n-\vect{P}_m) \cdot \vect{x}}|\langle m|{\cal O}|n\rangle|^2.
 \]
Using the delta function we can rewrite this as
\be
\label{thermalgusedelta}
\begin{split}
&G_\beta(\omega,\norm{x}) =e^{\beta \omega /2} \sum_{m,n}  \delta(\omega-E_n+E_m) e^{-\beta E_m/2} e^{-\beta E_n/2} 
e^{i (\vect{P}_n-\vect{P}_m) \cdot \vect{x}}|\langle m|{\cal O}|n\rangle|^2 \\
&= e^{\beta \omega /2} \sum_{m,n}  \delta(\omega-E_n+E_m) e^{-\beta E_m/2} e^{-\beta E_n/2} 
e^{i |\vect{P}_n-\vect{P}_m| \norm{x} \cos \theta_{n m}} |\langle m|{\cal O}|n\rangle|^2,
\end{split}
\ee
where $\theta_{n m}$ is the angle between $\vect{P}_n - \vect{P}_m$ and $\vect{x}$. 

Now, in a relativistic QFT, we have the  ``spectrum condition'': in addition to $E\geq 0$ we also have that $E\geq |\vect{P}|$ for all states in the theory. Since 
by the triangle inequality, $ |\vect{P}_m + \vect{P}_n|  \leq |\vect{P}_m| + |\vect{P}_n|$, we see from the expansion \eqref{thermalgusedelta} that, viewed as a function of $\norm{x}$, the function $G_{\beta}(\omega, \norm{x})$  can be continued in the positive imaginary direction till $\text{Im}(\norm{x}) = {\beta \over 2}$.

This concludes our proof that for large $\norm{k}$ and fixed $\omega$, we have
\[
G_{\beta}(\omega, \norm{k}) \underset{\norm{k} \rightarrow \infty}{\longrightarrow} e^{-{\beta \norm{k} \over 2}}.
 \]
In fact, this observation is very important in reconstructing local bulk observables in the presence of a black hole as discuss in subsection \ref{comparelit}.

\subsection{Uplifting of fields and local bulk observables}
We are now ready to construct local bulk observables outside the horizon of the black hole. Let us consider bulk modes in the background of an AdS black hole (black brane) that we defined in section \ref{modesbrane} and which have the form $f_{\omega,\vect{k}}(t,\vect{x},z) = e^{-i\omega t+i\vect{k}\vect{x}} \psi_{\omega,\vect{k}}(z)$ and normalized so that $\psi_{\omega,\vect{k}} \sim {1 \over \normfact} z^{\Delta}$ as $z\rightarrow 0$. Then we construct the operator
\be
\label{uplifta}
\boxed{
\phi_{\text{CFT}}(t,\vect{x},z) = \int_{\omega>0}{d\omega d^{d-1}\vect{k} \over (2 \pi)^d} \, \left[{\cal O}_{\omega,\vect{k}} \, f_{\omega,\vect{k}}(t,\vect{x},z) + {\cal O}_{\omega,\vect{k}}^\dagger \, f_{\omega,\vect{k}}^*(t,\vect{x},z)\right]}
\ee
We emphasize once more, that this is an operator in the CFT. Notice that, if instead of the modes $f_{\omega,\vect{k}}(t,\vect{x},z)$ which go to $z^{\Delta}$ near the boundary, we work with the modes $\hatbb{f}_{\omega,\vect{k}}(t,\vect{x},z) $ which multiply canonically normalized oscillators in the bulk, then we can rewrite the expansion as
\be
\label{upliftb}
\phi_{\text{CFT}}(t,\vect{x},z) = \int_{\omega>0} {d\omega d^{d-1}\vect{k} \over (2 \pi)^d} \, \left[\hatbb{{\cal O}}_{\omega,\vect{k}} \, \hatbb{f}_{\omega,\vect{k}}(t,\vect{x},z) + \hatbb{{\cal O}}_{\omega,\vect{k}}^\dagger \, \hatbb{f}_{\omega,\vect{k}}^*(t,\vect{x},z) \right],
\ee
where the operators $\hatbb{\cal O}_{\omega,\vect{k}}$ are precisely the ones defined by equation \eqref{rescaled}. The point is that the normalization factor relating ${\cal O}_{\omega,\vect{k}}$ to $\hatbb{\cal O}_{\omega,\vect{k}}$ is precisely the inverse of that relating $f_{\omega,\vect{k}}(z)$ to $\hatbb{f}_{\omega,\vect{k}}$, hence the two expansions \eqref{uplifta} and \eqref{upliftb} are consistent.
Written in the form \eqref{upliftb}, it should be clear that at large $N$ the operator $\phi_{\rm CFT}(t,\vect{x},z)$ behaves like a free local field on the black hole background. In particular we have
\be
\label{bulkrecon}
Z_{\beta}^{-1}{\rm Tr}\left[e^{-\beta H} \phi_{\text{CFT}}(t_1,\vect{x}_1,z_1).....
\phi_{\text{CFT}}(t_n,\vect{x}_n,z_n)\right]_{\text{CFT}}  = \langle \phi_{bulk}(t_1,\vect{x}_1,z_1)...\phi_{bulk}(t_n,\vect{x}_n,z_n)\rangle_{\rm HH},
\ee
where the RHS is the correlator of a free scalar field in the black hole background (region $\front$) which is evaluated in the Hartle-Hawking vacuum.
Given this equality, we are entitled to identify the operator $\phi_{\rm CFT}(t,\vect{x},z)$ that we constructed with the local bulk operator in region $\front$ of 
the black hole. 

In particular, we have
\be
\label{localityout}
 [\phi_{\rm CFT}(t_1,\vect{x}_1,z_1)\,,\,\phi_{\rm CFT}(t_2,\vect{x}_2,z_2)] = 0
\ee
for two point $(t_1,\vect{x}_1,z_1)\,,\,(t_2,\vect{x}_2,z_2)$ which are spacelike separated {\it with respect to the metric} \eqref{threebrane}. This holds as an operator equation 
inserted inside the thermal trace, possibly together with the insertion of an $N^0$ number of other operators of the form $\phi_{\rm CFT}$. More generally, it holds as an operator
equation modulo caveats like those mentioned in section \ref{factorcav}.

\subsection{Comparison with previous studies \label{comparelit}}
Before we conclude this section, it is useful to compare our construction with that of \cite{Hamilton:2005ju}. The authors of that paper were working directly in position space, and attempted to write the bulk field as
\be
\label{bhkernel}
\phi_{\text{CFT}}(t,\vect{x},z) = \int dt' d^{d-1}\vect{x}' {\cal O}(t', \vect{x}') K(t,\vect{x},z;,t', \vect{x}')
\ee
for an appropriate choice of ``kernel'' $K$ . However, they found 
that attempting to compute $K$ directly led to a divergence, and so they were forced to complexify the boundary.

From the discussion above, we can see the origin of this divergence. Consider a mode in the bulk, with a given frequency $\omega$, and momentum $\vect{k}$ that solves the wave equation. Near the horizon we have the expansion
\be
\label{psinearh}                          
\hat{f}_{\omega,\vect{k}} \rightarrow e^{-i \omega t+i\vect{k}\vect{x}} \left( e^{i\delta_{\omega,k}}e^{i\omega z_*} + e^{-i \delta_{\omega,k}} e^{-i\omega z_*}\right),
\ee
where $z_*$ is the tortoise coordinate, in which horizon is at $z_*\rightarrow -\infty$. (For a precise definition of the tortoise coordinate, we refer the reader to section \ref{secadseternal}.) 

If we write the bulk field as
\be
\label{nearbhop}
\phi_{\text{CFT}}(t,\vect{x}, z) = \int_{\omega>0} {d \omega  d^{d-1} \over (2 \pi)^d} \vect{k}  {1 \over \sqrt{2\omega}}
\left( a_{\omega,\vect{k}}\hat{f}_{\omega,\vect{k}}(t,\vect{x},z) + {\rm c.c.}\right),
\ee
then merely the fact that $\phi_{\text{CFT}}$ must satisfy the canonical commutation relations near the horizon tells us that we must have
\be
\label{nearbhcommut}
[a_{\omega,\vect{k}},a^\dagger_{\omega',\vect{k}'}] = \delta(\omega-\omega')\delta^{d-1}(\vect{k}-\vect{k}').
\ee
Our analysis above then tells us that for $\beta \norm{k} >> 1$ and $\norm{k} >> \omega$, the relation between  ${\cal O}_{k, \omega}$ with $a_{k, \omega}$ must asymptote to
\[
a_{\omega, \vect{k}} = {\cal O}_{\omega,\vect{k}} e^{\beta \norm{k} \over 4}
 \]
However, if apart from doing this, we also try and Fourier transform this
expression to write $\phi_{\text{CFT}}(t,\vect{x},z)$ as an integral transform of ${\cal O}(t, \vect{x})$ as in \eqref{bhkernel}, it is clear that we will get a divergence. This is because, in the large-spacelike momenta region above, this Fourier transform
will look like
\[
 \int {\cal O}(\vect{x'},t') e^{-i \vect{k} \cdot \vect{x'} + i \omega t'} d t d^{d-1} \vect{x}  e^{\beta \norm{k} \over 4} \hat{f}_{\omega, \vect{k}}(t,\vect{x},z) d^{d -1} \vect{k} d \omega
 \]
Clearly, the integral over $\vect{k}$ diverges at least near the horizon of the black hole where it is clear that $\hat{f}_{\omega, \vect{k}}$ has no compensatory decaying exponential factor. 

However, this divergence is just fictitious.  The problem, of course, is that in doing this integral we have no way of taking into account the fact that the operator ${\cal O}_{\omega, \vect{k}}$ has a ``natural norm'' that is exponentially suppressed for large spacelike momenta. In momentum space, we do not have to deal with this problem.

It is worth making one more comment on this exponentially suppressed norm. The claim above (a) regarding spacelike modes in a thermal state, (b) the near-horizon expansion \eqref{psinearh}, (c) the commutation relations \eqref{nearbhcommut}, and finally (d) the claim that boundary correlators are limits of bulk correlators
\[
\langle {\cal O}(t_1, \vect{x}_1) {\cal O}(t_2, \vect{x}_2) \rangle_\beta =   \lim_{z_1, z_2 \rightarrow 0}\left[\normfact^2 z_1^{-\Delta} z_2^{-\Delta} \langle \phi_{\text{CFT}}(t_1,\vect{x}_1, z_1)   \phi_{\text{CFT}}(t_2,\vect{x}_2, z_2)\rangle_\beta\right],
 \]
If we consider the modes $\hat{f}_{\omega, k}(t, \vect{x}, z)$ normalized so that their near horizon expansion in \eqref{psinearh}, then near the boundary for large spacelike momenta, they must behave like
\be
\hat{f}_{\omega, k}(t, \vect{x}, z) \underset{z \rightarrow 0}{\longrightarrow} z^{\Delta} e^{-{\beta \norm{k} \over 4}} e^{-i \omega t+ i \vect{k} \cdot \vect{x}}
 \ee
These ideas, and this result, are examined and verified in a very concrete setting in the BTZ black hole background in Appendix \ref{appendix2dthermal}. 

%%% Local Variables: 
%%% mode: latex
%%% TeX-master: "infalling_paper"
%%% End: 

\section{Looking beyond the black hole horizon \label{sec:behind}}

In the previous section we reviewed the boundary construction of local bulk observables which lie outside the horizon of the black hole. What about points behind the horizon? 

For the reconstruction of local bulk observables in region $\front$ it was important to identify the modes of the scalar field in region $\front$ with the Fourier modes of the boundary operator.  In section \ref{semiblack} we explained that in order to write operators in region $\black$ (inside the black hole) we need modes {\it both} from region $\front$ and region $\other$. This raises the question: what should play the role of the modes of region $\other$ in the boundary field theory?

In fact the reader can easily persuade herself that, without these modes, it is not possible to write down a local field operator behind the horizon. Heuristically, the reason for this is as follows. If we try and extend the modes that we have constructed for region I, past the horizon, it is only the ``left moving'' modes that can be smoothly analytically continued past the horizon. To obtain a local field theory, we need independent right-movers behind the horizon, and these must come from some analogue of region $\other$.\footnote{The very perceptive reader might note that our condition of imposing normalizability at the boundary actually led to a constraint between left and right movers. This is true, but our condition was imposed on a time-like boundary. Behind the horizon, if we were to impose such a constraint, it would set constraints on a space-like boundary leading to a loss of locality.}

Now, as we have pointed out above, at late times, the geometry of a collapsing black hole is well approximated by the geometry of an eternal black hole. It is also known that an eternal black hole in anti-de Sitter space can be described by {\em two} CFTs --- one on each boundary of this geometry --- that are placed in a particular entangled state. This construction was first explored in \cite{Maldacena:2001kr}. In this picture, it is clear what plays the role of the required right-moving modes in the eternal black-hole. These are simply the right-moving modes that are constructed in region $\other$ and come from the second copy of the CFT. This story is consistent with the fact that, at finite temperature, the CFT is well described in the thermofield doubled formalism of Takahashi and Umezawa \cite{takahashi1996thermo}. 

In this thermofield doubled formalism, we double the Hilbert space of the theory and consider the pure state
\be
\label{thermostate}
|\Psi_{\text{tfd}}\rangle = {1\over \sqrt{Z_\beta}}\sum e^{-{\beta E \over 2}} |E\rangle \otimes |E\rangle,
\ee
where the states of the original theory are labeled by the energy $E$ (and perhaps other quantum numbers).  Corresponding to this ``doubling'' of the Hilbert space, every operator ${\cal O}$ in the original Hilbert space can be given a partner $\widetilde{\cal O}$ that acts on the second copy of the space. The point of this construction is that thermal
expectation values of operators in the original (single-copy) Hilbert space, can equivalently be computed as standard expectation values {\it on the  pure state} $|\Psi_{\text{tfd}}\rangle$ in the doubled Hilbert space
\[
 Z_{\beta}^{-1}{\rm Tr}\left[e^{-\beta H} {\cal O}(t_1,\vect{x}_1)...{\cal O}(t_n,\vect{x}_n) \right] = \langle \Psi_{\text{tfd}}| {\cal O}(t_1,\vect{x}_1)...{\cal O}(t_n,\vect{x}_n)|\Psi_{\text{tfd}}\rangle
\]
Moreover, correlators involving both ${\cal O}$ and $\widetilde{\cal O}$ in the doubled Hilbert space can be related to correlators
in the thermal, single-copy Hilbert space, with the replacement 
\[\widetilde{\cal O}\left(t,\vect{x})\rightarrow {\cal O}(t+i\beta/2,\vect{x}\right)\]
where the analytic continuation in complex time is defined  along the ``principal sheet''  as described in section \ref{therman}. More precisely, we have
\[
 Z_{\beta}^{-1}{\rm Tr}\left[e^{-\beta H} {\cal O}(t_1,\vect{x}_1)...{\cal O}\left(t_k+i\beta /2,\vect{x}_k\right)...\right] =
 \langle \Psi_{\text{tfd}}| {\cal O}(t_1,\vect{x}_1)...\widetilde{\cal O}(t_k,\vect{x}_k)...|\Psi_{\text{tfd}}\rangle
\]
It is an easy exercise to verify that this identification of $\widetilde{\cal O}(t,\vect{x})$ in the thermofield doubled Hilbert space, with ${\cal O}\left(t+i\beta/2,\vect{x}\right)$ inside the thermal correlators of the single Hilbert space, is consistent with the commutator $[{\cal O},\widetilde{\cal O}]=0$ in the doubled Hilbert space.

\vskip10pt

To summarize, it is well known that if we have a single theory {\it placed in a thermal density matrix}, we can map the thermal correlators to correlators in a {\it doubled theory} placed in the specific entangled pure state \eqref{thermostate}. And moreover, operators acting on the ``second copy'' of the doubled theory correspond to analytically continued versions of the thermal correlators of the original theory.  

However, besides this story about {\it thermal correlators}, we know that in the original single theory there is a large number of {\it pure states} which approximate the thermal density matrix to an excellent approximation. For these pure states, what is the meaning of the thermofield doubled formalism and of the second copy of operators $\widetilde{\cal O}$? We answer this question in the next subsection.

\subsection{Approximating a  pure state as a thermofield doubled state}
While the thermofield formalism is used quite frequently, and quite possibly the answer to our question above is clear to experts, we were surprised not to find it stated clearly in the literature.  We believe that the correct answer is as follows. 

Consider a pure-state $|\Psi\rangle$ in a theory with many degrees of freedom that thermalizes. We try to ``coarse-grain'' this state and identify some degrees of freedom that are easily accessible to measurement, and others that are not. The fact that the observables of the theory admit a division into ``easily accessible to measurements'' (coarse-grained observables) and those that are not (fine-grained observables), is a {\it prerequisite} to be able to talk about  ``thermalization of a pure state''. If we start with a closed quantum system, we can talk about ``thermalization of a pure state'' if there is some sense that we can divide the full system into two parts, one of which plays the role of the 
``environment/heat-bath'' and the other of the ``system in thermal equilibrium'' and if moreover the dynamics is such that under time evolution the pure state $|\Psi\rangle$, {\it when projected on the small system}, eventually looks approximately like a thermal density matrix. Clearly this is an approximate notion, which can only be made precise if there is some parameter which controls the validity of these approximations. In the case relevant to us, the full quantum system is the order $O(N^2)$ degrees of freedom of the large $N$ gauge theory, which may be in some specific pure state $|\Psi\rangle$, while the role of the ``sub-system'' that reaches thermal equilibrium is the $O(1)$ degrees of freedom corresponding to light gauge invariant operators. The parameter which controls the validity of the splitting into coarse- and fine-grained spaces is $N$.

For example, in the case that we have been considering above, the observer may be able to measure the eigenvalue of the ``occupation level'' $n_{\omega,\vect{k}} = \hat{\cal O}^{\dagger}_{\omega, \vect{k}} \hat{\cal O}_{\omega, \vect{k}}$ of the modes \eqref{rescaled} but this does not completely specify the state of the system; for that, the observer would need to measure several other operators including those that correspond to ``stringy'' and trans-Planckian degrees of freedom in the bulk. Using this intuition, we divide the Hilbert space of the CFT into a direct-product of a ``coarse-grained'' part and a ``fine-grained'' part
\be
\label{hcoarsefine}
{\cal H} = {\cal H}_{\text{coarse}} \otimes {\cal H}_{\text{fine}}.
 \ee

There are many such possible divisions of the gauge theory into coarse- and fine-grained degrees of freedom. However, as long as the coarse-grained degrees of freedom do not probe too much information --- we quantify this phrase below ---  about the state we will see that our interpretation below holds. 

We now consider some operator in the theory. For any generalized field ${\cal O}(t, \vect{x})$, what our low energy observer really measures is
\be
\label{coarseprojector}
{\cal O}_{c}(t,\vect{x}) = {\cal P}_{\text{coarse}} {\cal O}(t,\vect{x}),
\ee
where ${\cal P}_{\text{coarse}}$ is the ``projector'' onto the coarse-grained Hilbert space which traces out the fine grained degrees of freedom. We should mention that a division of the Hilbert space into ``coarse'' and ``fine'' degrees of freedom is often
performed in studies of ``dissipation'' in quantum systems. As far as we know, this was first attempted by von Neumann \cite{von2010proof} and our ``coarse'' operators are precisely the ``macroscopic'' operators of that paper. 

Now, let us turn to the pure state that we are considering, which is close to a thermal state. Corresponding to the Hilbert space decomposition above,  this state can be written as
\be
\label{entangledcoarsefine}
| \Psi \rangle = \sum_{i,j} \alpha_{ij} | \Psi^{c}_i \rangle \otimes | \Psi^{f}_j \rangle,
\ee
i.e. as an entangled state between the coarse- and fine-grained parts of the Hilbert space. Here the states $|\Psi^{c}_i\rangle$ run over an orthonormal basis of states in the coarse-grained Hilbert space, and $|\Psi^{f}_j\rangle$ run over
an orthonormal basis in the fine-grained Hilbert space.

In fact, what the low-energy observer sees directly is not
this entangled state, but rather a density matrix. Since the pure state is close to being thermal, we even know the eigenvalues of this density matrix. In some basis, the coarse-grained density matrix can be written
\be
\label{thermaldensity}
\rho_{c} = {1 \over Z^c_{\beta}}\text{diag}\left(e^{-\beta E_i} \right)
\ee
where $E_i$ are the eigenvalues of the coarse-grained Hamiltonian and $Z^c_{\beta} = \sum_i e^{-\beta E_i}$ is the coarse-grained partition function.\footnote{
 In in any given pure state $|\Psi \rangle$, the exact coarse-grained density matrix may be different from the exactly thermal one above, but these differences are unimportant in what follows, so we ignore them.}

What is the form of the density matrix in the fine-grained Hilbert space? In fact, as is well known, we can find a basis for the {\em fine-grained} Hilbert space so that the density matrix looks exactly like \eqref{thermaldensity} with zeroes in the other places. 

The argument is very simple, and we repeat it here for the reader's convenience. We need to ``diagonalize'' the matrix $\alpha_{i j}$ in \eqref{entangledcoarsefine}. This process, called a  ``singular value decomposition'', involves writing $\alpha_{i j}$ as
\[
\alpha_{i j} = \sum_m U_{i m} D_{m m} V_{m j}
 \]
where $U$ is a unitary matrix that acts in ${\cal H}_{\text{coarse}}$ and $V$ is a unitary matrix in  ${\cal H}_{\text{fine}}$, while $D$ is a rectangular-diagonal matrix i.e. a matrix of the form
\be
\label{rectangularmat}
D = \begin{pmatrix}
D_{11} & 0 & 0 & 0 &\ldots \\
0 & D_{22} & 0 & 0 & \ldots \\
 0 & 0 & D_{33} & 0 & \ldots
\end{pmatrix}
 \ee
This means that we can re-write \eqref{entangledcoarsefine} as
\be
\label{maximalentanglement}
| \Psi \rangle = \sum_i D_{i i} | \hat{\Psi}_i^{c} \rangle \otimes | \hat{\Psi}_i^{f} \rangle
\ee
where now, both $|\hat{\Psi}_i^{c}\rangle$ and $|\hat{\Psi}_i^f\rangle$ are orthonormal. In fact from \eqref{thermaldensity}, we already know that $D_{i i} = {1 \over \sqrt{Z^c_{\beta}}} e^{-{\beta E_i \over 2}}$.

Now, \eqref{maximalentanglement} immediately tells us that the fine-grained density matrix must also look like
\[
\rho_f = \sum_i D_{i i} |\hat{\Psi}_i^{f} \rangle \langle \hat{\Psi}_i^f |.
 \]
Let us return to the generalized free field ${\cal O}(t,\vect{x})$ that we have been discussing. As we mentioned, the low energy observer can only measure ${\cal O}_{c}(t,x)$ defined in \eqref{coarseprojector}. However, in the coarse-grained Hilbert space, we should be able
to write\footnote{Note, that we have kept the time evolution of the coarse operator arbitrary. In particular, the division \eqref{hcoarsefine} does not correspond to an analogous separation of the Hamiltonian, which may well couple the coarse and fine degrees of freedom. In studies of quantum dissipation, this coupling is sometimes quantified by means of an ``influence functional.'' It would be interesting
to understand this in detail for the boundary theory.}
\[
{\cal O}_{c}(t,\vect{x}) = \sum_{i_1, i_2} {\omega}_{i_1 i_2}(t,\vect{x})|\hat{\Psi}^c_{i_1} \rangle \langle \hat{\Psi}_{i_2}^c|.
 \]
Now, we {\em define} an operator that acts in the same way on the fine-grained Hilbert space
\be
\label{tildeodef}
\widetilde{\cal O}(t,\vect{x}) = \sum_{i_1, i_2} \omega^*_{i_1 i_2}(t,\vect{x})|\hat{\Psi}^f_{i_1} \rangle \langle \hat{\Psi}_{i_2}^f|. 
 \ee
The complex conjugation on $\omega$ is consistent with the original conventions of \cite{takahashi1996thermo}, and implies that the map between ${\cal O}$ and $\widetilde{\cal O}$ is {\em anti-linear.} This is necessary to ensure that the map is invariant under a change of basis that leaves \eqref{maximalentanglement} invariant.

What we have shown is that, even in a pure state for the gauge theory, for each generalized free-field ${\cal O}(t,\vect{x})$ and its corresponding modes ${\cal O}_{\omega, \vect{k}}$, we spontaneously obtain doubled operators $\widetilde{\cal O}(t,\vect{x})$, with corresponding modes $\widetilde{\cal O}_{\omega, \vect{k}}$! We explore this emergence further in section \ref{sec:applications} in a concrete example.\footnote{The coarse-fine splitting of the Hilbert space described in this section holds to an excellent approximation in the CFT at large-N. The construction described here, also
provides very important intuition including the ``state dependence''
of the $\widetilde{\cal O}(t,\vect{x})$ operators. However, we refer
the reader to \cite{Papadodimas:2013,Papadodimas:2013b} for a somewhat more
refined construction.}

These doubled operators are precisely what we need to construct local operators in region \black. However, before we attack that question we first want to mention a relation between the correlators of these operators and analytically continued correlators of ordinary operators.

From this point in this paper, we will drop the subscript ${\cal O}_c(t, \vect{x})$; it is understood that correlators now refer to the coarse grained operators only.

\subsection{Relation to analytically continued thermal correlators}
In this subsection, we will point out below that  $\widetilde{\cal O}$ operators  can---in a sense that we make precise ---be thought of as analytically continued versions of the ordinary operators. 

Let us say that we wish to compute a two-point correlator involving both ${\cal O}$ and $\widetilde{\cal O}$ in the state $| \Psi \rangle$ above
\be
\label{o1o2rel}
\begin{split}
&\langle \Psi | {\cal O}(t_1, \vect{x}_1) \widetilde{\cal O}(t_2, \vect{x}_2) | \Psi \rangle =  (Z^c_{\beta})^{-1}\sum_{i_1, i_2} e^{-{\beta (E_{i_1} + E_{i_2}) \over 2}} \langle \hat{\Psi}^{c}_{i_1} | {\cal O}(t_1, \vect{x}_1) | \hat{\Psi}^{c}_{i_2} \rangle  \langle \hat{\Psi}^{f}_{i_1} | \widetilde{\cal O}(t_2, \vect{x}_2) | \hat{\Psi}^{f}_{i_2} \rangle \\
&= (Z^c_{\beta})^{-1} \sum_{i_1, i_2} e^{-{\beta (E_{i_1} + E_{i_2}) \over 2}} \langle \hat{\Psi}^{c}_{i_1} | {\cal O}(t_1, \vect{x}_1) | \hat{\Psi}^{c}_{i_2} \rangle  \langle \hat{\Psi}^{c}_{i_1} | {\cal O}(t_2, \vect{x}_2) | \hat{\Psi}^{c}_{i_2} \rangle^* \\
&=  (Z^c_{\beta})^{-1} \sum_{i_1, i_2} e^{-{\beta (E_{i_1} + E_{i_2}) \over 2}} \langle \hat{\Psi}^{c}_{i_1} | {\cal O}(t_1, \vect{x}_1) | \hat{\Psi}^{c}_{i_2} \rangle  \langle \hat{\Psi}^{c}_{i_2} | {\cal O}(t_2, \vect{x}_2) | \hat{\Psi}^{c}_{i_1} \rangle \\
&= (Z^c_{\beta})^{-1} \sum_{i_1, i_2} e^{-\beta E_{i_1} } \langle \hat{\Psi}^{c}_{i_1} | {\cal O}(t_1, \vect{x}_1) | \hat{\Psi}^{c}_{i_2} \rangle  \langle \hat{\Psi}^{c}_{i_2} | {\cal O}(t_2 + {i \beta \over 2}, \vect{x}_2) | \hat{\Psi}^{c}_{i_1} \rangle \\
&= (Z^c_{\beta})^{-1}{\rm Tr}_c\left[e^{-\beta H} {\cal O}(t_1, \vect{x}_1) {\cal O}(t_2 + {i \beta \over 2}, \vect{x}_2) \right],
\end{split}
\ee
where the last trace is over the coarse grained Hilbert space.
Note that the operator ordering is important. The analytically continued operator always appears on the right. As we discussed above, when we discussed thermal correlators,  if this operator had appeared on the left, the correlator above would naively have diverged and we would have had to define it by giving a prescription for how to cross the branch cut. In fact, the ordering in \eqref{o1o2rel} corresponds to always remaining in the {\em principal branch} of the function ${\cal F}$ described in section \ref{therman}.

\subsection{Reconstructing the full eternal black hole from a single CFT}
We now describe how to reconstruct, from the CFT, local bulk operators behind the horizon. We emphasize that we are working in a {\em single CFT} in a {\em pure state.} 
As explained in the previous subsection, for every local operator ${\cal O}(t,\vect{x})$ we define the operator
\[
{\cal O}(t, \vect{x}) \rightarrow \widetilde{\cal O} (t,\vect{x}).
\]
and its Fourier modes
\[
\widetilde{\cal O}_{\omega,\vect{k}} = \int dt d^{d-1}\vect{x} \,e^{i\omega t-i\vect{k} \vect{x}}\, \widetilde{\cal O}(t,\vect{x})
\]
The complex conjugation in \eqref{tildeodef} implies that the  modes $\widetilde{\cal O}_{\omega,\vect{k}}$ generate an algebra that is identical to
that of the modes ${\cal O}^{\dagger}_{\omega,\vect{k}}$, while at the same time these two sets of modes commute with each other. In the previous 
section we identified the gauge theory modes ${\cal O}_{\omega,\vect{k}}$ with the Schwarzschild modes $a_{\omega,\vect{k}}$ in region I of the eternal black hole. It is clear that the modes $\widetilde{\cal O}_{\omega,\vect{k}}$ and $\widetilde{\cal O}_{-\omega,\vect{k}}$ --- which we emphasize act on the same Hilbert space, of the single CFT --- play the role of the bulk modes $\widetilde{a}^{\dagger}_{\omega,\vect{k}}$ and $\widetilde{a}_{\omega,\vect{k}}$  in region III of the eternal black hole. This means that we identify
\[
\begin{split}
\hatbb{\widetilde{\cal O}}_{\omega,\vect{k}}\qquad&\Leftrightarrow \qquad a^{\dagger}_{\omega,\vect{k}},  \\
\hatbb{\widetilde{\cal O}}_{-\omega,\vect{k}}\qquad&\Leftrightarrow \qquad a_{\omega,\vect{k}},  
\end{split}
\]
where both identifications assume $\omega > 0$ and $\hatbb{\widetilde{\cal O}}$ is defined in analogy to \eqref{rescaled}.

Having established this identification, it is now straightforward to write down local bulk operators of all four regions of the eternal black hole. For region I, we already did that in the previous section and we had
\[
\phi_{\text{CFT}}^{\rm I}(t,\vect{x},z) = \int_{\omega>0}{d\omega d^{d-1}\vect{k} \over (2 \pi)^d}\,\left[{\cal O}_{\omega,\vect{k}} f_{\omega,\vect{k}}(t,\vect{x},z) + {\cal O}_{\omega,\vect{k}}^\dagger f^*_{\omega,\vect{k}}(t,\vect{x},z)\right]
\] 
For region III, we have a similar expansion. We define the analogue of the AdS-Schwarzschild coordinates $t,\vect{x},z$ for region III, in terms of the Kruskal coordinates $U,V$ by the relations
\be
\label{kruskalcb}
\begin{split}
& u = t-z_*\qquad,\qquad v=t+z_* \\ &
U = e^{-{d u \over 2 \zhor}}\qquad,\qquad V = -e^{{d v \over 2 \zhor}}
\end{split}
\ee
with $z_*$ defined in terms of $z$ exactly as above equation \eqref{lightconec}. Then the CFT operator 
\be
\label{finalregionother}
\phi_{\text{CFT}}^{\other}(t,\vect{x},z) = \int_{\omega>0}{d\omega d^{d-1}\vect{k} \over (2 \pi)^d}\,\left[\widetilde{\cal O}_{\omega,\vect{k}} f_{\omega,\vect{k}}(t,\vect{x},z) + \widetilde{\cal O}_{\omega,\vect{k}}^\dagger f^*_{\omega,\vect{k}}(t,\vect{x},z)\right]
\ee
plays the role of the local bulk field in region $\other$. Notice that while we parameterize the points in region III by the same set of coordinates $t,\vect{x},z$, they are obviously distinct points from those in region I, since their $U,V$ values are different. 

Notice that the relations \eqref{kruskalcb} for region $\other$ are similar, but not quite the same, as the relations \eqref{kruskalc} for region $\front$. In particular, the different signs in the second line imply that as Kruskal time increases, $t$ decreases in region $\other$. Related to this, we have the fact that although \eqref{finalregionother} and \eqref{uplifta} look similar, $\widetilde{O}_{\omega, k}$ is actually a ``creation operator.'' (See appendix \ref{rindlerq} for a similar situation in the simpler case of Rindler space quantization.)

It is now straightforward to use the expansions in regions I and III to define a local field in regions II and IV. We focus on region II which is our main interest.
We can parametrize region $\black$ by $t,\vect{x},z$, with $\zhor<z<\infty$ and $-\infty<t<\infty$, which is now a spacelike coordinate. Remember that in these coordinates the singularity is at $z\rightarrow \infty$. The time $t$ in this region increases as we approach the horizon between $\front$ and $\black$. 
We look for solutions of the Klein Gordon equation of the form 
\[
g_{\omega,\vect{k}}(t,\vect{x},z) = e^{-i\omega t+i\vect{k}\vect{x}} \chi_{\omega,\vect{k}}(z)
\]
We get a 2nd order ODE for $\chi_{\omega,\vect{k}}(z)$. What is important is that in region II there are no boundary conditions that we need to impose (unlike what happened in region $\front$ or $\other$ where we imposed the ``normalizable'' boundary conditions at infinity). So it seems that in region $\black$ we have twice the number of modes as in region $\front$ (or $\other$). Of course this was to be expected since region $\black$ lies in the causal future of both regions $\front$ and $\other$ and the information of both regions $\front$ and $\other$ eventually enters region $\black$. So it is normal to have twice as many modes in $\black$.

Hence for any choice of $\omega,\vect{k}$ we get two linearly independent solutions $\chi_{\omega,\vect{k}}^{(1,2)}(z)$. We can choose the linear combination of these solutions so that as $z\rightarrow \zhor$ (from larger values, i.e. from inside the black hole) we have
\[
\chi_{\omega,\vect{k}}^{(1)}(z) \sim c(\omega, \vect{k}) (z-\zhor)^{-i \omega},\qquad \chi_{\omega,\vect{k}}^{(2)}(z) \sim c(-\omega, \vect{k})(z-\zhor)^{i \omega}\quad {\rm for}\quad z\rightarrow \zhor^+,
\]
where $c(\omega, \vect{k})$ also appeared in \eqref{nearhpsi}. 
Now imposing the continuity of the field
at the horizon between $\front$ and $\black$ and also between $\other$ and $\black$ we find that the field in region $\black$ has the expansion
\begin{equation}
\boxed{\label{finalbehind}
\phi^{\rm II}_{\text{CFT}}(t,\vect{x},z) =
\int_{\omega>0} {d\omega d^{d-1}\vect{k}  \over (2 \pi)^d}\left[ {\cal O}_{\omega,\vect{k}}\, g_{\omega,\vect{k}}^{(1)}(t,\vect{x},z) + \widetilde{\cal O}_{\omega,\vect{k}} \,g_{\omega,\vect{k}}^{(2)}(t,\vect{x},z)+ {\rm h.c.}
\right]}
\end{equation}
where $g^{(1,2)}_{\omega,\vect{k}}(t,\vect{x},z) = e^{-i\omega t+i \vect{k}\vect{x}}\chi^{(1,2)}(z)$. A similar construction can be performed for the field in region $\white$.

To summarize, using the boundary modes ${\cal O}_{\omega,\vect{k}}$ and $\widetilde{\cal O}_{\omega,\vect{k}}$ which were described above, we can reconstruct the local quantum field $\phi$ everywhere in the Kruskal extension and that at large $N$ the correlators of these operators satisfy the analogue of \eqref{bulkrecon}, but now operators can be inside or outside the horizon
\begin{equation}
\label{bulkreconb}
Z_{\beta}^{-1}{\rm Tr}\left(e^{-\beta H} \phi_{\text{CFT}}(t_1,\vect{x}_1,z_1) \ldots \phi_{\text{CFT}}(t_n,\vect{x}_n,z_n)\right)_{\text{CFT}}  = \langle \phi_{\text{bulk}}(t_1,\vect{x}_1,z_1)...\phi_{\text{bulk}}(t_n,\vect{x}_n,z_n)\rangle_{\rm HH}
\end{equation}
The expansions \eqref{finalbehind} can be written more explicitly for the case of the BTZ black hole, where the mode wave functions $g_{\omega,\vect{k}}^{(1,2)}$ can be found analytically.

In particular, for any two points $P_1,P_2$ in any of the two regions I,II we have
\be
\label{localityb}
 [\phi_{\rm CFT}(P_1)\,,\,\phi_{\rm CFT}(P_2)] = 0
\ee
when $P_1,P_2$ are spacelike separated with respect to the causal structure of the AdS-Kruskal diagram. Here, to write the operator $\phi_{\rm CFT}(P)$, we use either \eqref{uplifta} or \eqref{finalbehind} depending
on whether $P$ is in region I or II. Also, as before, this equation holds as an operator equation inside the thermal trace/or on a heavy
typical pure state, to leading order at large $N$ and modulo the caveats in section \ref{factorcav}.

Notice, that while it naively seems so, the claim \eqref{localityb} is not in conflict with the idea from black hole complementarity, that the Hilbert space of the interior of the black hole is not completely
independent from the Hilbert space outside. The point is that \eqref{localityb} holds only in the 
large $N$ limit and up to the aforementioned caveats. So while the commutator of {\it two simple} measurements of the scalar field, inside and outside the horizon, is zero in 
the large $N$ limit, if we consider the commutator of an operator inside the horizon with a very complicated measurement outside (which effectively corresponds to measuring $N^{a}\,\,\text{with}~ a>0$ instances of $\phi$), we have no reason to expect the commutator to vanish --- {\it not even in the large $N$ limit}. Hence the Hilbert
spaces inside and outside are not completely independent. These issues will be discussed in more detail in section \ref{sec:applications}.

%%% Local Variables: 
%%% mode: latex
%%% TeX-master: "infalling_paper"
%%% End: 

\section{Applications \label{sec:applications}}
Our construction above tells us that bulk correlators can be written in terms of boundary correlators. This has a significant advantage: it translates questions about quantum gravity in the bulk, which are mysterious, to questions about conformal field theory correlators that are well defined. 

There has been significant recent discussion of the information paradox. In an important paper, Mathur \cite{Mathur:2009hf} sharpened the contradiction between semi-classical evolution and unitarity in quantum gravity, by using the strong subadditivity of entropy. This argument was recently re-emphasized in \cite{Almheiri:2012rt}. Both papers pointed out that one way to resolve the contradiction would be to abandon the standard assumption that the horizon of the black hole is featureless. We can use our construction to obtain hints about the nature of the horizon, and also to the resolution of the information paradox. We do this in turn below. 

\subsection{Nature of the horizon \label{subsechorizonnat}}

First, let us address the issue of what happens to the infalling observer when he crosses the horizon. This is the same as the question of what happens to correlators of $\phi_{\text{CFT}}$ for points near the horizon in a typical heavy pure state. If the theory has a good large $N$ expansion in the thermal state then, as we show below, these correlators are the same (to leading order in ${1 \over N}$) to those in the thermal ensemble. In the thermal ensemble, it is clear that correlators of $\phi_{\text{CFT}}$ are the same as those predicted by 
semi-classical GR. 

The reader should not be concerned about the fact that on one side of the horizon we have both the left and right moving modes of a field constructed just from ${\cal O}$,  while on the other side the modes of ${\widetilde{\cal{O}}}$ appear. 
This is analogous to the fact that when we quantize a field in Minkowski space using Rindler coordinates (see Appendix \ref{rindlerq} for notation and a review), in region I only the modes $a_{\omega,k}$ appear, while in region II the modes $\widetilde{a}_{\omega,k}$ also appear.  Nevertheless, in the state corresponding to the Minkowski vacuum correlation functions are perfectly regular across the horizon. 
In
precisely the same fashion, the definition \eqref{tildeodef} ensures that correlators of $\widetilde{\cal{O}}$, in a typical pure state close to the thermal state,  have the same properties as those of ${\cal{O}}$, which we discussed in detail in section \ref{subsec:thermalfourier}. From here, it is easy to show that, in our construction, correlation functions are regular across the horizon.

So, we will conclude below that
\begin{quote}
For AdS duals of conformal field theories with a good ${1 \over N}$ expansion at finite temperature, our construction within AdS/CFT predicts that an observer falling through the horizon of a big black hole will measure the correlators predicted by semi-classical GR. In particular, he will not observe anything special about the horizon by doing either high energy\footnote{But not too high. Here we are talking about experiments at energy scale $E$, measured in the local frame of the infalling observer, where $E$ can be anything, as long as it does not scale with $N$ (or $\lambda$ --- in the case of the ${\cal N}=4$ SYM).} or low energy experiments. 
\end{quote}

This conclusion should be contrasted with the fuzzball proposal, which makes a different claim about the nature of the horizon.  We are aware that there are different perspectives on the fuzzball proposal. So, for the sake of precision, we will consider the following statement:
\begin{quote}
{\it (Extrapolated) Fuzzball Proposal:} The geometry is modified near the region where one expects to find the horizon, and it is this modification which contains information about the microstate that resolves the information paradox. It is possible to see these geometrical differences by doing experiments at the scale $T$--- the temperature of the black hole.\footnote{Regardless of the ability of the infalling observer to actually conduct experiments at this energy scale in his lifetime, this is a statement about correlation functions that we can examine.}
\end{quote}
It appears to us that this interpretation of the proposal is necessary if we are to resolve the information paradox using this proposal: the fuzzball proposal resolves the Hawking paradox by finding a mechanism to encode information about the black hole microstate in the Hawking radiation. This is also the statement that we seem to find in the recent literature \cite{Mathur:2009hf,Mathur:2012np, Mathur:2011uj, Avery:2012tf,Chowdhury:2012vd}. 

Since the geometry can be measured by correlators of the metric, and of other {\em light fields}, the fuzzball proposal must imply the correlators of light operators differ between the different states of the ensemble, so that if we do our construction in one state, we get one geometry whereas in another state we get another geometry. More precisely, we take a typical pure state $|\Psi\rangle$ in the CFT and evaluate the correlator  
\begin{equation}
\label{finalcorrelator}
\langle \Psi| \phi_{\text{CFT}}(P_1)...\phi_{\text{CFT}}(P_n) |\Psi\rangle
\end{equation}
where the points $P_1,...P_n$ correspond to bulk points along the trajectory of an infalling observer, then the proposal seems to be that this correlator gives us information about the microstate $|\Psi \rangle$.

Our construction tells us that (a)  $\phi_{\text{CFT}}$ can be written in terms of the boundary operator ${\cal O}$ and (b) to accurately evaluate
correlators involving insertions of $\phi_{\text{CFT}}$ with finite momentum and frequency, we do not need a parametric enhancement of this frequency 
on the boundary. The latter property is discussed in more detail in section \ref{sec:subtleties}.  So, our construction allows us to translate the extrapolated fuzzball proposal into the  following statement in the CFT:
\begin{quote}
{\it Implication of the (Extrapolated) Fuzzball Proposal}: By measuring low point correlators of light operators, we can distinguish between the various microstates of the CFT that constitute a black hole. 
\end{quote}
We will now show that this is impossible if the CFT admits a large $N$ expansion for thermal correlators. If such a large $N$ expansion exists--- as we believe it does for, say, ${\cn = 4}$ Super Yang-Mills at strong coupling--- then our construction, or the idea that bulk correlators can be written as non-singular transforms of CFT correlators is in contradiction with the fuzzball proposal.

Let us now consider a 2-point function for simplicity (higher-point functions can be considered similarly)
\[
\langle \Psi | {\cal O}(x_1) {\cal O}(x_2) | \Psi \rangle.
 \]
 The question is whether we can use this 2-point information to extract some information about the state $|\Psi\rangle$.  
Let us say that we can measure this correlator to order ${1 \over N^{\alpha}}$, where $\alpha$ is some fixed power (that does not scale like $N$). Then we have
\begin{align}
\label{corrfourpt} \langle \Psi | {\cal O}(x_1) {\cal O}(x_2) | \Psi \rangle &= \langle \Psi(\infty) {\cal O}(x_1) {\cal O}(x_2)  
\Psi(0) \rangle \\ \label{corrope} &= \sum_Q C_{{\cal O }{\cal O}}^Q \langle \Psi(\infty) {Q(x_2) \over |x_1 - x_2|^{2 \Delta_{\cal O} - \Delta_Q}} \Psi(0) \rangle \\ \label{corrthreept1}
&= \sum_Q C_{{\cal O}{\cal O}}^Q C_{\Psi \Psi Q}  {1 \over |x_1 - x_2|^{2 \Delta_{\cal O} - \Delta_Q} |x_2|^{\Delta_Q}} \\ \label{corrsumq}
&\equiv \sum_Q G_Q(x_1, x_2).
\end{align}
Here in \eqref{corrfourpt} we have written the 2-point correlator in the state $|\Psi\rangle$ as a 4-point correlator in the vacuum. In \eqref{corrope} we have used the OPE expansion to write the operator product of the two ${\cal O}$ operators as a sum over all other operators $Q$ in the theory, with coefficients that are fixed by conformal invariance up to 3-point coefficients, which are pure numbers, denoted by $C_{{\cal O}{\cal O}}^Q$ above.\footnote{More precisely, this is true only for scalar operators. We have suppressed tensor operators only to lighten the notation. The inclusion of these does not alter the argument in any way.} We have then used conformal invariance again to evaluate the remaining 3-point function, and finally $G_Q$ is merely a short-hand for the contribution of the operator $Q$ to the initial correlator.

Now the key point is that at any given value of $x_1, x_2$, if this correlator has a large $N$ expansion, and if we are measuring it to an accuracy ${1 \over N^{\alpha}}$, then the number of 
operators $Q$ that can contribute to this order must itself be bounded by $N^{\alpha}$. One might have hoped to disentangle the contribution of different operators $Q$ by separating the points $x_1$ and $x_2$ by some order $N$ distance, but as we saw in our construction above we do not need to do so to reconstruct local operators near the horizon if we are measuring bulk correlators that are separated by $O(1)$ near the horizon.

So the correlator really can, at most, contain information about the product $C_{{\cal O}{\cal O}}^Q C_{\Psi \Psi Q}$ for some $\kappa N^{\alpha}$ operators, where $\kappa$ is some  order $1$ number. However, the  black hole consists of $e^{S}$ states. Since $S \propto N$ (we remind the reader that, in our notation, $N$ is a measure of the central charge and not the rank of the boundary gauge group) and the temperature and the leading constants do not scale with $N$ in our setup, and so we can loosely say that we need $e^{N}$ pieces of information to identify the black hole microstate. Clearly this is impossible with the information we can glean from the correlator. To distinguish the different microstates of the black hole, we would need to measure correlators to an accuracy $e^{-N}$. At this level it is not clear (and probably not true) that the correlators have a geometric interpretation.

 It is useful to consider the example of 2-charge solutions \cite{Lunin:2001fv,Lunin:2002bj} which can be identified with the Ramond ground states
 of the  $D1-D5$ CFT. In that case also, most of the classical solutions are string scale. In the AdS$_5$ case, to obtain information about string scale objects, we would need to measure correlators to an accuracy $e^{-\lambda}$. However, what our construction here tells us is that to recover information about a {\it big} black hole, we will need to measure correlators to an accuracy that scales exponentially with $N$, not just with $\lambda$. While string scale geometries might possibly make sense, by switching duality frames, the degeneracy of big black holes looks like it might come from Planck scale geometries.  We do not understand how to discern geometries at the Planck scale.

\subsection{Recovering Information, small corrections, firewalls and complementarity}
We now turn to the issue of the information paradox itself. Indeed, this is the key issue in all the recent discussions on whether the nature of the horizon needs to be modified. For example, several regular solutions, with the same charges as the black hole, have been found in several theories (see \cite{Skenderis:2008qn} for a review and references to the very extended literature). These solutions are extremely interesting. However, by themselves, they are not enough to justify the extrapolated fuzzball proposal that we considered above. Instead, Mathur \cite{Mathur:2009hf} provided an indirect argument for why the nature of the horizon had to be 
modified to resolve the information paradox. 

A very similar argument has recently been made in \cite{Almheiri:2012rt}. However, starting with the same argument as Mathur,  Almheiri et al. stopped at  the (weaker) conclusion that the nature of the horizon must be modified. Since any such modification would generically cause the infalling observer to burn up, these authors spoke of ``firewalls'' rather than fuzzballs. 
 Since our conclusion suggests the opposite --- that the nature of the horizon is {\em not} modified --- we must explain how this is consistent with 
the preservation of information. We briefly review the information loss arguments below. We then describe why we disagree with these arguments and furthermore how our proposal suggests a natural resolution to the information paradox.  
\subsubsection{Strong Subadditivity and the Information Paradox}
\paragraph{Review of the ``Hawking Theorem''}
The argument that the information paradox cannot be resolved if we retain
the hypothesis that the horizon is featureless was termed the ``Hawking theorem'' in \cite{Mathur:2009hf}. 

Consider a toy model of the Hawking process where a pair of entangled qubits are produced at the horizon at each step; one of them is emitted into the radiation outside and the other falls into the black hole. So, at each 
stage we produce the pair
\be
\label{pairproduction}
{1 \over \sqrt{2}} \left( |0 \rangle |0 \rangle + | 1 \rangle | 1 \rangle \right).
\ee
Of these, the second bit goes out of the black hole, whereas the first bit falls inside. We will call the bit that goes outside $B$, while the bit that falls inside is $C$. Then, from \eqref{pairproduction}, we can abstract the lesson that $B$ and $C$ are maximally entangled up to small corrections
\be
\label{bcpure}
S_{BC} \approx 0.
\ee
This more general conclusion also follows from the semi-classical 
observation that, in a certain frame, the state of the black-hole is merely the vacuum, and Hawking radiation is obtained by performing a Bogoliubov transformation on this state. 

If we consider the density matrix of the bit outside, in this case, it is given by
\be
\label{rhosingle}
\rho_B =  {1 \over 2} \begin{pmatrix}1&0\\0&1 \end{pmatrix} \equiv {1 \over 2} I.
\ee
The entanglement entropy of the bit with the black hole is given by
\[
S_B = -\tr \left( \rho_B \log \rho_B \right) = \ln 2.
 \]
However, in general we can abstract away the rule that
\be
\label{bthermal}
S_{B} > 0, \quad S_{C} > 0.
\ee
Since $S_{BC} \approx 0$, we have $S_{C} \approx S_{B}$. 

Apart from the systems $B$ and $C$ designated above, let us also consider the system $A$, which comprises all the radiation that has been emitted by the black hole up to the step under consideration. If we assume that the black hole started in a pure state, then {\em eventually} the entanglement entropy of this radiation with the black hole should start decreasing. 
So, for a very old black hole, it must be true that
\be
\label{BpurifiesA}
S_{AB} < S_A.
\ee
This is the statement that when the bit $B$ goes and joins the radiation $A$, it decreases its entropy. 

Mathur \cite{Mathur:2009hf} pointed out that strong subadditivity tells us that \eqref{BpurifiesA}, \eqref{bcpure} and \eqref{bthermal} appear to be in contradiction. For three {\em separate} systems $A, B, C$, it is possible to show that
\be
\label{strongsubadditivitym}
S_{A} + S_{C} \leq S_{AB} + S_{BC}
\ee
Since we have $S_{A} > S_{AB}$ and $S_{C} > S_{BC}$, we seem to have a contradiction. 

The same argument was made in a slightly different way by \cite{Almheiri:2012rt}. They instead used the identity
\be
\label{strongsubadditivityp}
S_{ABC} + S_{B} \leq S_{AB} + S_{BC},
\ee
which is equivalent to \eqref{strongsubadditivitym}. Since $S_{BC} = 0$ (see \eqref{bcpure}), we have $S_{ABC} = S_{A}$. Then from \eqref{BpurifiesA}, we get
\be
S_{A} + S_{B} < S_{A}  \Rightarrow S_{B} < 0 \, ?
\ee
This clearly contradicts \eqref{bthermal}.

\paragraph{The Resolution\\}
The resolution to this paradox is automatically provided by our construction of the black hole interior in section \ref{sec:behind}:
\begin{quote}
Our construction makes it clear that for the black hole it is incorrect to apply \eqref{strongsubadditivitym} or \eqref{strongsubadditivityp} since $A, B, C$ {\em cannot} be treated as separate subsystems at the level of
accuracy where \eqref{BpurifiesA} is true. 
\end{quote}
Note that bit $C$ that falls in, is an excitation of the $\widetilde{\cal O}$ operators. However, as we explained in detail above, the $\widetilde{\cal O}$ operators arise because we coarse-grained our initial Hilbert space.

Now, at the start of black hole evaporation this is a perfectly good description since we can easily accommodate the
outgoing radiation $A$ in our coarse grained Hilbert space.\footnote{We should mention that, in a strict sense, our construction works for a big black hole in AdS. With standard reflecting boundary conditions, this black hole never evaporates. To make it evaporate, we need to couple the boundary 
theory to some other system which collects the ``glueballs'' as they form; such a system could mimic a boundary, which absorbs the radiation that reaches it. However, the reader who does not want to think of such a construction should note that the moral --- obtaining more and more fine details
of the system causes the semi-classical description in terms of a spacetime to break down --- is very robust and  should carry over to flat space black holes.}
As more and more bits emerge, we need to enlarge our coarse-grained Hilbert space to
describe them. 

As we do this, beyond a point, our construction of the $\widetilde{\cal O}$ operators breaks down. In a discrete Hilbert space, this happens when the number of rows in the matrix in \eqref{rectangularmat} (the dimension of ${\cal H}_{\text{coarse}}$) become larger than the number of columns (the dimension of ${\cal H}_{\text{fine}}).$ 

Indeed, it is precisely at this point --- when the density matrix of the radiation outside is of the same dimensionality as the rest of the system --- that we expect \eqref{BpurifiesA} to start being true. 

So, our construction points out that if we wish to make our description of particles outside the black hole so precise that it can keep track of the quantum state of all the particles that have been emitted by an old black hole then, at this level of precision, the semi-classical picture of spacetime is invalid. 

 This {\em does not} imply a breakdown of effective field theory. In fact an infalling observer using effective field theory would measure the operator that we constructed in \eqref{finalbehind} and its correlators. We can rephrase the ``strong subadditivity'' paradox in terms of correlation functions. However, it is clear that the number of operators in these correlation functions would have to scale with $N$. It is for such correlators, with $O(N)$ insertions that we cannot --- and should not expect to be able to ---  use effective field theory. 

We should mention that a variant of our ideas was discussed in \cite{Bousso:2012as,Susskind:2012uw}, where it was written in the form $A = R_B$ indicating that the interior of the black-hole was, in some sense, a ``scrambled'' version of the exterior. The authors of \cite{Bousso:2012as,Susskind:2012uw} decided that this led to problems involving ``quantum cloning'' (quantum information appears to be duplicated), and with ``chronology protection''. Our idea is subtly different: when the CFT is obsered at a coarse-grained level, it is possible to write down a semi-classical spacetime that reproduces these observations, with a very specific combination of the fine-grained degrees of freedom playing the part of the interior of the black hole. Both these regions are rewritings of parts of the same CFT Hilbert space. If we insist on a higher level of accuracy, then it is not quantum mechanics that breaks down but rather the {\em interpretation of CFT correlators} in terms of a semi-classical and local spacetime. 

\subsubsection{Can small correlations unitarize Hawking radiation?}
Once we have resolved the strong subadditivity paradox, there is still
apparently an information puzzle that makes no reference to the ingoing bit (i.e. to system $C$) : how can the large number of apparently
thermal bits that are emitted by the black hole lead to a pure state. In fact the resolution to this has
been understood for a long time \cite{Page:1993df}. We review this puzzle here and its resolution also.  The reader who is already familiar with this
can skip straight to the example in \ref{sec:toy}.
\paragraph{The Paradox\\}
Naively if we assume that the {\em same process} of Hawking radiation repeats $K$ times (where $K$ is the number of bits emitted), then the density matrix of the radiation outside will look like
\[
\rho_K = (\rho_B)^K,
 \]
where, by $\rho_B^K$ we really mean
\[
\rho_0^K \equiv \rho \otimes \rho \otimes \rho \ldots {\text{K~times}},
 \]
i.e. it is a $2^K \times 2^K$ dimensional matrix.

The entropy of this density matrix is
\be
\label{entangleK}
S_{\text{hawk}} = -\tr(\rho_K \ln \rho_K) = K \ln 2.
\ee

If we modify the density matrix \eqref{rhosingle} by a small amount --- such corrections would be expected through quantum and stringy effects --- but continue to assume that each Hawking emission is {\em exactly independent} then the entanglement entropy computed above does not change appreciably.

More precisely, if we take each individual density matrix to be
\[
\rho_{\text{str}} = \rho_1 + \epsilon \rho_{\text{corr}},
 \]
then
\[
S_{\text{hawk}} - \left[- \tr(\rho_{\text{str}}) \ln (\rho_{\text{str}})\right] \sim O(\epsilon).
 \]
This conclusion continues to hold even if we allow $\rho_{\text{corr}}$ at each step to be different but {\em uncorrelated}.

The paper \cite{Mathur:2012np} explains that ordinary objects like coal or burning paper avoid this paradox since in those cases successive emissions are not uncorrelated. So, for example, if the computer on which we are typing this were to go up in flames then successive portions of the computer would emit distinct and identifiable thermal emissions. First, the keyboard would burn, and by collecting that radiation, an observer could recover some information about the keyboard, even if the keyboard were only a negligible fraction of the mass of the full computer. 

However, this is simply the statement that small subsystems of everyday objects are {\em not} maximally entangled with the rest of the object. As we will show in a toy example below, by introducing very small off-diagonal elements (which, however, may link the first bit emitted to the last bit emitted), we can unitarize Hawking radiation. 

\paragraph{The Resolution\\}

Just to be specific, let us consider a 4-dimensional asymptotically flat black hole of Schwarzschild radius $R_{bh}$ and take $l_{\rm p}$ to be the Planck scale. In the lifetime of this black hole, it will emit about $N \approx \left({ R_{\text{bh}} \over l_{\text{p}}}\right)^2$ quanta.  Clearly, for the first $K << N$ of these quanta, the correction is unimportant and the density matrix is well approximated by the Hawking matrix. 

Let us imagine that the density matrix that is produced at each step is actually given by
\be
\label{unitarizingcorr}
\rho_{\text{exact}} = \rho_{\text{hawk}} + 2^{-N} \rho_{\text{corr}},
\ee
where $\rho_{\text{corr}}$ is a density matrix that, in the natural basis of observables, has elements that are $O(1)$. This correction is exponentially suppressed and can have several origins. Moreover, since it is exponentially suppressed in ${R_{\text{bh}} \over l_{\text{p}}}$, it cannot be detected at any order in perturbation theory: this is an inherently non-perturbative correction. 

This correction is reminiscent of the ``second saddle point'' discussed in \cite{Maldacena:2001kr}. It would be nice to see if such a semi-classical correction can be identified directly in Lorentzian space. However, even if such a semi-classical perspective does not exist, and this correction is visible in the gauge theory but cannot be interpreted in terms of a simple geometric process, that would not be a contradiction. 

Now, it is clear from \eqref{unitarizingcorr} that for the first $K$ emissions, where $K << N$, the density matrix of the radiation is given by the Hawking matrix to an excellent approximation. However, as $K$ grows large and becomes comparable to $N$, we see that the individual elements of the Hawking density matrix become so small, that they are comparable to the size of the corrections. At this point, the corrections can no longer be neglected.

In fact, with the numerical pre-factors that we have inserted in front, the corrections are precisely of the correct order to unitarize the process. Note that after $N$ steps, the Hawking density matrix looks like
\[
\rho_{\text{hawk}} = {1 \over 2^N} I_{2^N \times 2^N},
 \]
where $I_{2^N \times 2^N}$ is the identity matrix in $2^N$ dimensions. 

For the full density matrix to be unitary, we must have, {\em in some basis}, 
\[
\rho_{\text{exact}} = \begin{pmatrix}1&0&0&\ldots&0\\
0&0&0&\ldots&0\\
\ldots&\ldots&\ldots&\ldots&\ldots \\
0&0&0&\ldots&0\\
\end{pmatrix}.
 \]
So, in this basis---which may be quite unnatural from the point of observables accessible to an observer at low energies---the correction matrix must look like
\be
\label{rhocorrunusualbasis}
\rho_{\text{corr}} 
%= \rho_{\text{exact}} 
= \begin{pmatrix}2^N-1&0&0&\ldots&0\\
0&-1&0&\ldots&0\\
\ldots&\ldots&\ldots&\ldots&\ldots \\
0&0&0&\ldots&-1\\
\end{pmatrix}.
\ee
We now show that this is not inconsistent with the statement below \eqref{unitarizingcorr}--- that $\rho_{\text{corr}}$ has all elements of $O(1)$ in a basis of natural observables. 

For \eqref{rhocorrunusualbasis} to hold, we must have
\[
\tr(\rho_{\text{corr}}^2) = 2^{2 N}.
 \]
However, even if all the elements of $\rho_{\text{corr}}$ in an ordinary basis are $O(1)$, we have
\[
\tr(\rho_{\text{corr}}^2)= \sum_{i j} \rho_{\text{corr}}^{i j} \rho_{\text{corr}}^{i j} = \Or[2^{2N}],
 \]
since the sum over $i,j$ runs over $2^{2N}$ elements. 

\subsubsection{An Example \label{sec:toy}}
Let us now give an example where all the ideas are realized. In particular:
\begin{enumerate}
\item
We will construct a toy model that describes the emission of ``spins''; each emitted spin has an almost exactly thermal density matrix. 
\item
We will show how small corrections can unitarize this process so that after more than half the system has ``evaporated'', the density matrix that describes all the emitted spins together starts becoming pure.
\item
We will show how, for each spin that is emitted, we can also identify (within an effective description), a corresponding ``spin'' that remains within the system and is perfectly anti-correlated with the emitted spin. This is the analogue of the bit ``C'' that falls inside the black hole. However, we will show that 
this effective description breaks down at {\em precisely} in time to avoid any contradiction with the strong subadditivity of entropy.
\end{enumerate}

Consider a system of $N$ spins, each of which can be in two states--- $|0\rangle$, or $|1 \rangle$. The Hilbert space of the system is spanned by $2^N$ basis vectors, each of which can be represented by a single binary number between $0$ and $2^{N-1}$. We can use this as a convenient representation of our basis
\[
|0 \rangle_{b} \equiv |000\ldots00 \rangle, |1 \rangle_{b} \equiv |000\ldots01\rangle, |2 \rangle_{b} \equiv |000\ldots10\rangle, |3 \rangle_{b} \equiv |000\ldots11\rangle, \ldots,
 \]
where we have placed a $b$ in the subscript of the new basis to distinguish it from the old one.
Now, consider the state
\be
\label{typicalstate}
|\Psi\rangle = {1 \over 2^{N/2}} \sum_{j=0}^{2^N-1} p_j | j \rangle_b,
\ee
where $p_j$ is a number that can be either $1$ or $-1$. There are $2^{2^N-1}$ such states (since there are $2^N$ coefficients, but states that are the same up to an overall minus sign are equivalent) but let us consider a ``typical state'' in this set --- where we pick the $p_j$ coefficients with an equal probability to be either $1$ or $-1$.

In doing so, we are not considering the ensemble of states formed by the various choices of the $p_j$. We make some choice, and then consider the pure state $|\Psi\rangle$ corresponding to this choice. In speaking of a ``random'' choice above, we are merely pointing out that for our purposes the precise choice of the sequence $p_j$ is not important, and most choices will work.

Now, if we consider the reduced density matrix corresponding to the first spin, we find that we have
\begin{align}
\rho_1 &= {1 \over 2^{N}} \left(\sum_{j=0}^{2^{N-1}-1} p_{2 j}^2 |0 \rangle \langle 0| + p_{2 j+1}^2 |1 \rangle \langle 1| + p_{2 j} p_{2 j +1}\Big(
|0 \rangle \langle 1| +|1 \rangle \langle 0| \Big)\right) \\
&= {1 \over 2} \Bigg(|0 \rangle \langle 0| + |1 \rangle \langle 1| + \Or[2^{-{N\over2}}] \Big(|0 \rangle \langle 1|+|1 \rangle \langle 0| \Big) \Bigg).
\end{align}
Here, in the last step we have used the fact that successive $p_j$ coefficients are uncorrelated. Depending on the precise choice of the $p_j$ coefficients, that we started out with, we will get different values for ${1 \over 2^N} \sum p_{2 j} p_{2 j + 1}$. However, for a typical state, we expect this value to be of the order $2^{-{N \over 2}}$.\footnote{For a very few states, this value can be close to $1$, but these states are not ``typical'' and do not meet our purpose.}

Thus we see that for ``typical states'', the density matrix of the first spin is very close to the density matrix predicted by Hawking. The full density matrix is also of the form \eqref{unitarizingcorr}. However, the corrections become important once we start considering (in this example) ${N \over 2}$ spins. 

Thus if the dynamics of this spin chain naturally leads it to one of these typical states, then its evaporation will appear much like Hawking radiation. By a basis change, we can transform $\rho_{\text{corr}}$ into precisely the form \eqref{rhocorrunusualbasis}.

However, this is not sufficient. Indeed, as we pointed out above, a key feature of Hawking radiation is that for each emitted photon, there is also an ``ingoing photon'' that can perhaps be observed by an observer inside the black hole. The two photons are maximally entangled; each of them individually looks thermal but taken together they form a pure state. We referred to this feature in \eqref{pairproduction} above.

From our construction of the $\widetilde{O}$ operators above, we know how to construct the ingoing bit. To make this more precise, let us assume that in a ``coarse-grained'' description we can observe the first $p$ bits of the $N$ qubits in the toy model above. We assume that the other $N-p$ bits are much harder to observe and enter only in the ``fine grained'' description of the system.  In the same notation as above, we can write down a basis for both of these spaces
\[
\begin{split}
\text{coarse}: \quad |i\rangle_{\text{c}}, ~ \text{with}~ 0 \leq i \leq 2^p - 1\\
\text{fine}:\quad |j\rangle_{\text{f}}, ~\text{with}~ 0 \leq j \leq 2^{N-p} - 1.
\end{split}
 \]
The subscripts $c$ and $f$ emphasize that these are in a different vector space, from the space we used to describe the full system above. 
Now the state $|\Psi\rangle$ can clearly be written as
\[
| \Psi \rangle = \sum_{i, j} a_{i j} |i, j \rangle = \sum_i |i \rangle_{\text{c}} \otimes | \phi_i \rangle_{\text{f}},
 \]
where $|i, j\rangle \equiv |i\rangle_{\text{c}} \otimes |j\rangle_{\text{f}}$ and 
\[
|\phi_i \rangle_{\text{f}} \equiv \sum_j a_{i j} | j \rangle_{\text{f}}.
 \]
Let us call $S_1$ the operator acting on the first spin as the Pauli matrix $\sigma_3 = \begin{pmatrix} & 1 & 0 \\ & 0 & -1\end{pmatrix}$.
With this notation, the operator $\widetilde{S}_1$ in this basis can be written as
\be
\label{tildes1}
\widetilde{S}_1 = {\cal I}_{\text{c}} \otimes \left( \sum_{i=2^{p-1}}^{2^p - 1} | \phi_i \rangle  \langle \phi_i | - \sum_{i = 0}^{2^{p-1} - 1} | \phi_i \rangle \langle \phi_i | \right),
\ee
where ${\cal I}_{\text{c}}$ is the identity on the coarse-grained space. 

Note that the definition of $\widetilde{S}_1$ depends not only on the state $| \Psi \rangle$ but also on our division of the space into a coarse-grained and a fine-grained part. This operator $\widetilde{S}_1$ commutes with all operators in the coarse-grained space as is evident from \eqref{tildes1}
\[
[\widetilde{S}_1, S_m] = 0, ~\text{for}~1 \leq m \leq p.
 \]
Moreover, in the state $|\Psi \rangle$ measurements of $S_1$ are precisely anti-correlated with measurements of $\widetilde{S}_1$.  Similarly, we can define operators $\widetilde{S}_2, \ldots \widetilde{S}_p$ that are anti-correlated with $S_2 \ldots S_p$ respectively, and also commute with all the ordinary operators.

It is also important to recognize where this process breaks down. Once we expand our description of the coarse-grained system to $p = {N \over 2}$, then the construction above is no longer possible; there is simply not ``enough space'' in the fine-grained space. 
%Nevertheless, we could still define effective operators $\widetilde{S}_m$ that behave approximately like the operators above. For example, even for $p > {N \over 2}$, we could retain the definition of $\widetilde{S}_1$ from the case where $p = {N \over 2}$. This $\widetilde{S}_1$ will not commute with, say, $S_{{N \over 2} + 1}$. However, to detect this lack of commutation in the state $|\Psi\rangle$, we will have to make measurements that are precise up to $2^{-{N \over 2}}$.

Before we conclude this section, we would like to emphasize that this toy model holds important lessons for what happens in the real black hole. First it shows us how exponentially suppressed corrections can conspire, in an equally large Hilbert space, to restore unitarity to a seemingly thermal density matrix. Second, this system also gives us a toy model of black hole complementarity.  As long as less than half the spins have been measured, it is possible to pair each emitted spin with a perfectly anti-correlated and independent degree of freedom in the remaining spins. This is analogous to the statement that as long as we are measuring a suitably limited number of observables, there is no difficulty in describing the interior of the black hole as a separate space where semi-classical dynamics holds. However, once we start probing the system very finely by going either to very high energies, or with very heavy operators, the semi-classical spacetime description starts to break down. This is precisely what 
happens 
in the spin model above.  

%%% Local Variables: 
%%% mode: latex
%%% TeX-master: "infalling_paper"
%%% End: 

\section{Various technicalities \label{sec:subtleties}}

In this section, we discuss some subtle points of our construction.

\subsection{Stability of the reconstructed bulk operators/``spread of the transfer function''}

A possible objection to our construction is the following: in Schwarzschild coordinates, an infalling observer never crosses the horizon, since the point of crossing formally has Schwarzschild time $t\rightarrow \infty$. Moreover, the  ``time'' of the gauge theory is naturally identified with the Schwarzschild time in the bulk. These statements together may be taken as an indication that in order to reconstruct local operators at and behind the horizon, we need information from the gauge theory {\it for all times}. If true, this could be a serious problem since we expect the late time behavior of pure states to have significant variance due to Poincare recurrences and other kinds of statistical fluctuations, including ${1 \over N}$ 
corrections that might pile up over time.

The same conclusion seems to follow if one (wrongly) assumes that the support of the operator $\phi_{\rm CFT}(t,\vect{x},z)$ is mostly localized on the intersection of the boundary of AdS with the spacelike cone emanating from the point $(t,\vect{x},z)$. This (wrong) assumption seems to suggest that as the point moves close to the horizon, the support of the operator spreads over the entire boundary and moreover it becomes difficult to understand what happens when the point moves behind the horizon.

Fortunately these arguments are not really correct and we will explain that the reconstructed bulk observables $\phi_{\rm CFT}(t,\vect{x},z)$ only require knowledge of the boundary fields for (approximately) a finite interval in boundary time, which can be made parametrically smaller, in the large $N$ limit, from time scales related to fluctuations and recurrences.

We believe that our momentum space formulation helps us get a sharp sense of the accuracy of our construction. In particular, the question above can be converted to the following questions:
\begin{enumerate}
\item
From a measurement of boundary correlators, with a given accuracy and over a certain time-scale, how accurately can we extract the modes ${\cal O}_{\omega,\vect{k}}$ that we require above.
\item
Can a small inaccuracy in the ${\cal O}_{\omega,\vect{k}}$ modes get blown up to a large inaccuracy in the reconstructed bulk fields $\phi_{\rm CFT}$.
\end{enumerate}

Let us start with the first question.  Let us say that we  measure the correlator on the boundary for a {\em finite time interval} $[0,T]$. (It is trivial to do a similar analysis for the spatial coordinates, but for simplicity, we focus on time here.) Moreover, the observer samples this correlator at short-time intervals $t_{uv}, 2 t_{u v}, \ldots$ for a total number of samples ${T \over t_{uv}}$. (We will take this to be an integer to lighten the notation.) Using this procedure how accurately  can we discern the various frequencies? 

\begin{figure}
\begin{center}
\includegraphics[height=5cm]{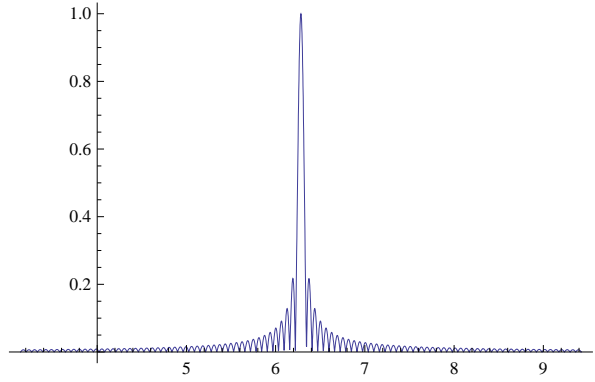}
\caption{$R_{\omega}$ with $\nu = 1$ \label{spreadfuncplot}}
\end{center}
\end{figure}

Let us say that the original correlator is $C(t) = \int C_{\omega} e^{-i \omega t}$. By doing a discrete Fourier transform using the sample points above, we can measure the frequency modes
\be
\label{measuredc}
\hat{C}\left({2 \pi \nu \over T} \right) =  {t_{u v} \over T} \sum_{j = 0}^{{T \over t_{u v}} - 1} \int C_{\omega} e^{-i \omega j \over T} e^{2 \pi i \nu j \over T} d \omega = \int C_{\omega} R_{\omega} d \omega 
\ee
with 
\be
R_{\omega} = {t_{uv} \over T} \frac{e^{i T (2 \pi  \nu -\omega )} -1}{e^{ i t_{uv} (2 \pi \nu - \omega)}-1 }.
\ee
We remind the reader (see figure \ref{spreadfuncplot}) that if $\omega t << 1$ and $\omega T >> 1$, then $R_{\omega}$ is a sharply peaked function with a maximum of $1$ at $\omega = 2 \pi \nu$.
 
On the other hand, if $\omega$ is a very low frequency --- signals at ultra-low frequencies can be produced by phenomena like Poincare recurrence --- then \eqref{measuredc} is negligible except at $\nu = 0$. So these frequencies do not contaminate the measurement of Fourier modes with higher frequencies.

Similarly, \eqref{measuredc} also tells us that if we wish to measure
the Fourier mode to some accuracy ${1 \over N^{\alpha}}$, then we can do
that by measuring the real-time signal to the same accuracy --- we do not
require any parametric enhancement of accuracy at low frequencies.

So we reach the expected conclusion it is possible measure the Fourier modes
of the operator ${\cal O}_{\omega, \vect{k}}$ between some IR-cutoff ${1 \over T}$ and some UV-cutoff ${1 \over t_{uv}}$ to any given accuracy
by sampling position correlators at intervals smaller than $t_{uv}$ and for a length of time larger than $T$, with the same accuracy.

Having now established that the Fourier modes ${\cal O}_{\omega, \vect{k}}$ 
can be measured in principle, the next question has to do with whether
low frequency modes become very important near the horizon. In order to do that we consider the bulk 2-point function of our local operators $\phi_{\rm CFT}(t,\vect{x},z)$.

\subsubsection{Outside the horizon}
First we start with points outside the horizon. We have
\begin{align*}
&{1 \over Z_{\beta}} {\rm Tr}\Big(e^{-\beta H} \phi_{\rm CFT}(t_1,\vect{x}_1,z_1)  \phi_{\rm CFT}(t_2,\vect{x}_2,z_2)\Big)  \\
&= (2\pi)^d \int_{\omega>0} {d\omega \over (2 \pi) 2 \omega} {d^{d-1}\vect{k} \over (2 \pi)^{d-1}}\, \begin{aligned}[t] \Bigg[&\hat{f}_{\omega,\vect{k}}(t_1,\vect{x}_1,z_1) \hat{f}^*_{\omega,\vect{k}}(t_2,\vect{x}_2,z_2) {e^{\beta \omega}
\over e^{\beta \omega}-1} \\ & + 
 \hat{f}^*_{\omega,\vect{k}}(t_1,\vect{x}_1,z_1) \hat{f}_{\omega,\vect{k}}(t_2,\vect{x}_2,z_2) {1
\over e^{\beta \omega}-1} \Bigg]
\end{aligned}
\end{align*}
Here we used the definition of the bulk operators \eqref{upliftb} and the thermal expectation values of the Fourier modes $\hat{\cal O}_{\omega,\vect{k}}$ that we discussed around equation \eqref{thermaloca}. Remember that the hatted modes $\hat{f}_{\omega,\vect{k}}(t,\vect{x},z)$ are those which are canonically normalized with respect to the Klein Gordon norm.

The question about the late-time sensitivity is a question about the region of this integral around $\omega=0$. First we notice that we have an explicit factor of ${1\over \omega}$. Second, the thermal factors ${e^{\beta\omega}\over e^{\beta\omega}-1}$ and ${1\over e^{\beta\omega}-1}$ both go like ${1\over \beta \omega}$ for low $\omega$. Third, the modes have the property that for fixed $(t,\vect{x},z)$ we have that $\hat{f}_{\omega,\vect{k}}(t,\vect{x},z)$ goes to zero linearly with $\omega$ for small $\omega$. This can be verified explicitly in the case of the modes on a BTZ background (see appendix \ref{appendix2dthermal}) but is also true for other black holes. All in all we find that if we keep the points fixed and consider low $\omega$, the integrand goes like $\omega^0$ and hence the integral converges.

Since the integral is convergent in the limit $\omega=0$, it means that the sensitivity of the operator $\phi_{\rm CFT}(t,\vect{x},z)$ to the region of small $\omega$ (or equivalently ``late-time'' in position space) is actually small. In order to demonstrate this let us consider the order of various limits a bit more carefully. Let us consider the reconstructed local bulk operator \eqref{uplifta}, but with the inclusion of an IR cutoff $\delta$ in frequency space
\[
\phi_{\rm CFT}^{\delta}(t,\vect{x},z) \equiv \int_{\omega>\delta}{d\omega d^{d-1}\vect{k} \over (2\pi)^d}\, \left[{\cal O}_{\omega,\vect{k}} \, f_{\omega,\vect{k}}(t,\vect{x},z) + {\cal O}_{\omega,\vect{k}}^\dagger \, f_{\omega,\vect{k}}^*(t,\vect{x},z)\right]
\]
The operator $\phi_{\rm CFT}^{\delta}$ is not exactly the same as our original operator $\phi_{\rm CFT}(t,\vect{x},z) =\phi^{\delta=0}_{\rm CFT}$ defined in \eqref{uplifta}. However, since the integral is convergent in the region of $\omega=0$, the difference between correlation functions of $\phi^\delta_{\rm CFT}$ and correlation functions of $\phi^{\delta=0}_{\rm CFT}$ goes to zero as $\delta\rightarrow 0$.

Coming back to the question about the sensitivity to the details of the pure state: suppose we want to reconstruct the bulk with some  ``resolution'' $\epsilon$ (which we take to be $N$-independent). For any given $\epsilon$ there is an IR cutoff $\delta$, such that the correlators of $\phi_{\rm CFT}^\delta(t,\vect{x},z)$ reproduce those of a local bulk field up to the accuracy $\epsilon$, {\it if the boundary correlators agreed with the thermal ones, down to the IR frequency cutoff $\delta$}. But for any given and fixed IR cutoff $\delta$ in frequency space, we can take $N$ to be large enough, so that the boundary correlators on a typical pure state agree with those of the thermal ensemble down to $\omega \approx \delta$.

Putting everything together, we find that for any desired ``resolution $\epsilon$'', it is possible to take $N$ to be large enough and to ensure that the details of the typical pure state become unimportant.

\subsubsection{Inside the horizon}

In order to reach the same conclusion for points behind the horizon, we need to demonstrate that the integral over $\omega$ in the 2-point function for points behind the horizon, is well-behaved around $\omega=0$. Here the situation is more interesting. The reconstructed bulk operator is given by \eqref{finalbehind}, that is
\[
 \phi^{\rm II}_{\text{CFT}}(t,\vect{x},z) =
\int_{\omega>0} {d\omega d^{d-1}\vect{k} \over (2 \pi)^d} \left[ {\cal O}_{\omega,\vect{k}}\, g_{\omega,\vect{k}}^{(1)}(t,\vect{x},z) + \widetilde{\cal O}_{\omega,\vect{k}} \,g_{\omega,\vect{k}}^{(2)}(t,\vect{x},z)+ {\rm h.c.}
\right]
\]
When computing the 2-point function of this operator we find several contributions coming from the 2-point functions of the  ``usual'' modes ${\cal O}_{\omega,\vect{k}}$, from the 2-point functions of the  ``tilde'' modes $\widetilde{\cal O}_{\omega,\vect{k}}$, as well as from cross-terms between the two.

If we focus on the contribution to the 2-point function coming from the terms $\langle {\cal O}_{\omega,\vect{k}}
{\cal O}^\dagger_{\omega',\vect{k}'}\rangle_\beta$ and $\langle {\cal O}_{\omega,\vect{k}}^\dagger {\cal O}_{\omega',\vect{k}'}\rangle_\beta$ we find that the integral over $\omega$ is divergent from the $\omega=0$ region. This is the Fourier-space manifestation of the arguments mentioned in the first two paragraphs of this section. It might seem that this small $\omega$ divergence would invalidate our claims.

However, instead of focusing only on {\it some} of the terms contributing to the 2-point function, if we consider the contribution from all terms together, i.e. including the ``tilde'' modes (and the cross terms) we find that the integral over small $\omega$ becomes convergent! The divergence encountered above when considering the modes ${\cal O}_{\omega,\vect{k}}$ alone, disappears when we consider both ${\cal O}_{\omega,\vect{k}}$ and $\widetilde{\cal O}_{\omega,\vect{k}}$ together!

This means that there is a specific way to reorganize the terms in the integral over $\omega$, so that the integral becomes manifestly convergent for small $\omega$. It turns out that if we group together the modes ${\cal O}_{\omega,\vect{k}}$ with the $\widetilde{\cal O}_{\omega,\vect{k}}$ in the natural ``Kruskal'' combinations
\[
 {\cal K}^{(1)}_{\omega, \vect{k}} = {{\cal O}_{\omega,\vect{k}} - e^{-{\beta \omega\over 2}} \widetilde{\cal O}_{\omega,\vect{k}} \over \sqrt{1-e^{-\beta \omega}}}\qquad,\qquad
{\cal K}^{(2)}_{\omega, \vect{k}} =  {\widetilde{\cal O}_{\omega,\vect{k}}^\dagger - e^{-{\beta \omega\over 2}}{\cal O}^{\dagger}_{\omega,\vect{k}} \over \sqrt{1-e^{-\beta \omega}}}
\]
then the integral is manifestly convergent for small $\omega$. The Kruskal creation operators are, of course, just given by the Hermitian conjugates
of the relations above.

As a matter of fact, this situation can also be studied in the case of Rindler space, where the modes are easy to write down explicitly. We present the analysis in appendix \ref{rindlerq}. In Rindler space, when computing the 2-point function for points behind the ``Rindler horizon'', we find the same small $\omega$ divergence, when considering the contribution from only one set of modes. 
We can check explicitly that the divergence disappears if the modes are grouped together in the ``Unruh modes''. The relation between the Unruh and the Rindler modes given in \eqref{unruh} is precisely the same as the relation between the Kruskal and the AdS-Schwarzschild modes above.

To summarize, for points behind the horizon the integral over small $\omega$ is still convergent, which means that the sensitivity of the bulk 2-point function to the low $\omega$ (or late time) boundary correlators is small. The argument about the order of limits mentioned in the previous subsection can be repeated
in identical form, and we conclude that the reconstructed black hole region II is --- in the large $N$ limit --- insensitive to the details of the specific pure state.

\subsubsection{What happens exactly on the horizon?}
At the horizon, the mode functions do not diverge but they start oscillating very rapidly. We can see this from the formula \eqref{psinearh}. Near the horizon, in terms of the coordinates $U$ and $V$ defined in section \ref{secadseternal}, the modes are simply
\[
\hat{f}_{\omega, \vect{k}} = e^{i \vect{k} \cdot \vect{x}} \left(e^{i \delta_{\omega, \vect{k}}} e^{-2 \pi \zhor \omega \over d} U^{2 i \zhor \omega \over d} + e^{-i \delta_{\omega, \vect{k}}} V^{-2 i z_0 \omega \over d} \right)
\]
Let us say that we have been able to measure boundary modes accurately up to some frequency ${2 d \over \zhor N^{\alpha}}$. ($\zhor$ is a natural scale since it also sets the temperature.)  Near the future horizon, which is at $U = 0$, modes below
this frequency become important only for 
\be
\label{breakdownregion}
\ln(U) \sim -N^{\alpha} \Rightarrow U \sim e^{-N^{\alpha}}.
\ee
It is only in this region that our construction starts to develop inaccuracies.

In fact the proper time taken by the infalling observer to cross the region \eqref{breakdownregion} is itself exponentially suppressed in $N$. It is unclear to us whether it is possible even, in principle, to speak of the experience of the infalling observer over such a short time scale. If we {\em average} over the experience of the observer as he crosses the black hole, over some time scale that is larger than this --- even a time scale that is power law suppressed in $N$ and scales like $\Or[{1 \over N^{\alpha}}]$ --- then we see that the tiny frequencies that we have neglected on the boundary do not cause any difficulties.

\subsection{Pure vs thermal states}

A tacit assumption in our analysis in sections \ref{sec:behind} and \ref{sec:outside} was that correlators of the operator $\phi_{\rm CFT}(t,\vect{x},z)$ in a ``typical'' heavy pure state are the same as its correlators in
the thermal ensemble. This is consistent with general intuition from statistical mechanics and it was on this basis that we concluded that an infalling observer will not notice anything special as he falls through the horizon, when the black hole is in a pure state.

The argument from the beginning of section \ref{subsechorizonnat} provides further weight to this expectation. In that section, we argued that if
correlators of light operators factorize on a typical heavy pure state
then these correlators cannot be used to distinguish between various typical pure states $|\Psi \rangle$ in the thermal ensemble. More precisely, our arguments suggest that by measuring these correlators to an accuracy ${1 \over N^{\alpha}}$, we could divide these pure states into $\Or[N^{\alpha}]$ different classes but still not pinpoint which one we are in since there are $\Or[e^{-N}]$ different states in the ensemble. 

But, how do we know that factorization holds in a pure state? Even if 
factorization holds for thermal correlators --- and we might be able
to make an argument for this by applying the usual power counting
arguments to the Feynman diagram expansion of these correlators in the 
Schwinger-Keldysh formalism --- it is not obvious that this carries over to pure states.

However, the argument for this is simple. The key point is that the question of how different an observable $A$ is, on a typical pure state, from the same observable on the thermal ensemble, is a question about the {\it variance} of the observable across the ensemble of all pure states. This can be estimated from computing the {\it thermal expectation value of the square of the observable}. 

In other words the variance of $A$ across different pure states can be related to the following quantity {\it which can be computed within the thermal ensemble}
\be
\label{variance}
{1 \over Z_{\beta}} {\rm Tr}(e^{-\beta H} A^2) - {1 \over Z_{\beta}^2} {\rm Tr}(e^{-\beta H}A)^2  
\ee

Let us apply this measure to the observable 
\[
A = {\cal O}_{\omega_1, \vect{k_1}}  {\cal O}_{\omega_2, \vect{k_2}}  {\cal O}_{\omega_3, \vect{k_3}}  {\cal O}_{\omega_4, \vect{k_4}}   -  {1 \over Z_{\beta}^2} {\rm Tr} \left(e^{-\beta H} {\cal O}_{\omega_1, \vect{k_1}}  {\cal O}_{\omega_2, \vect{k_2}} \right)  {\rm Tr} \left(e^{-\beta H} {\cal O}_{\omega_3, \vect{k_3}}  {\cal O}_{\omega_4, \vect{k_4}} \right)- \ldots,
\]
where $\ldots$ indicate the other
products of two point functions that enter here. Then the fact that 
the expectation value of $A$ and its powers vanishes in the thermal ensemble up to ${1 \over N}$ means that no ``appreciable'' class of pure states can have non-vanishing $A$.

One might also wonder about the significance of $\Or[1]$ variance of
the operators given in \eqref{thermaloca} and \eqref{thermalocab}. In fact
this variance has a natural semi-classical interpretation: it just indicates the statistical fluctuations in the thermal gas of particles that 
surround the black hole. These fluctuations do not modify the leading order geometry, since they are an $O(1)$ and not $O(N^2)$ effect and thus cannot backreact in the large $N$ limit. Moreover the amount of information 
contained in them, as we argued before, is parametrically smaller than that of the entropy of the black hole itself.

\subsection{Definition of tilde operators}
As the reader will have noticed, the construction of the $\widetilde{\cal O}$ operators in section \ref{sec:behind} suggests that the precise Heisenberg operators $\widetilde{\cal O}$ that an observer will encounter may differ quite significantly depending on which pure state the black hole is in. The important point is that this non-uniqueness is unimportant from an operational point of view. The bulk observer lives in a particular black hole microstate. By measuring correlators that involve a finite number of bulk fields $\phi$, the observer can only infer that the $\widetilde{\cal O}$ operators {\em effectively} satisfy the same algebra as the ${\cal O}$ operators. The reader may wish to consult the toy model in section \ref{sec:toy} for an explicit construction of the $\widetilde{\cal O}$ operators, where this dependence on the state and its operational insignificance can both be seen. 

The $\widetilde{\cal O}$ operators also depend on the specific division of the CFT Hilbert space into a coarse and a fine grained part. Once again, this dependence is not operationally significant. This property can also be seen in the toy model of section \ref{sec:toy}. 
\subsection{Do our operators describe the ``real'' infalling observer?}
A question that we have commonly encountered is: ``how do we know that this description corresponds to the `real' experience of the infalling observer.'' 

Before we address this, let us briefly emphasize a philosophical point, which is uncontroversial and even seemingly banal.  Let us say that we are given a quantum system ${\cal Q}$ and that in some approximation the accessible observables in this system can be re-organized into observables $\phi^{(i)}(\vect{x})$, where ${\vect{x}}$ are points on some manifold ${\cal M}$ and, and $i$ is some index that labels the operator in question.  Moreover if the correlators of these operators are the same as the correlators of perturbative fields propagating on ${\cal M}$, then the system ${\cal Q}$ is indistinguishable from the system describing perturbative fields on ${\cal M}$. 

In this paper, we have shown that the natural observables of a CFT with large $N$ factorization can be reorganized, at leading order, into the correlators of a non-local CFT operator $\phi_{\text{CFT}}(t, \vect{x}, z)$, which is labeled by points in AdS (or an AdS black hole). All the accessible dynamical processes of the CFT can be given a description in terms of these perturbative fields propagating on AdS (black hole).

Ultimately, ontological questions cannot really be settled by physics. However, we hold that such a situation is indistinguishable from a ``true'' perturbative field propagating in this spacetime. So our operators do describe the ``real'' infalling observer.

\subsection{Uniqueness of our construction and interactions}
We now turn to the issue of the uniqueness of our construction, which is something that we glossed over above. In fact, even at \eqref{finalpoincare}, the reader could have asked: ``what impels us to multiply the creation and annihilation operators with the modes inside anti-de Sitter space. Why can't we choose modes from some other spacetime.''  Of course, the AdS/CFT correspondence tells us that any other spacetime will not work. Already at this level, we see that if we would like the bulk theory to realize the symmetries of the boundary in a natural way (as isometries), then we should choose modes from anti-de Sitter space.

Furthermore, we believe that if we choose a different spacetime then we will not be able to correct our prescription at subleading orders in ${1 \over N}$ consistent with locality. We can see the difficulty immediately. While writing down \eqref{finalpoincare}, we emphasized that one reason it worked was because we did not have to worry about the ``spacelike'' modes ${\cal O}_{\omega,\vect{k}}$ at leading order in perturbation theory. However, if we go even to ${1 \over N}$, we cannot consistently neglect these modes and, in fact, their presence will lead to a conflict with locality as was explored in \cite{Kabat:2011rz}. Let us briefly describe how this problem can be fixed and how we can extend our construction to higher orders in ${1 \over N}$. 

Our argument is somewhat indirect. It is believed that if we take a consistent interacting CFT with various conditions on its operator spectrum (such as the presence of only a small number of operators at low dimensions) then we should be able to write down a bulk interaction in anti-de Sitter space that reproduces any set of boundary correlators \cite{Heemskerk:2009pn, Heemskerk:2010ty,
Fitzpatrick:2010zm,Fitzpatrick:2012cg,ElShowk:2011ag}. Of course, implementing this procedure in practice is quite difficult but schematically let us say that we have constructed the bulk Lagrangian that reproduces the boundary correlators. For the operator ${\cal O}$ under consideration, let us say that it is of the form:
\[
L_{\text{bulk}} = \int \sqrt{-g} \left[\partial_{\mu} \phi \partial^{\mu} \phi - V(\phi) \right],
 \]
where $V(\phi)$ has a perturbative expansion in powers of ${1 \over N}$. 

Then, as was discussed in \cite{Kabat:2011rz,Kabat:2012hp,Heemskerk:2012mn} it is possible to correct our prescriptions above perturbatively in the interaction. We now simply start solving the Heisenberg equations of motion:
\[
\Box \phi = V(\phi).
 \]
with the zeroth order solution $\phi_0$ taken to be \eqref{finalpoincare} or, in the black hole background, to be \eqref{finalbehind}. 
So, with the bulk Green function $G(x,y)$ where $x,y$ are bulk points, at first order we have:
\[
\phi_1(y) = \int G(x,y) V(\phi_0) d y,
 \]
where, of course, at this order only the lowest order terms in ${1 \over N}$ in $V(\phi_0)$ contribute. We can, of course, keep track of the higher order terms
and use the solution $\phi_1$ to repeat this procedure at any order in perturbation theory. The Heisenberg field that we obtain in this way is manifestly local. 

The point, of course, is that if we are not in AdS we cannot write down a bulk interaction consistent with boundary correlators and so we do not have an algorithm for extending our construction to higher orders in ${1 \over N}$. This is, of course, not a proof but is highly suggestive. We would like to explore this further in future work.\footnote{In fact the example of conformal gravity \cite{Maldacena:2011mk} is already subtle. In this example, it was shown that it is possible to reproduce correlators in AdS$_4$/CFT$_3$ with ordinary Hilbert-Einstein gravity in the bulk by consider another theory on  flat space (cut off in one direction at $z = 0$) but with the Lagrangian of conformal gravity. We expect that this is only a tree-level coincidence and cannot be extended to higher orders in ${1 \over N}$}.

%%% Local Variables: 
%%% mode: latex
%%% TeX-master: "infalling_paper"
%%% End: 

\section{Conclusions and further directions \label{sec:conclusion}}

Let us summarize the implications of our construction. Since the paper is already quite long we will be as brief as possible. 

In what follows it is important to keep in mind
that our precise results apply to a black brane or a big black hole\footnote{Although, for notational reasons, we presented the results for a black brane it should be clear that similar local bulk observables
can be defined for a big black hole in global AdS, by replacing the integrals over $\vect{k}$ with sums over spherical harmonics.} in AdS, neither of which evaporate. We leave it to the reader to decide how relevant the following conclusions are for  black holes in asymptotically flat space.

\vskip5pt
\noindent {\bf Reconstructing local bulk observables outside the black hole:} 
We revisited the construction of local operators near a big AdS black hole. At large $N$ it is straightforward to write CFT operators which reconstruct local bulk fields. We pointed out that a divergence which appears when attempting to write a transfer function in position space, is a harmless artifact --- effectively ``regularized'' by the general behavior of boundary thermal correlators --- and can be evaded by working with the operators in momentum space. In particular, there is no need for an analytic continuation along the spacelike directions of the boundary. The construction of local bulk observables outside the black hole seems to be robust under the inclusion of $1/N$ corrections.

\vskip10pt

\noindent {\bf Reconstructing the region behind the horizon:} We argued that in certain theories, including large $N$ gauge theories, there is a natural splitting of the Hilbert space into coarse- and fine-grained components. Typical pure states have the property that they ``self-thermalize'', i.e. the reduced density matrix of the coarse-grained Hilbert space is automatically driven by the dynamics very close to the thermal one. In such a situation, for every operator acting on the coarse-grained Hilbert space, it is natural to identify a ``partner'' operator acting on the fine-grained Hilbert space. These partner operators obey an identical algebra as the 
coarse-grained operators, and can be identified with the  ``tilde''  operators in the ``thermofield doubled Hilbert space'' formalism. Using these partner operators we can reconstruct local bulk observables behind the horizon of the black hole. We argue that, despite what one might naively expect,  local bulk observables behind the horizon are 
essentially supported over finite time scales on the boundary and hence --- in the large $N$ limit --- they are not too sensitive on the specific microstate. This implies that the ``interior geometry'' (as probed by low-energy experiments) looks the same for all pure states, in contradiction with (some versions of) the fuzzball proposal.

\vskip10pt

\noindent {\bf Fate of the infalling observer:} From these results it follows that a semi-classical observer falling cross the horizon of a big black hole,  will not measure anything special. In particular he will not see a firewall or a fuzzball. Our precise prediction is the following. If we first fix: i) the size of the black hole in AdS units,  ii) the trajectory of the infalling observer and the points at which he will measure the local fields, iii) the number of measurements he can make and the accuracy he has iv) the type (mass) of fields that he can measure {\it and then} send $N$ to infinity (i.e. we do not scale any of the previous quantities with $N$), then the observer will not notice any difference from semi-classical GR. Whether this is consistent or not with the fuzzball proposal depends on its precise definition. However, to the extent that the fuzzball proposal posits a departure from
semi-classical GR at energy scales of the order of the temperature of the black hole, our construction seems to contradict it.

\vskip10pt

\noindent {\bf Lessons about the information paradox:}  We argued that, contrary to the claims of \cite{Mathur:2012np} and \cite{Almheiri:2012rt}, {\it small corrections to the Hawking computation can restore
unitarity}. The corrections suggested by our construction are small in the sense that low point correlators of perturbative fields are almost exactly the same as those computed by semi-classical quantum gravity.  The intuitive reason that this is consistent with the unitarity of the Hawking process is that the number of emitted Hawking particles is large, so even small correlations between them can carry away the information. 
The works of \cite{Mathur:2012np} and \cite{Almheiri:2012rt}, claimed that this is not possible by invoking the strong subadditivity theorem applied to particles located {\it both inside and outside}  the black hole.
Our construction of local bulk observables, explicitly demonstrates that the semi-classical Hilbert spaces corresponding to the interior and exterior of the black hole {\it are not independent}
and hence it is not permissible to use the strong subadditivity theorem.
We discussed this in the context of a simple qubit toy model. Our construction suggests a form of ``black hole complementarity'' from the point of view of the boundary conformal field theory, which would be interesting to understand in more detail.

\vskip10pt

\noindent {\bf Future work:} Clearly there are many questions that need to be explored in more detail. An obvious one is to understand how to extend the reconstruction of the bulk in perturbation theory in $1/N$. We would like to further investigate the properties of the tilde operators, the splitting of the Hilbert space into coarse- and fine-grained components and the conditions under which an isolated quantum system can undergo thermalization. We hope to revisit this question in future work.
It would be nice to find a more realistic dynamical toy-model that would capture the essential features of complementarity. Finally, given that we have constructed local bulk observables behind the horizon, it would obviously be interesting to study what happens when the bulk point approaches the black hole singularity.

%%% Local Variables: 
%%% mode: latex
%%% TeX-master: "infalling_paper"
%%% End: 

\section{Acknowledgments}
We would like to thank  Ofer Aharony, Luis Alvarez-Gaume, Nima
Arkani-Hamed, Jan de Boer, Borun Chowdhury, Sheer El-Showk, Ben
Freivogel, Rajesh Gopakumar, Monica Guica, Dan Harlow, Gary Horowitz,
Veronika Hubeny, Norihiro Iizuka, Tom Hartman, Elias Kiritsis, Neil
Lambert, Klaus Larjo, Gilad Lifschytz, Loganayagam,  David Lowe, Juan Maldacena, Don
Marolf, Paul McFadden, Shiraz Minwalla, Boris Pioline, Joe Polchinski,
Mukund Rangamani, Ashoke Sen, Masaki Shigemori, Joan Simon, Sandip
Trivedi, Erik Verlinde, Spenta Wadia, Edward Witten and Xi Yin for helpful discussions. We would
also like to thank all the participants of the discussion meeting on
string theory at ICTS, Bangalore, in June 2012 and members of the quantum information group at HRI. KP would like to thank the organizers and participants
of the ``Bits, Branes and Black Holes'' workshop at KITP 2012, the Cosmology and Complexity 2012 meeting in Hydra, 
the Amsterdam summer string workshop 2012, the ``Holographic Way'' workshop in
Nordita and the Crete Center for Theoretical Physics for hospitality. SR is partially supported by a
Ramanujan fellowship of the Department of Science and Technology
(India).  SR gratefully acknowledges the hospitality of Brown
University, where part of this work was done.

\appendix

\section{Thermal correlators in 2d CFTs}
\label{appendix2dthermal}
\subsection{Modes in a BTZ black hole}
In our discussion above we frequently referred to modes in the black
hole background. In this appendix, we explore these modes in the
background of a BTZ black hole. This will allow us to check several
claims including their behaviour near the horizon and their growth for
large spacelike momenta that we made above. 

Let us start with the usual BTZ coordinates
\[
ds^2 = -(r^2-r_h^2)dt^2 + (r^2-r_h^2)^{-1}dr^2 + r^2 dx^2.
\]
The horizon is at $r_h$ and $x$ runs from $-\infty$ to $+\infty$ i.e. we are looking at a planar BTZ black hole. The temperature of the black hole is
\[
\beta = {2\pi \over r_h}
\]
We consider a massive scalar whose dual operator has dimension $\Delta$, obeying the equation
\[
(\Box -m^2)\phi=0
\]
Solving the KG equation with an ansatz of the form $e^{-i\omega t}e^{i k x}\psi(r)$ we find two linearly independent solutions
\[
\psi_1(r)=\left({r^2\over r_h^2}\right)^a\left({r^2\over r_h^2}-1\right)^b \, _2F_1\left(1+a+b-{\Delta \over 2},a+b+{\Delta \over 2},1+2b,1-{r^2\over r_h^2}\right)
\]
\[
\psi_2(r) =\left({r^2\over r_h^2}\right)^{a}\left({r^2\over r_h^2}-1\right)^{-b}\, _2F_1\left(1+a-b-{\Delta
\over 2},a-b+{\Delta \over 2},1-2b,1-{r^2 \over r_h^2} \right)
\]
where 
\[
a = {i k \over 2 r_h}\qquad b= {i \omega \over 2r_h}
\]
For any given $\omega,k$ only a specific linear combination of these modes is normalizable at infinity. Using hypergeometric identities we find that the normalizable combination is
\[
\psi_1(r) -\left[{\Gamma(1+2b)\Gamma(a-b+{\Delta\over 2})
\Gamma(-a-b+{\Delta\over 2})\over \Gamma(1-2b) \Gamma(-a+b+{\Delta \over 2})
\Gamma(a+b+{\Delta \over 2})}\right]\psi_2(r)
\]
Using some hypergeometric identities and fixing the overall normalization we can rewrite the normalizable mode as
\[
\hat{\psi}_{\omega,k}(r) = {1\over \Gamma(\Delta)\sqrt{r_h}}\sqrt{\Gamma(a+b+\Delta/2)\Gamma(a-b+\Delta/2)
\Gamma(-a+b+\Delta/2)\Gamma(-a-b+\Delta/2)\over \Gamma(2b)\Gamma(-2b)}
\]
\[
\times \left({r^2\over r_h^2}\right)^a \left({r^2\over r_h^2}-1 \right)^{-a-{\Delta\over 2}}
{ \, _2 F_1\left(a-b+\Delta/2,a+b+\Delta/2,\Delta,{r_h^2 \over r_h^2-r^2}\right)}
\]
The overall normalization was fixed so that the modes are canonically normalized with respect to the Klein Gordon norm, which is why we use the hatted notation --- as discussed in section
\ref{modesbrane}. In particular, near the horizon we have the expansion\footnote{The overall factor of $r_h^{-1/2}$ in the near horizon normalization, is automatically fixed if we 
require the modes defined by \eqref{psinearhapp}, \eqref{btzmodes} to have canonical Klein-Gordon norm. }
\be
\label{psinearhapp}                          
\hat{\psi}_{\omega,k}(r) = r_h^{-1/2} \left(e^{i\delta_{\omega,k}}e^{i\omega r_*} + e^{-i \delta_{\omega,k}} e^{-i\omega r_*}\right)
\ee
where $r_*$ is the tortoise coordinate (in which horizon is at $r_*\rightarrow -\infty$) 
\[
 r_* = {1\over 2 r_h}\log\left(r-r_h\over r+r_h\right)
\]
and the ``phase shift'' is
\[
e^{i \delta_{\omega,k}} = 4^{b} \sqrt{\Gamma(-2b) \,\Gamma(a+b+\Delta/2)\,\Gamma(-a+b+\Delta/2)\over\Gamma(2b)\,
\Gamma(-a-b+\Delta/2)\,\Gamma(a-b+\Delta/2)}
\]
To be more precise the complete bulk mode is
\be
\label{btzmodes}
\hat{f}_{\omega,k}(t,x,r) = \hat{\psi}_{\omega,k}(r) e^{-i\omega t} e^{i k x}
\ee
here both the frequency $\omega$ and the momentum $k$ are continuous.

For the bulk operator field we have
\be
\label{bulkfieldbtz}
\phi(t,x,r) = 
\int_{\omega>0} {d\omega\,dk  \over (2 \pi)^2} {1\over \sqrt{2\omega}}
\left(a_{\omega,k}\hat{f}_{\omega,k}(t,x,r) + {\rm h.c.}\right)
\ee
with
\[
[a_{\omega,k},a^\dagger_{\omega',k'}] = \delta(\omega-\omega')\delta(k-k')
\]
In the AdS Hartle-Hawking state we have
\[
\begin{split}
&\langle a_{\omega,k} \, a^\dagger_{\omega',k'}\rangle_{\rm HH} = {e^{\beta \omega}\over e^{\beta \omega}-1} \delta(\omega-\omega')\delta(k-k')\\
&\langle a_{\omega,k}^\dagger \,a_{\omega',k'}\rangle_{\rm HH} = {1\over e^{\beta \omega}-1} \delta(\omega-\omega')\delta(k-k')
\end{split}
\]
Using these formulas we can compute the bulk 2-point function and then by taking the bulk points to the boundary, we can recover the boundary 2-point function
\[
G_\beta(t,x;t',x') = \lim_{r,r'\rightarrow \infty}  \left[\normfact^2 r^{\Delta}(r')^{\Delta} \langle \phi(t,x,r) \phi(t',x',r')\rangle_{\rm HH}\right],
\]
where $\normfact$ is given in \eqref{normfactdef}. 
In momentum space, and using time- and space-translational invariance we have
\[
G_\beta(\omega,k) = \int dt\,dx\, e^{i\omega t - i k x} G_\beta(t,x;0,0)
\]
From the previous results we find
\be
\label{btzfourp}
G_\beta(\omega,k) = 
\normfact^2 {1\over 2 \pi (2\omega)}{ e^{\beta \omega} \over e^{\beta \omega}-1}\left({2\pi \over \beta}\right)^{2\Delta-1}\left|{\Gamma\left(i{\beta(\omega+k)\over 4\pi}+{\Delta\over 2}\right)\Gamma\left(i{\beta(\omega-k)\over 4\pi}+{\Delta\over 2}\right)
\over  \Gamma(\Delta)\Gamma\left(i{\beta\omega\over 2\pi}\right)}\right|^2\qquad,\qquad \omega>0
\ee
and
\be
\label{btzfourn}
G_\beta(-\omega,k) = 
\normfact^2 {1\over 2 \pi (2\omega)}{ 1 \over e^{\beta \omega}-1}\left({2\pi \over \beta}\right)^{2\Delta-1}\left|{\Gamma\left(i{\beta(\omega+k)\over 4\pi}+{\Delta\over 2}\right)\Gamma\left(i{\beta(\omega-k)\over 4\pi}+{\Delta\over 2}\right)
\over  \Gamma(\Delta)\Gamma\left(i{\beta\omega\over 2\pi}\right)}\right|^2\quad,\qquad \omega>0
\ee
Actually although the two forms above are useful to see the qualitative
properties of the solution we can rewrite them in a simpler form using 
\be
\left| \Gamma\left(i{\beta\omega\over 2\pi}\right) \right|^2 = {2 \pi^2 \over \beta |\omega|} {1 \over e^{\beta |\omega| \over 2} - e^{-\beta |\omega| \over 2}}
\ee
This leads to the expression
\be
G_{\beta}(\omega, k) = {\normfact^2 \over (2 \pi)^2}  e^{\beta \omega \over 2} 
\left(2 \pi \over \beta \right)^{2 \Delta - 2} \left|{\Gamma\left(i{\beta(\omega+k)\over 4\pi}+{\Delta\over 2}\right)\Gamma\left(i{\beta(\omega-k)\over 4\pi}+{\Delta\over 2}\right) \over  \Gamma(\Delta)}\right|^2,
\ee
and this expression is valid for $\omega$ both positive and negative.

In the next subsection we will rederive these expressions from the boundary CFT, using the constraints from 2d conformal invariance. Here we notice that these results manifestly 
satisfy the general properties we mentioned in section \ref{sec:outside}:

i) It is obvious from the expressions above that the KMS condition is satisfied 
\[
 G(-\omega,k) = e^{-\beta \omega} G(\omega,k)
\]

ii) It can be checked that the 2-point function $G(\omega,k)$ is exponentially suppressed for large spacelike momenta, that is for fixed $\omega$ and large $k$ we have
\[ 
 G(\omega,k) \underset{|k| \rightarrow \infty}{ \lessapprox} e^{-{\beta k \over 2}}
\]
as expected from our general arguments in section \ref{subsec:thermalfourier}.

iii) It can be checked that in the limit of low temperature ($\beta\rightarrow \infty$) the thermal Wightman function $G_\beta(\omega,k)$ reduces to the zero-temperature one 
that we found in section \ref{sec:emptyads}, that is
$$
G_\beta(\omega,k) \approx N_\Delta \,\theta(\omega)\,\theta(\omega^2-k^2)\, (\omega^2 - k^2)^{\Delta-1} ,\qquad {\rm for}\quad \beta\rightarrow \infty
$$
\subsection{Boundary correlators in 2 dimensions}
The two
dimensional CFT correlator at finite temperature, can be completely fixed using conformal invariance and we will rederive the results \eqref{btzfourp}, \eqref{btzfourn} directly from the boundary.
We start with the Euclidean correlator. We put the CFT on ${\mathbb R}^1\times {\mathbb S}^1$ where the perimeter of the circle is $\beta$. We then have
\be
\label{exact2dcorrelator}
\langle {\cal O}(\tau,x) {\cal O}(0,0)\rangle_{\beta} = \left({2\pi \over \beta} \right)^{2\Delta} \left[2 \cosh\left({2\pi x \over \beta}\right)- 2 \cos\left({2\pi \tau \over \beta}\right)\right]^{-\Delta}
 \ee
By taking the short distance expansion we can check that the normalization is correct.
However, we can now use modular invariance and understand this to be
the thermal correlator of a CFT on flat space at an inverse
temperature $\beta$.  

Now, let us continue to Lorentzian space, following the same logic as in section \ref{sec:emptyads}. (See also \cite{Skenderis:2008dg}.)
We find that
the time-ordered correlator is given by
\[
\langle T\left\{{\cal O}(t,x), {\cal O}(0,0) \right\} \rangle_{\beta} = \left({2 \pi \over \beta} \right)^{2 \Delta}  \left[2 \cosh \left({2 \pi x \over \beta} \right) - 
2 \cosh \left( {2 \pi (1 - i \epsilon) t \over \beta} \right) \right]^{-\Delta}
 \]
and for the Wightman correlator, we have
\[
\begin{split}
&\langle {\cal O}(t,x), {\cal O}(0,0) \rangle_{\beta} = \left({2 \pi \over \beta} \right)^{2 \Delta}  \left[2 \cosh \left({2 \pi x \over \beta} \right) - 
2 \cosh \left( {2 \pi (t - i \epsilon) \over \beta} \right)  \right]^{-\Delta} \\
&= \left({2 \pi \over \beta} \right)^{2 \Delta} 2^{-2 \Delta} \left[ \sinh {\pi (u - i \epsilon) \over \beta} \sinh{\pi (v - i \epsilon) \over \beta} \right]^{-\Delta}
\end{split}
 \]
where we defined $u=t-x,\,v=t+x$. So, the basic integral that we are interested in is
\be
\label{ibetadef}
I_{\beta}(k_{+}) = \left({2 \pi \over \beta} \right)^{ \Delta} 2^{- \Delta} \int_{-\infty}^{\infty} e^{i k_{+} u}  \left(\sinh{\pi (u - i \epsilon) \over \beta} \right)^{-\Delta} d u
\ee

Let us understand the qualitative properties of this integral. We see that the integral has branch cuts running along the segments $(-\infty + i n \beta + i \epsilon, i n \beta + i \epsilon)$ for integer $n$. Now, if $k_{+} > 0$, then we can move the $u$ contour down to the first branch cut, which is at ${\rm Im}~u = - i \beta$. So, for negative $k_{+}$, we get a contribution proportional to
$e^{-k_+ \beta}$ i.e. the integral is exponentially damped for large negative $k_{+}$.  On the other hand, for positive $k_{+}$, there is no such damping. 

To evaluate the integral precisely, we write it as
\be
\begin{split}
I_{\beta}(k_{+}) &= \left({2 \pi \over \beta} \right)^{\Delta} 2^{-\Delta} \left[e^{i \pi \Delta} \int_{-\infty}^{0} d u \, e^{i k_{+} u} \left| \sinh {\pi u \over \beta} \right|^{-\Delta} + \int_0^{\infty} d u \, e^{i k_{+} u} \left| \sinh {\pi u \over \beta} \right|^{-\Delta} \right],\\
\end{split}
\ee
where the phase factors come from the $i \epsilon$ prescription explained in section \ref{sec:emptyads}.
This integral can be performed analytically to obtain
\be
\begin{split}
I_{\beta}(k_{+})
&=  \left({2 \pi \over \beta} \right)^{ \Delta} {\pi^2 \beta \over 4} {\csc (\pi \Delta )  \over  \Gamma (\Delta )} \left[ e^{i \pi \Delta} \frac{ \Gamma \left(\frac{1}{2} \left(\frac{i k_{+} \beta}{\pi }+\Delta \right)\right)}{\Gamma \left(\frac{i k_{+}
   \beta }{2 \pi }-\frac{\Delta }{2}+1\right) } 
+ \frac{  \Gamma \left(\frac{1}{2} \left(\Delta -\frac{i k_{+} \beta }{\pi }\right)\right)}{\Gamma \left(-\frac{i k_{+} \beta }{2 \pi }-\frac{\Delta }{2}+1\right)} \right] \\
&= \left({2 \pi \over \beta} \right)^{ \Delta} {\pi^3 \beta \over 4} \frac{ \left(\coth \left(\frac{1}{2} (k_{+} \beta +i \pi  \Delta )\right)+1\right) \text{csch}\left(\frac{1}{2} (k_{+} \beta -i   \pi  \Delta )\right)}{\Gamma \left(-\frac{i k_{+} \beta }{2 \pi }-\frac{\Delta }{2}+1\right) \Gamma \left(\frac{i k_{+}\beta }{2 \pi }-\frac{\Delta }{2}+1\right) \Gamma (\Delta )} \\
&= \left({2 \pi \over \beta} \right)^{ \Delta - 1} {\pi^2 \over 2} \frac{ e^{i \pi \Delta + k_{+} \beta \over 2} \Gamma \left(\frac{\Delta }{2} + \frac{i k_{+} \beta }{2 \pi } \right) \Gamma \left(\frac{\Delta }{2} - \frac{i k_{+}\beta }{2 \pi }\right)}{ \Gamma (\Delta )}.
\end{split}
\ee

Since the full answer for the Green function is given by
\be
G_{\beta}(\omega, k) =  I_{\beta}({k + \omega \over 2}) I_{\beta}({k - \omega \over 2}),
\ee
and since, in this case $\normfact^2 = 2 \pi (2 \pi)^2$, 
we see that our boundary calculation matches precisely with the answer
from the bulk up to a momentum and temperature independent pre-factor.
% that we did not keep track of above.
%were not careful about above.
%is independent of both the momentum and the temperature.
%numerical pre-factor that is independent of both the momentum and temperature. 

%%% Local Variables: 
%%% mode: latex
%%% TeX-master: "infalling_paper"
%%% End: 

\section{Quantization in Rindler space \label{rindlerq}}

\subsection{Expansion in Rindler modes}
We start with $d+1$ dimensional Minkowski space
\[
ds^2 = -dt^2 + dz^2 + d\vect{x}^2
\]
and consider a massless scalar field obeying
\[
 \Box \phi = 0
\]

\subsubsection{Region I}

We first expand the field in modes in region I. In that region the Rindler coordinates $(\tau,\sigma,\vect{x})$ are defined by $t= \sigma \sinh \tau,\, z = \sigma \cosh t$. The metric looks like
\[
ds^2 = -\sigma^2 d\tau^2 + d\sigma^2 + d\vect{x}^2
\]
The field in region I has the expansion
\[
\phi(\tau,\sigma,\vect{x}) = \int_{\omega>0}{d\omega d^{d-1}\vect{k} \over (2 \pi)^d} \left[{1\over \sqrt{2\omega}} a_{\omega,\vect{k}} e^{-i\omega \tau+i \vect{k} \vect{x}}{2 K_{i\omega}(|\vect{k}| \sigma)
\over |\Gamma(i\omega)|} + {\rm h.c.}
\right]
\]
Notice that the Bessel function $K_{i\omega}(|\vect{k}| \sigma)$ is real. 

We define the lightcone coordinates in Minkowski space
\[
U = t-z
\]
\[
V= t+z
\]
In region I we have $U<0,V>0$. In terms of the Rindler coordinates in region I we have
\[
U = - \sigma e^{-\tau}\]
\[
V = \sigma e^{\tau}
\]
Considering the field expansion near the horizon between regions I and II we find
\[
\phi  \underset{U \rightarrow 0}{\approx} \int_{\omega>0}{d\omega d^{d-1}\vect{k} \over (2 \pi)^d}{1\over \sqrt{2\omega}} a_{\omega\vect{k}} e^{i \vect{k} \vect{x}} \left[V^{-i\omega} e^{-i\delta} + (-U)^{i\omega}e^{i\delta}\right]
\]
\[
+ {1\over \sqrt{2\omega}} a_{\omega,\vect{k}}^\dagger e^{-i \vect{k} \vect{x}} \left[V^{i\omega} e^{i\delta} + (-U)^{-i\omega}e^{-i\delta}\right]
\]
where the phase shift is
\begin{equation}
\label{phase}
e^{i\delta} = \left({|\vect{k}|\over 2}\right)^{i\omega} {\Gamma(-i\omega) \over |\Gamma(i\omega)|}
\end{equation}
Due to the Riemann-Lebesgue lemma, when $U\rightarrow 0$ terms like $U^{-i \omega}$ can be discarded. So what is important is to keep the terms that depend on $V$ only, and we have
\begin{equation}
\label{boundaryIandII}
\phi  \underset{U \rightarrow 0}{\approx}\int_{\omega>0}{d\omega d^{d-1}\vect{k} \over (2 \pi)^d} {1\over \sqrt{2\omega}} \left[a_{\omega,\vect{k}} e^{i \vect{k} \vect{x}} V^{-i\omega} e^{-i\delta} + 
 a_{\omega,\vect{k}}^\dagger e^{-i \vect{k} \vect{x}} V^{i\omega} e^{i\delta} \right]
\end{equation}

\subsubsection{Region III}

Here the coordinates are $t= -\sigma \sinh \tau,\,\,z=-\sigma \cosh \tau$. We notice that the Rindler time $\tau$ runs opposite of the Minkowski time $t$. Also,
notice that due to our parametrization we still have $\sigma>0$ even though $z<0$ in region III. We expand the field as
\[
\phi(\tau,\sigma,\vect{x}) = \int_{\omega>0}{d\omega d^{d-1}\vect{k} \over (2 \pi)^d} \left[{1\over \sqrt{2\omega}} \wa_{\omega,\vect{k}} e^{i\omega \tau-i \vect{k} \vect{x}}{2 K_{i\omega}(|\vect{k}| \sigma)
\over |\Gamma(i\omega)|} + {\rm h.c.} \right]
\]
Notice that we have defined the modes $\wa_{\omega,\vect{k}}$ to multiply the function which is the conjugate of the one before.
We want to expand the field near the horizon between regions III and II. Again in Minkowski lightcone coordinates we have
\[
U = \sigma e^{-\tau}
\]
\[
V = -\sigma e^\tau
\]
The horizon is now at $V\rightarrow 0$. Following the same steps as for region I we find that near the horizon the non-vanishing terms are
\begin{equation}
\label{boundaryIIandIII}
\phi \underset{V \rightarrow 0}{\approx} \int_{\omega>0}{d\omega d^{d-1}\vect{k} \over (2 \pi)^d} {1\over \sqrt{2\omega}}\left[ \wa_{\omega,\vect{k}} e^{-i \vect{k} \vect{x}} U^{-i\omega} e^{-i\delta}
+\wa^\dagger_{\omega,\vect{k}} e^{i\vect{k}\vect{x}} U^{i\omega}e^{i\delta}   \right]
\end{equation}
the phase factor is again given by \eqref{phase}.

\subsubsection{Region II}

Here we choose the Rindler coordinates as $t =\sigma  \cosh \tau,\,\, z = \sigma \sinh \tau$. The horizon between I and II is at $\tau\rightarrow +\infty$. In terms of the lightcone coordinates
we have
\[
U = \sigma e^{-\tau}
\]
\[
V = \sigma e^{\tau}
\]
We write the general expansion 
as
\[
\phi = \int_{\omega>0}{d\omega d^{d-1}\vect{k} \over (2 \pi)^d} {1\over \sqrt{2\omega}}\left[A_{\omega,\vect{k}}e^{-i \omega \tau + i \vect{k}\vect{x}} J_{i\omega}(|\vect{k}|\sigma) +
B_{\omega,\vect{k}} e^{-i \omega t + i \vect{k} \vect{x}} J_{-i \omega}(|\vect{k}|\sigma) + {\rm h.c.} \right]
\]
For the $J$ Bessel functions as $\sigma\rightarrow 0$ we have
\[
J_{i\omega}(|\vect{k}|\sigma) \approx {1\over \Gamma(1+i\omega)} \left({|\vect{k}|\sigma \over 2}\right)^{i\omega} +\ldots
\]
Now we are looking at the expansion closed to the horizon between I and II. There we have $U\rightarrow 0$, 
so we keep only the terms which depend on $V$ and we find
\[
\phi \underset{U \rightarrow 0}{\approx} \int_{\omega>0}{d\omega d^{d-1}\vect{k} \over (2 \pi)^d} {1\over \sqrt{2\omega}}
\left[
{B_{\omega,\vect{k}}\over \Gamma(1-i\omega)} e^{i\vect{k} \vect{x}}\left({|\vect{k}|\over 2}\right)^{-i\omega} V^{-i\omega}+{\rm h.c.}
\right]
\]
Comparing with the expansion \eqref{boundaryIandII} we find that
\[
B_{\omega,\vect{k}} = {\Gamma(1-i\omega) \Gamma(i\omega) \over |\Gamma(i\omega)|}a_{\omega,\vect{k}}= -i\sqrt{\pi \omega \over \sinh(\pi \omega)} a_{\omega,\vect{k}}
\]
While, looking at the horizon between II and III we find
\[
\phi \underset{V \rightarrow 0}{\approx} \int_{\omega>0} {d\omega d^{d-1} \vect{k}  \over (2 \pi)^d} {1\over \sqrt{2\omega}} \left[{A_{\omega,\vect{k}}\over \Gamma(1+i\omega)}e^{i\vect{k}\vect{x}}
\left({|\vect{k}|\over 2}\right)^{i\omega}U^{i\omega} + {\rm h.c.}
\right]
\]
Comparing with \eqref{boundaryIIandIII} we find
\[
A_{\omega,\vect{k}} = {\Gamma(1+i\omega)\Gamma(-i\omega) \over |\Gamma(i\omega)|}\widetilde{a}_{\omega,\vect{k}}^\dagger
=i\sqrt{\pi \omega \over \sinh(\pi \omega)} \widetilde{a}_{\omega,\vect{k}}^\dagger
\]
Putting everything together we find the expansion in region II
\be
\label{regionIIrindler}
\phi(\tau,\sigma,\vect{x}) = \int_{\omega>0}{d\omega d^{d-1}\vect{k}  \over (2 \pi)^d} {1\over \sqrt{2\omega}}\sqrt{\pi \omega\over \sinh(\pi \omega)} \left[i\,\widetilde{a}^\dagger_{\omega,\vect{k}}
\,e^{-i\omega \tau+ i \vect{k} \vect{x}}\, J_{i\omega}(|\vect{k}|\sigma) 
- i a_{\omega,\vect{k}} e^{-i\omega \tau + i \vect{k} \vect{x}} J_{-i\omega}(|\vect{k}|\sigma)+ {\rm h.c.}\right]
\ee 	
\subsubsection{Bogoliubov transformation}

Now we express the Rindler modes in terms of the Unruh modes $d^{1,2}_{\omega,\vect{k}}$. We have
\be
\label{unruh}
d^1_{\omega,\vect{k}} = {a_{\omega,\vect{k}} - e^{-\pi \omega}\, \widetilde{a}_{\omega,\vect{k}}^\dagger \over \sqrt{1-e^{-2\pi \omega}}}\qquad,\qquad 
d^2_{\omega,\vect{k}} = {\widetilde{a}_{\omega,-\vect{k}} - e^{-\pi \omega}\, a_{\omega,-\vect{k}}^\dagger \over \sqrt{1-e^{-2\pi \omega}}}
\ee
or inverting
\be
\label{unruhinv}
a_{\omega,\vect{k}} = {d^1_{\omega,\vect{k}} + e^{-\pi \omega}(d^2_{\omega,-\vect{k}})^\dagger\over \sqrt{1-e^{-2\pi \omega}}}\qquad,\qquad
\widetilde{a}_{\omega,\vect{k}} = {d^2_{\omega,-\vect{k}}+ e^{-\pi \omega} (d^1_{\omega,\vect{k}})^\dagger \over  \sqrt{1-e^{-2\pi \omega}}}
\ee
The Minkowski vacuum $|0\rangle$ is defined by $d^{1,2}_{\omega,\vect{k}}|0\rangle =0$. The Rindler mode occupation levels are
\[
 \langle 0| a_{\omega,\vect{k}} a^\dagger_{\omega',\vect{k}'}|0\rangle ={e^{2\pi \omega} \over e^{2\pi \omega}-1}\delta(\omega-\omega')\delta^{d-1}(\vect{k}-\vect{k}')\,\,,\,\,
  \langle 0| a_{\omega,\vect{k}}^\dagger a_{\omega',\vect{k}'}|0\rangle ={1 \over e^{2\pi \omega}-1}\delta(\omega-\omega')\delta^{d-1}(\vect{k}-\vect{k}')
\]
and similarly for the $\widetilde{a}_{\omega,\vect{k}}$ modes.

\subsection{2-point function in terms of Rindler modes}

\subsubsection{Region I}

We can now write the usual Wightman 2-point function of a scalar field in terms of the Rindler modes. For  points in region I we have
\[
 \langle 0| \phi(\tau_1,\sigma_1,\vect{x}_1) \phi(\tau_2,\sigma_2,\vect{x}_2)|0\rangle = (2\pi)^d \int_{\omega>0}{d\omega d^{d-1}\vect{k}  \over (2 \pi)^d}\]
 \[{1\over 2\omega}
 \Bigg[{e^{2\pi \omega}\over e^{2\pi \omega}-1} {4K_{i\omega}(|\vect{k}|\sigma_1)K_{i\omega}(|\vect{k}|\sigma_2)
\over |\Gamma(i\omega)|^2}e^{-i\omega \tau_{12}+i\vect{k}\,\vect{x}_{12}}+{1\over e^{2\pi \omega}-1} {4K_{i\omega}(|\vect{k}|\sigma_1)K_{i\omega}(|\vect{k}|\sigma_2)
\over |\Gamma(i\omega)|^2}e^{i\omega \tau_{12}-i\vect{k}\,\vect{x}_{12}}\Bigg]\]
We are interested in the convergence of this integral in the region $\omega=0$. We have the explicit factor of ${1\over 2\omega}$ in front, the thermal occupation factors 
give another factor of ${1\over \omega}$. The Bessel function $K_{i\omega}(z)$ goes to a non-zero constant as $\omega$ goes to zero, for fixed $z$. Finally the factor ${1\over |\Gamma(i\omega)|^2}$
goes like $\omega^2$ for small $\omega$. All in all, the integrand goes like $\omega^0$ for small $\omega$ and hence the integral converges when $\omega\rightarrow 0$.

\subsubsection{Region II}

Now let us consider two points behind the Rindler horizon i.e. in region II, using the expansion \eqref{regionIIrindler}. We have contributions of several bilinears made out of $a_{\omega,\vect{k}}$ and $\widetilde{a}_{\omega,\vect{k}}$. If focus on only the contributions from the bilinears $\langle 0|a_{\omega,\vect{k}}
a_{\omega',\vect{k}'}^\dagger|0\rangle$ and $\langle 0|a_{\omega,\vect{k}}^\dagger
a_{\omega',\vect{k}'}|0\rangle$ of the non-tilde operators, we find the terms
\[
\begin{split}
 (2\pi)^d \int_{\omega>0}{d\omega d^{d-1}\vect{k}  \over (2 \pi)^d} {\pi \over \sinh(\pi \omega)}\Bigg[ &{e^{2\pi \omega} \over e^{2\pi \omega}-1} J_{i\omega}(|\vect{k}|\sigma_1)
 J_{-i\omega}(|\vect{k}|\sigma_2)e^{-i\omega \tau_{12}+i \vect{k}\,\vect{x}_{12}}\\
& + {1 \over e^{2\pi \omega}-1} J_{-i\omega}(|\vect{k}|\sigma_1)
 J_{i\omega}(|\vect{k}|\sigma_2)e^{i\omega \tau_{12}-i \vect{k}\,\vect{x}_{12}}\Bigg]
\end{split}
\]
For small $\omega$ (and fixed $\sigma_1,\sigma_2$) the Bessel functions go to nonzero constants, hence the integrand goes like ${1\over \omega^2}$ and the integral
diverges as
\[
 \int_{\omega>0} {d\omega \over \omega^2}
 \]
This seems to suggest that the small $\omega$ region has a very large contribution to the 2-point function, however we know that this cannot be the correct answer.

After all, the 2-point function that we are considering the the standard Wightman function of a massless scalar field, which is obviously finite for two points in region II. Hence it must be that, while the contributions from these two terms mentioned above are formally divergent, the total contribution from all terms --- that is from the non-tildes and from cross terms--- must be finite. There must be cancellations between the terms that we have considered and the remaining terms. 

In other words, if we regroup the terms correctly {\it before} doing the $\omega$ integral, the resulting expression must be manifestly finite as we integrate down to $\omega=0$. It turns out that regrouping the $a_{\omega,\vect{k}}$ and $\widetilde{a}_{\omega,\vect{k}}$ into the ``Unruh combinations'' \eqref{unruh} makes the integral manifestly convergent. Indeed, substituting from \eqref{unruhinv} into \eqref{regionIIrindler} we find that the field in region II can be written as
\[
\phi(\tau,\sigma,\vect{x})= \int_{\omega>0} {d\omega d^{d-1}\vect{k}  \over (2 \pi)^d} {1\over \sqrt{2\omega}}\sqrt{\pi \omega\over \sinh(\pi \omega)}\left[e^{-i\omega \tau + i \vect{k} \, \vect{x}}\,i\, \left({
e^{-\pi \omega}J_{i\omega}(|\vect{k}|z) - J_{-i\omega}(|\vect{k}|\sigma)\over \sqrt{1-e^{-2\pi \omega}}}\right)d^1_{\omega,\vect{k}} + {\rm h.c.} \right]
 \]
\[
+{\rm terms\,\,involving\,\,}d^2_{\omega,\vect{k}}
\]
We have that
\[
e^{-\pi \omega}J_{i\omega}(z) - J_{-i\omega}(z) = -\sinh (\pi \omega)H^2_{i\omega}(z)
\]
Hence we find
\[
 \phi(\tau,\sigma,\vect{x}) = \int_{\omega>0} {d\omega d^{d-1}\vect{k} \over (2 \pi)^d} {\sqrt{\pi}\over 2}e^{\pi \omega\over 2}\left[-\,i \,e^{-i\omega \tau + i \vect{k} \,\vect{x}} 
 H^2_{i\omega}(|\vect{k}|\sigma) d^1_{\omega,\vect{k}}
 + {\rm h.c.} \right]
\]
\[
+{\rm terms\,\,involving\,\,}d^2_{\omega,\vect{k}}
\]
On the Minkowski vacuum we have $d^{1,2}_{\omega,k} |0\rangle = 0$. So the 2-point function for points in region II becomes
\[
\langle 0| \phi(\tau_1,\sigma_1,\vect{x}_1) \phi(\tau_2,\sigma_2,\vect{x}_2) |0\rangle = (2\pi)\int_{\omega>0}{d\omega d^{d-1} \vect{k} \over (2 \pi)^d} {\pi \over 4} e^{\pi \omega}
e^{-i\omega t_{12}+i \vect{k}\, \vect{x}_{12}} H^2_{i \omega}(|\vect{k}|\sigma_1)\,H^2_{i\omega}(|\vect{k}|\sigma_2)^* +\]
\[
+{\rm terms\,\,involving\,\,}d^2_{\omega,\vect{k}}
\]
In this form we notice that the contribution from
$d^1_{\omega,\vect{k}}$ is manifestly finite when integrating all the
way down to $\omega=0$. The same is true for the contribution from $d^2_{\omega,\vect{k}}$.

%%% Local Variables: 
%%% mode: latex
%%% TeX-master: "infalling_paper"
%%% End: 

\bibliographystyle{JHEP}
\bibliography{references}

\providecommand{\href}[2]{#2}\begingroup\raggedright\begin{thebibliography}{10}

\bibitem{Mathur:2012np}
S.~D. Mathur, {\it {The information paradox: conflicts and resolutions}},
  \href{http://xxx.lanl.gov/abs/1201.2079}{{\tt arXiv:1201.2079}}.

\bibitem{Mathur:2009hf}
S.~D. Mathur, {\it {The Information paradox: A Pedagogical introduction}},
  {\em Class.Quant.Grav.} {\bf 26} (2009) 224001,
  [\href{http://xxx.lanl.gov/abs/0909.1038}{{\tt arXiv:0909.1038}}].

\bibitem{Mathur:2008nj}
S.~D. Mathur, {\it {Fuzzballs and the information paradox: A Summary and
  conjectures}},  \href{http://xxx.lanl.gov/abs/0810.4525}{{\tt
  arXiv:0810.4525}}.

\bibitem{Lunin:2002bj}
O.~Lunin, S.~D. Mathur, and A.~Saxena, {\it {What is the gravity dual of a
  chiral primary?}},  {\em Nucl. Phys.} {\bf B655} (2003) 185--217,
  [\href{http://xxx.lanl.gov/abs/hep-th/0211292}{{\tt hep-th/0211292}}].

\bibitem{Lunin:2001fv}
O.~Lunin and S.~D. Mathur, {\it {Metric of the multiply wound rotating
  string}},  {\em Nucl. Phys.} {\bf B610} (2001) 49--76,
  [\href{http://xxx.lanl.gov/abs/hep-th/0105136}{{\tt hep-th/0105136}}].

\bibitem{Almheiri:2012rt}
A.~Almheiri, D.~Marolf, J.~Polchinski, and J.~Sully, {\it {Black Holes:
  Complementarity or Firewalls?}},
  \href{http://xxx.lanl.gov/abs/1207.3123}{{\tt arXiv:1207.3123}}.

\bibitem{Bousso:2012as}
R.~Bousso, {\it {Complementarity Is Not Enough}},
  \href{http://xxx.lanl.gov/abs/1207.5192}{{\tt arXiv:1207.5192}}.

\bibitem{Nomura:2012sw}
Y.~Nomura, J.~Varela, and S.~J. Weinberg, {\it {Complementarity Endures: No
  Firewall for an Infalling Observer}},
  \href{http://xxx.lanl.gov/abs/1207.6626}{{\tt arXiv:1207.6626}}.

\bibitem{Mathur:2012jk}
S.~D. Mathur and D.~Turton, {\it {Comments on black holes I: The possibility of
  complementarity}},  \href{http://xxx.lanl.gov/abs/1208.2005}{{\tt
  arXiv:1208.2005}}.

\bibitem{Susskind:2012rm}
L.~Susskind, {\it {Singularities, Firewalls, and Complementarity}},
  \href{http://xxx.lanl.gov/abs/1208.3445}{{\tt arXiv:1208.3445}}.

\bibitem{Bena:2012zi}
I.~Bena, A.~Puhm, and B.~Vercnocke, {\it {Non-extremal Black Hole Microstates:
  Fuzzballs of Fire or Fuzzballs of Fuzz ?}},
  \href{http://xxx.lanl.gov/abs/1208.3468}{{\tt arXiv:1208.3468}}.

\bibitem{Giveon:2012kp}
A.~Giveon and N.~Itzhaki, {\it {String Theory Versus Black Hole
  Complementarity}},  \href{http://xxx.lanl.gov/abs/1208.3930}{{\tt
  arXiv:1208.3930}}.

\bibitem{Banks:2012nn}
T.~Banks and W.~Fischler, {\it {Holographic Space-Time Does Not Predict
  Firewalls}},  \href{http://xxx.lanl.gov/abs/1208.4757}{{\tt
  arXiv:1208.4757}}.

\bibitem{Ori:2012jx}
A.~Ori, {\it {Firewall or smooth horizon?}},
  \href{http://xxx.lanl.gov/abs/1208.6480}{{\tt arXiv:1208.6480}}.

\bibitem{Brustein:2012jn}
R.~Brustein, {\it {Origin of the blackhole information paradox}},
  \href{http://xxx.lanl.gov/abs/1209.2686}{{\tt arXiv:1209.2686}}.

\bibitem{Susskind:2012uw}
L.~Susskind, {\it {The Transfer of Entanglement: The Case for Firewalls}},
  \href{http://xxx.lanl.gov/abs/1210.2098}{{\tt arXiv:1210.2098}}.

\bibitem{Marolf:2012xe}
D.~Marolf and A.~C. Wall, {\it {Eternal Black Holes and Superselection in
  AdS/CFT}},  \href{http://xxx.lanl.gov/abs/1210.3590}{{\tt arXiv:1210.3590}}.

\bibitem{Hossenfelder:2012mr}
S.~Hossenfelder, {\it {Comment on the black hole firewall}},
  \href{http://xxx.lanl.gov/abs/1210.5317}{{\tt arXiv:1210.5317}}.

\bibitem{Nomura:2012cx}
Y.~Nomura, J.~Varela, and S.~J. Weinberg, {\it {Black Holes, Information, and
  Hilbert Space for Quantum Gravity}},
  \href{http://xxx.lanl.gov/abs/1210.6348}{{\tt arXiv:1210.6348}}.

\bibitem{Hwang:2012nn}
D.-i. Hwang, B.-H. Lee, and D.-h. Yeom, {\it {Is the firewall consistent?:
  Gedanken experiments on black hole complementarity and firewall proposal}},
  \href{http://xxx.lanl.gov/abs/1210.6733}{{\tt arXiv:1210.6733}}.

\bibitem{Larjo:2012jt}
K.~Larjo, D.~A. Lowe, and L.~Thorlacius, {\it {Black holes without firewalls}},
   \href{http://xxx.lanl.gov/abs/1211.4620}{{\tt arXiv:1211.4620}}.

\bibitem{braunstein2009v1}
S.~Braunstein, {\it {Black hole entropy as entropy of entanglement, or it's
  curtains for the equivalence principle}},
  \href{http://xxx.lanl.gov/abs/0907.1190}{{\tt arXiv:0907.1190}}.

\bibitem{Avery:2012tf}
S.~G. Avery, B.~D. Chowdhury, and A.~Puhm, {\it {Unitarity and fuzzball
  complementarity: 'Alice fuzzes but may not even know it!'}},
  \href{http://xxx.lanl.gov/abs/1210.6996}{{\tt arXiv:1210.6996}}.

\bibitem{Chowdhury:2012vd}
B.~D. Chowdhury and A.~Puhm, {\it {Is Alice burning or fuzzing?}},
  \href{http://xxx.lanl.gov/abs/1208.2026}{{\tt arXiv:1208.2026}}.

\bibitem{Maldacena:1997re}
J.~M. Maldacena, {\it {The Large N limit of superconformal field theories and
  supergravity}},  {\em Adv.Theor.Math.Phys.} {\bf 2} (1998) 231--252,
  [\href{http://xxx.lanl.gov/abs/hep-th/9711200}{{\tt hep-th/9711200}}].

\bibitem{Banks:1998dd}
T.~Banks, M.~R. Douglas, G.~T. Horowitz, and E.~J. Martinec, {\it {AdS dynamics
  from conformal field theory}},
  \href{http://xxx.lanl.gov/abs/hep-th/9808016}{{\tt hep-th/9808016}}.

\bibitem{Balasubramanian:1999ri}
V.~Balasubramanian, S.~B. Giddings, and A.~E. Lawrence, {\it {What do CFTs tell
  us about anti-de Sitter spacetimes?}},  {\em JHEP} {\bf 03} (1999) 001,
  [\href{http://xxx.lanl.gov/abs/hep-th/9902052}{{\tt hep-th/9902052}}].

\bibitem{Bena:1999jv}
I.~Bena, {\it {On the construction of local fields in the bulk of AdS(5) and
  other spaces}},  {\em Phys.Rev.} {\bf D62} (2000) 066007,
  [\href{http://xxx.lanl.gov/abs/hep-th/9905186}{{\tt hep-th/9905186}}].

\bibitem{Hamilton:2006fh}
A.~Hamilton, D.~N. Kabat, G.~Lifschytz, and D.~A. Lowe, {\it {Local bulk
  operators in AdS/CFT: A Holographic description of the black hole interior}},
   {\em Phys.Rev.} {\bf D75} (2007) 106001,
  [\href{http://xxx.lanl.gov/abs/hep-th/0612053}{{\tt hep-th/0612053}}].

\bibitem{Hamilton:2006az}
A.~Hamilton, D.~N. Kabat, G.~Lifschytz, and D.~A. Lowe, {\it {Holographic
  representation of local bulk operators}},  {\em Phys.Rev.} {\bf D74} (2006)
  066009, [\href{http://xxx.lanl.gov/abs/hep-th/0606141}{{\tt
  hep-th/0606141}}].

\bibitem{Hamilton:2005ju}
A.~Hamilton, D.~N. Kabat, G.~Lifschytz, and D.~A. Lowe, {\it {Local bulk
  operators in AdS/CFT: A Boundary view of horizons and locality}},  {\em
  Phys.Rev.} {\bf D73} (2006) 086003,
  [\href{http://xxx.lanl.gov/abs/hep-th/0506118}{{\tt hep-th/0506118}}].

\bibitem{Hamilton:2007wj}
A.~Hamilton, D.~N. Kabat, G.~Lifschytz, and D.~A. Lowe, {\it {Local bulk
  operators in AdS/CFT and the fate of the BTZ singularity}},
  \href{http://xxx.lanl.gov/abs/0710.4334}{{\tt arXiv:0710.4334}}.

\bibitem{VanRaamsdonk:2009ar}
M.~Van~Raamsdonk, {\it {Comments on quantum gravity and entanglement}},
  \href{http://xxx.lanl.gov/abs/0907.2939}{{\tt arXiv:0907.2939}}.

\bibitem{VanRaamsdonk:2010pw}
M.~Van~Raamsdonk, {\it {Building up spacetime with quantum entanglement}},
  {\em Gen.Rel.Grav.} {\bf 42} (2010) 2323--2329,
  [\href{http://xxx.lanl.gov/abs/1005.3035}{{\tt arXiv:1005.3035}}].

\bibitem{VanRaamsdonk:2011zz}
M.~Van~Raamsdonk, {\it {A patchwork description of dual spacetimes in
  AdS/CFT}},  {\em Class.Quant.Grav.} {\bf 28} (2011) 065002.

\bibitem{Czech:2012bh}
B.~Czech, J.~L. Karczmarek, F.~Nogueira, and M.~Van~Raamsdonk, {\it {The
  Gravity Dual of a Density Matrix}},  {\em Class.Quant.Grav.} {\bf 29} (2012)
  155009, [\href{http://xxx.lanl.gov/abs/1204.1330}{{\tt arXiv:1204.1330}}].

\bibitem{Heemskerk:2012mn}
I.~Heemskerk, D.~Marolf, and J.~Polchinski, {\it {Bulk and Transhorizon
  Measurements in AdS/CFT}},  \href{http://xxx.lanl.gov/abs/1201.3664}{{\tt
  arXiv:1201.3664}}.

\bibitem{ElShowk:2011ag}
S.~El-Showk and K.~Papadodimas, {\it {Emergent Spacetime and Holographic
  CFTs}},  \href{http://xxx.lanl.gov/abs/1101.4163}{{\tt arXiv:1101.4163}}.

\bibitem{'tHooft:1990fr}
G.~'t~Hooft, {\it {The black hole interpretation of string theory}},  {\em
  Nucl.Phys.} {\bf B335} (1990) 138--154.

\bibitem{Susskind:1993if}
L.~Susskind, L.~Thorlacius, and J.~Uglum, {\it {The Stretched horizon and black
  hole complementarity}},  {\em Phys.Rev.} {\bf D48} (1993) 3743--3761,
  [\href{http://xxx.lanl.gov/abs/hep-th/9306069}{{\tt hep-th/9306069}}].

\bibitem{Kiem:1995iy}
Y.~Kiem, H.~L. Verlinde, and E.~P. Verlinde, {\it {Black hole horizons and
  complementarity}},  {\em Phys.Rev.} {\bf D52} (1995) 7053--7065,
  [\href{http://xxx.lanl.gov/abs/hep-th/9502074}{{\tt hep-th/9502074}}].

\bibitem{Balasubramanian:1999zv}
V.~Balasubramanian and S.~F. Ross, {\it {Holographic particle detection}},
  {\em Phys.Rev.} {\bf D61} (2000) 044007,
  [\href{http://xxx.lanl.gov/abs/hep-th/9906226}{{\tt hep-th/9906226}}].

\bibitem{Giddings:2001pt}
S.~B. Giddings and M.~Lippert, {\it {Precursors, black holes, and a locality
  bound}},  {\em Phys.Rev.} {\bf D65} (2002) 024006,
  [\href{http://xxx.lanl.gov/abs/hep-th/0103231}{{\tt hep-th/0103231}}].

\bibitem{Maldacena:2001kr}
J.~M. Maldacena, {\it {Eternal black holes in anti-de Sitter}},  {\em JHEP}
  {\bf 0304} (2003) 021, [\href{http://xxx.lanl.gov/abs/hep-th/0106112}{{\tt
  hep-th/0106112}}].

\bibitem{Hubeny:2002dg}
V.~E. Hubeny, {\it {Precursors see inside black holes}},  {\em Int.J.Mod.Phys.}
  {\bf D12} (2003) 1693--1698,
  [\href{http://xxx.lanl.gov/abs/hep-th/0208047}{{\tt hep-th/0208047}}].

\bibitem{Kraus:2002iv}
P.~Kraus, H.~Ooguri, and S.~Shenker, {\it {Inside the horizon with AdS / CFT}},
   {\em Phys.Rev.} {\bf D67} (2003) 124022,
  [\href{http://xxx.lanl.gov/abs/hep-th/0212277}{{\tt hep-th/0212277}}].

\bibitem{Levi:2003cx}
T.~S. Levi and S.~F. Ross, {\it {Holography beyond the horizon and cosmic
  censorship}},  {\em Phys.Rev.} {\bf D68} (2003) 044005,
  [\href{http://xxx.lanl.gov/abs/hep-th/0304150}{{\tt hep-th/0304150}}].

\bibitem{Fidkowski:2003nf}
L.~Fidkowski, V.~Hubeny, M.~Kleban, and S.~Shenker, {\it {The Black hole
  singularity in AdS / CFT}},  {\em JHEP} {\bf 0402} (2004) 014,
  [\href{http://xxx.lanl.gov/abs/hep-th/0306170}{{\tt hep-th/0306170}}].

\bibitem{Barbon:2003aq}
J.~Barbon and E.~Rabinovici, {\it {Very long time scales and black hole thermal
  equilibrium}},  {\em JHEP} {\bf 0311} (2003) 047,
  [\href{http://xxx.lanl.gov/abs/hep-th/0308063}{{\tt hep-th/0308063}}].

\bibitem{Kaplan:2004qe}
J.~Kaplan, {\it {Extracting data from behind horizons with the AdS / CFT
  correspondence}},  \href{http://xxx.lanl.gov/abs/hep-th/0402066}{{\tt
  hep-th/0402066}}.

\bibitem{Balasubramanian:2004zu}
V.~Balasubramanian and T.~S. Levi, {\it {Beyond the veil: Inner horizon
  instability and holography}},  {\em Phys.Rev.} {\bf D70} (2004) 106005,
  [\href{http://xxx.lanl.gov/abs/hep-th/0405048}{{\tt hep-th/0405048}}].

\bibitem{Festuccia:2005pi}
G.~Festuccia and H.~Liu, {\it {Excursions beyond the horizon: Black hole
  singularities in Yang-Mills theories. I.}},  {\em JHEP} {\bf 0604} (2006)
  044, [\href{http://xxx.lanl.gov/abs/hep-th/0506202}{{\tt hep-th/0506202}}].

\bibitem{Balasubramanian:2005mg}
V.~Balasubramanian, J.~de~Boer, V.~Jejjala, and J.~Simon, {\it {The Library of
  Babel: On the origin of gravitational thermodynamics}},  {\em JHEP} {\bf
  0512} (2005) 006, [\href{http://xxx.lanl.gov/abs/hep-th/0508023}{{\tt
  hep-th/0508023}}].

\bibitem{Balasubramanian:2005qu}
V.~Balasubramanian, P.~Kraus, and M.~Shigemori, {\it {Massless black holes and
  black rings as effective geometries of the D1-D5 system}},  {\em
  Class.Quant.Grav.} {\bf 22} (2005) 4803--4838,
  [\href{http://xxx.lanl.gov/abs/hep-th/0508110}{{\tt hep-th/0508110}}].

\bibitem{Festuccia:2006sa}
G.~Festuccia and H.~Liu, {\it {The Arrow of time, black holes, and quantum
  mixing of large N Yang-Mills theories}},  {\em JHEP} {\bf 0712} (2007) 027,
  [\href{http://xxx.lanl.gov/abs/hep-th/0611098}{{\tt hep-th/0611098}}].

\bibitem{Balasubramanian:2007qv}
V.~Balasubramanian, B.~Czech, V.~E. Hubeny, K.~Larjo, M.~Rangamani, {\em
  et.~al.}, {\it {Typicality versus thermality: An Analytic distinction}},
  {\em Gen.Rel.Grav.} {\bf 40} (2008) 1863--1890,
  [\href{http://xxx.lanl.gov/abs/hep-th/0701122}{{\tt hep-th/0701122}}].

\bibitem{Marolf:2008mf}
D.~Marolf, {\it {Unitarity and Holography in Gravitational Physics}},  {\em
  Phys.Rev.} {\bf D79} (2009) 044010,
  [\href{http://xxx.lanl.gov/abs/0808.2842}{{\tt arXiv:0808.2842}}].

\bibitem{Balasubramanian:2008da}
V.~Balasubramanian, J.~de~Boer, S.~El-Showk, and I.~Messamah, {\it {Black Holes
  as Effective Geometries}},  {\em Class.Quant.Grav.} {\bf 25} (2008) 214004,
  [\href{http://xxx.lanl.gov/abs/0811.0263}{{\tt arXiv:0811.0263}}].

\bibitem{deBoer:2009un}
J.~de~Boer, S.~El-Showk, I.~Messamah, and D.~Van~den Bleeken, {\it {A Bound on
  the entropy of supergravity?}},  {\em JHEP} {\bf 1002} (2010) 062,
  [\href{http://xxx.lanl.gov/abs/0906.0011}{{\tt arXiv:0906.0011}}].

\bibitem{Horowitz:2009wm}
G.~Horowitz, A.~Lawrence, and E.~Silverstein, {\it {Insightful D-branes}},
  {\em JHEP} {\bf 0907} (2009) 057,
  [\href{http://xxx.lanl.gov/abs/0904.3922}{{\tt arXiv:0904.3922}}].

\bibitem{Balasubramanian:2011dm}
V.~Balasubramanian and B.~Czech, {\it {Quantitative approaches to information
  recovery from black holes}},  {\em Class.Quant.Grav.} {\bf 28} (2011) 163001,
  [\href{http://xxx.lanl.gov/abs/1102.3566}{{\tt arXiv:1102.3566}}].

\bibitem{Avery:2011nb}
S.~G. Avery, {\it {Qubit Models of Black Hole Evaporation}},
  \href{http://xxx.lanl.gov/abs/1109.2911}{{\tt arXiv:1109.2911}}.

\bibitem{Simon:2011zza}
J.~Simon, {\it {Extremal black holes, holography and coarse graining}},  {\em
  Int.J.Mod.Phys.} {\bf A26} (2011) 1903--1971,
  [\href{http://xxx.lanl.gov/abs/1106.0116}{{\tt arXiv:1106.0116}}].

\bibitem{Bousso:2012mh}
R.~Bousso, B.~Freivogel, S.~Leichenauer, V.~Rosenhaus, and C.~Zukowski, {\it
  {Null Geodesics, Local CFT Operators and AdS/CFT for Subregions}},
  \href{http://xxx.lanl.gov/abs/1209.4641}{{\tt arXiv:1209.4641}}.

\bibitem{Luscher:1974ez}
M.~Luscher and G.~Mack, {\it {Global Conformal Invariance in Quantum Field
  Theory}},  {\em Commun.Math.Phys.} {\bf 41} (1975) 203--234.

\bibitem{Aharony:1999ti}
O.~Aharony, S.~S. Gubser, J.~M. Maldacena, H.~Ooguri, and Y.~Oz, {\it {Large N
  field theories, string theory and gravity}},  {\em Phys. Rept.} {\bf 323}
  (2000) 183--386, [\href{http://xxx.lanl.gov/abs/hep-th/9905111}{{\tt
  hep-th/9905111}}].

\bibitem{Minwalla:1997ka}
S.~Minwalla, {\it {Restrictions imposed by superconformal invariance on quantum
  field theories}},  {\em Adv. Theor. Math. Phys.} {\bf 2} (1998) 781--846,
  [\href{http://xxx.lanl.gov/abs/hep-th/9712074}{{\tt hep-th/9712074}}].

\bibitem{Unruh:1976db}
W.~Unruh, {\it {Notes on black hole evaporation}},  {\em Phys.Rev.} {\bf D14}
  (1976) 870.

\bibitem{birrell1984quantum}
N.~Birrell and P.~Davies, {\em {Quantum fields in curved space}}.
\newblock Cambridge Univ Press, 1986.

\bibitem{Hayden:2007cs}
P.~Hayden and J.~Preskill, {\it {Black holes as mirrors: Quantum information in
  random subsystems}},  {\em JHEP} {\bf 0709} (2007) 120,
  [\href{http://xxx.lanl.gov/abs/0708.4025}{{\tt arXiv:0708.4025}}].

\bibitem{Susskind:1993mu}
L.~Susskind and L.~Thorlacius, {\it {Gedanken experiments involving black
  holes}},  {\em Phys.Rev.} {\bf D49} (1994) 966--974,
  [\href{http://xxx.lanl.gov/abs/hep-th/9308100}{{\tt hep-th/9308100}}].

\bibitem{Sekino:2008he}
Y.~Sekino and L.~Susskind, {\it {Fast Scramblers}},  {\em JHEP} {\bf 0810}
  (2008) 065, [\href{http://xxx.lanl.gov/abs/0808.2096}{{\tt
  arXiv:0808.2096}}].

\bibitem{takahashi1996thermo}
Y.~Takahashi and H.~Umezawa, {\it Thermo field dynamics},  {\em International
  Journal of Modern Physics B} {\bf 10} (1996) 1755--1805.

\bibitem{von2010proof}
J.~von Neumann, {\it Proof of the ergodic theorem and the h-theorem in quantum
  mechanics},  {\em The European Physical Journal H} {\bf 35} (2010), no.~2
  201--237, [\href{http://xxx.lanl.gov/abs/1003.2133}{{\tt arXiv:1003.2133}}].

\bibitem{Papadodimas:2013}
K.~Papadodimas and S.~Raju, {\it {State-Dependent Bulk-Boundary Maps and Black
  Hole Complementarity}},  \href{http://xxx.lanl.gov/abs/1310.6335}{{\tt
  arXiv:1310.6335}}.

\bibitem{Papadodimas:2013b}
K.~Papadodimas and S.~Raju, {\it {The Black Hole Interior in AdS/CFT and the
  Information Paradox}},  \href{http://xxx.lanl.gov/abs/1310.6334}{{\tt
  arXiv:1310.6334}}.

\bibitem{Mathur:2011uj}
S.~D. Mathur, {\it {What the information paradox is {\it not}}},
  \href{http://xxx.lanl.gov/abs/1108.0302}{{\tt arXiv:1108.0302}}.

\bibitem{Skenderis:2008qn}
K.~Skenderis and M.~Taylor, {\it {The fuzzball proposal for black holes}},
  {\em Phys.Rept.} {\bf 467} (2008) 117--171,
  [\href{http://xxx.lanl.gov/abs/0804.0552}{{\tt arXiv:0804.0552}}].

\bibitem{Page:1993df}
D.~N. Page, {\it {Average entropy of a subsystem}},  {\em Phys.Rev.Lett.} {\bf
  71} (1993) 1291--1294, [\href{http://xxx.lanl.gov/abs/gr-qc/9305007}{{\tt
  gr-qc/9305007}}].

\bibitem{Kabat:2011rz}
D.~Kabat, G.~Lifschytz, and D.~A. Lowe, {\it {Constructing local bulk
  observables in interacting AdS/CFT}},  {\em Phys.Rev.} {\bf D83} (2011)
  106009, [\href{http://xxx.lanl.gov/abs/1102.2910}{{\tt arXiv:1102.2910}}].

\bibitem{Heemskerk:2009pn}
I.~Heemskerk, J.~Penedones, J.~Polchinski, and J.~Sully, {\it {Holography from
  Conformal Field Theory}},  {\em JHEP} {\bf 0910} (2009) 079,
  [\href{http://xxx.lanl.gov/abs/0907.0151}{{\tt arXiv:0907.0151}}].

\bibitem{Heemskerk:2010ty}
I.~Heemskerk and J.~Sully, {\it {More Holography from Conformal Field Theory}},
   {\em JHEP} {\bf 1009} (2010) 099,
  [\href{http://xxx.lanl.gov/abs/1006.0976}{{\tt arXiv:1006.0976}}].

\bibitem{Fitzpatrick:2010zm}
A.~Fitzpatrick, E.~Katz, D.~Poland, and D.~Simmons-Duffin, {\it {Effective
  Conformal Theory and the Flat-Space Limit of AdS}},  {\em JHEP} {\bf 1107}
  (2011) 023, [\href{http://xxx.lanl.gov/abs/1007.2412}{{\tt
  arXiv:1007.2412}}].

\bibitem{Fitzpatrick:2012cg}
A.~L. Fitzpatrick and J.~Kaplan, {\it {AdS Field Theory from Conformal Field
  Theory}},  {\em JHEP} {\bf 1302} (2013) 054,
  [\href{http://xxx.lanl.gov/abs/1208.0337}{{\tt arXiv:1208.0337}}].

\bibitem{Kabat:2012hp}
D.~Kabat, G.~Lifschytz, S.~Roy, and D.~Sarkar, {\it {Holographic representation
  of bulk fields with spin in AdS/CFT}},  {\em Phys.Rev.} {\bf D86} (2012)
  026004, [\href{http://xxx.lanl.gov/abs/1204.0126}{{\tt arXiv:1204.0126}}].

\bibitem{Maldacena:2011mk}
J.~Maldacena, {\it {Einstein Gravity from Conformal Gravity}},
  \href{http://xxx.lanl.gov/abs/1105.5632}{{\tt arXiv:1105.5632}}.

\bibitem{Skenderis:2008dg}
K.~Skenderis and B.~C. van Rees, {\it {Real-time gauge/gravity duality:
  Prescription, Renormalization and Examples}},  {\em JHEP} {\bf 0905} (2009)
  085, [\href{http://xxx.lanl.gov/abs/0812.2909}{{\tt arXiv:0812.2909}}].

\end{thebibliography}\endgroup
\end{document}